\documentclass[journal]{IEEEtran}
\usepackage{cite}
\usepackage{amsmath,amssymb,amsfonts}
\usepackage{algorithmic}
\usepackage{graphicx,color}
\usepackage{textcomp}\usepackage{gensymb}
\usepackage{hyperref}
\usepackage{acronym}
\usepackage{balance}
\acrodef{ISAC}{integrated sensing and communication}    
\acrodef{ADC}{analog-to-digital converter}
\acrodef{AP}{access point}
\acrodef{DAC}{digital-to-analog converter}
\acrodef{AGV}{automated guided vehicle}
\acrodef{AI}{artificial intelligence}
\acrodef{AMR}{autonomous mobile robot}
\acrodef{API}{application programming interface}
\acrodef{ASIC}{application-specific integrated circuit}
\acrodef{BS}{base station}
\acrodef{CAMARA}{(api framework, aggregated level api)}
\acrodef{CIR}{channel impulse response}
\acrodef{CRB}{Cramér-Rao bound}
\acrodef{MCRB}{misspecified Cramér-Rao bound}
\acrodef{CSI}{channel state information}
\acrodef{CNN}{convolutional neural network}
\acrodef{DFTS-OFDM}{discrete fourier transform spread orthogonal frequency division multiplexing}
\acrodef{DL}{deep learning}
\acrodef{DIP}{direct path}
\acrodef{IDP}{indirect path}
\acrodef{ID}{identifier}
\acrodef{D-MIMO}{distributed multiple input multiple output}
\acrodef{DU}{distributed unit}
\acrodef{EO}{environmental object}
\acrodef{AoA}{angle-of-arrival}
\acrodef{ToA}{time-of-arrival}
\acrodef{ETSI}{European telecommunications standards institute}
\acrodef{FFT}{fast fourier transform}
\acrodef{FLOP}{floating point operation}
\acrodef{FR}{frequency range}
\acrodef{GPU}{graphics processing unit}
\acrodef{HAPS}{high altitude platform station}
\acrodef{KPI}{key performance indicator}
\acrodef{KVI}{key value indicator}
\acrodef{LCM}{life-cycle management}
\acrodef{LEO}{low earth orbit}
\acrodef{LiDAR}{light detection and ranging}
\acrodef{LoS}{line-of-sight}
\acrodef{MAC}{medium access control}
\acrodef{MIMO}{multiple input multiple output}
\acrodef{MISO}{multiple input single output}
\acrodef{ML}{machine learning}
\acrodef{MLOps}{machine learning operations}
\acrodef{mmWave}{millimeter wave}
\acrodef{NEF}{network exposure function}
\acrodef{NGMA}{next-generation multiple access}
\acrodef{NLoS}{non-line-of-sight}
\acrodef{NOMA}{non-orthogonal multiple access}
\acrodef{NTN}{non-terrestrial network}
\acrodef{OFDM}{orthogonal frequency division multiplexing}
\acrodef{PA}{power amplifier}
\acrodef{PAPR}{peak-to-average power ratio}
\acrodef{PHY}{physical layer}
\acrodef{QoS}{quality of service}
\acrodef{RAN}{radio access network}
\acrodef{RAT}{radio access technology}
\acrodef{RCS}{radar cross section}
\acrodef{RF}{radio frequency}
\acrodef{RIS}{reconfigurable intelligent surface}
\acrodef{RSMA}{rate-splitting multiple access}
\acrodef{RTO}{research and technology organization}
\acrodef{SME}{small and medium-sized enterprise}
\acrodef{Rx}{receiver}
\acrodef{TRx}{transceiver}
\acrodef{SCF}{sensing control function}
\acrodef{SC}{single carrier}
\acrodef{SeMF}{sensing management function}
\acrodef{SLAM}{simultaneous localization and mapping}
\acrodef{SLA}{service level agreement}
\acrodef{SNS}{smart networks and services}
\acrodef{SNR}{signal-to-noise ratio}
\acrodef{SPCTM}{sensing policy, consent, and transparency management}
\acrodef{UC}{use case}
\acrodef{SPF}{sensing processing function}
\acrodef{STRIDE}{spoofing, tampering, repudiation, information disclosure, denial-of-service, elevation of privilege}
\acrodef{SU}{sensing unit}
\acrodef{TR}{technical report}
\acrodef{Tx}{transmitter}
\acrodef{UAV}{unmanned aerial vehicle}
\acrodef{UE}{user equipment}
\acrodef{X-haul}{crosshaul (integration of fronthaul and backhaul)}
\acrodef{CIA}{confidentiality, integrity, availability}
\acrodef{DoS}{denial-of-service}
\acrodef{LINDDUN}{linkability, identifiability, non-repudiation, detectability, disclosure of information, unawareness, non-compliance}
\acrodef{PII}{personally identifiable information}
\acrodef{TEE}{trusted execution environment}
\acrodef{CAPIF}{common api framework}
\acrodef{NWaaP}{network-as-a-platform}
\acrodef{GNP}{generic network platform}
\acrodef{EDCA}{evolved data collection architecture}
\acrodef{FMCW}{frequency-modulated continuous-wave}
\acrodef{SeRS}{sensing radio signals}
\acrodef{SUSF}{sensing unit selection function}
\acrodef{IQ}{in-phase and quadrature}
\acrodef{NEF}{network exposure function}
\acrodef{SEAL}{service enabler architecture layer}
\acrodef{LO}{local oscillator}
\acrodef{PN}{phase noise}
\acrodef{PTRS}{phase tracking reference signal}
\acrodef{CP}{control plane}
\acrodef{DP}{data plane}
\acrodef{PoC}{proof-of-concept}
\acrodef{DoF}{degrees of freedom}
\acrodef{AWGN}{additive white Gaussian noise}
\acrodef{PPBP}{power-per-beam-pair}
\acrodef{MPJPE}{mean per joint position error}
\acrodef{FoV}{field of view}

\def\BibTeX{{\rm B\kern-.05em{\sc i\kern-.025em b}\kern-.08em
    T\kern-.1667em\lower.7ex\hbox{E}\kern-.125emX}}
\AtBeginDocument{\definecolor{ojcolor}{cmyk}{0.93,0.59,0.15,0.02}}

\IEEEoverridecommandlockouts
\usepackage{xcolor}
\usepackage{comment}
\usepackage{graphicx}
\usepackage{amsmath,amssymb}
\usepackage{multirow}
\usepackage[caption=false,font=footnotesize]{subfig}

\usepackage{enumitem}
\setlist{nosep}
\setlist[itemize]{noitemsep, topsep=0pt}
 \usepackage{textcomp} 
\usepackage{parskip}
\makeatother
\usepackage{booktabs,array}
\usepackage{calc} 
\usepackage{etoolbox}
\makeatletter
\patchcmd\longtable{\par}{\if@noskipsec\mbox{}\fi\par}{}{}
\makeatother
\usepackage{footnote}
\makesavenoteenv{longtable}
\usepackage{soul}

\usepackage{amsthm}

\newtheorem{remark}{Remark}
\usepackage{bookmark}
\usepackage{cleveref}

\newcommand{\rev}[1]{{#1}}

\begin{document}
\bstctlcite{IEEEexample:BSTcontrol}

\newcommand{\txrm}{{{\rm{T}}}}
\newcommand{\rxrm}{{{\rm{R}}}}
\newcommand{\comrm}{{{\rm{com}}}}
\newcommand{\radrm}{{{\rm{rad}}}}
\newcommand{\thn}[1]{ {#1^{\rm{th} } } }

\newcommand{\complexset}[2]{ \mathbb{C}^{#1 \times #2}  }
\newcommand{\complexsett}{ \mathbb{C}  }
\newcommand{\realset}[2]{ \mathbb{R}^{#1 \times #2}  }
\newcommand{\nset}{ \mathbb{N}  }
\newcommand{\zset}{ \mathbb{Z}  }

\newcommand{\trp}{^\top}
\newcommand{\herm}{^\mathrm{H}}

\newcommand{\boldzero}{{ {\boldsymbol{0}} }}
\newcommand{\boldone}{{ {\boldsymbol{1}} }}
\newcommand{\boldonet}{{ {\boldsymbol{1}}\trp }}
\newcommand{\Imatrix}{{ \boldsymbol{\mathrm{I}} }}

\newcommand{\realp}[1]{ \Re \left\{#1\right\}  }
\newcommand{\imp}[1]{ \Im \left\{#1\right\}  }
\newcommand{\vecc}[1]{ {\rm{vec}}\left(#1\right)  }
\newcommand{\veccs}[1]{ {\rm{vec}}\big(#1\big)  }
\newcommand{\veccb}[1]{ {\rm{vec}}\Big(#1\Big)  }

\newcommand{\ycom}{y^{{\rm{com}}}}
\newcommand{\hhcom}{\hh^{{\rm{com}}}}
\newcommand{\hhlos}{\hh^{{\rm{LOS}}}}
\newcommand{\hhnlos}{\hh^{{\rm{NLOS}}}}
\newcommand{\hnm}{ h_{n,m} }

\newcommand{\boldYcom}{\mathbf{Y}^{{\rm{com}}}}
\newcommand{\sigmacsq}{{ \sigma^2_{\rm{c}} }}

\newcommand{\ns}{ N_{\mathrm{s}} }

\newcommand{\aaa}{\mathbf{a}}
\newcommand{\cc}{ \mathbf{c} }
\newcommand{\bb}{ \mathbf{b} }
\newcommand{\nn}{ \mathbf{n} }
\newcommand{\hh}{ \mathbf{h} }
\newcommand{\hhfreqc}{\hh_{\rm{c,freq}}}
\newcommand{\uu}{ \mathbf{u} }

\newcommand{\bxi}{ \boldsymbol{\xi} }
\newcommand{\bxir}{ \boldsymbol{\xi}_{\rm{R}} }

\newcommand{\ffb}{ \mathbf{f} }
\newcommand{\ww}{ \mathbf{w} }
\newcommand{\vv}{ \mathbf{v} }

\newcommand{\Tsym}{ T_{\rm{sym}} }
\newcommand{\Tcp}{ T_{\rm{cp}} }
\newcommand{\Ttot}{ T_{\rm{tot}} }
\newcommand{\Ttotw}{ \widebar{T}_{\rm{tot}} }

\newcommand{\Rmax}{ R_{\rm{max}} }
\newcommand{\Rmaxcp}{ R_{\rm{max, CP}} }
\newcommand{\vmax}{ v_{\rm{max}} }
\newcommand{\conj}[1]{ {#1}^{\ast} }

\newcommand{\uutau}{ \uu_{\tau} }
\newcommand{\uutautilde}{ \widetilde{\uu}_{\tau} }
\newcommand{\uutaut}{ \uutau\trp }
\newcommand{\uutautildet}{ \uutautilde\trp }

\newcommand{\uunu}{ \uu_{\nu} }
\newcommand{\uunutilde}{ \widetilde{\uu}_{\nu} }
\newcommand{\uutaunu}{ \uu_{\tau \nu} }

\newcommand{\uunut}{ \uunu\trp }
\newcommand{\uunutildet}{ \uunutilde\trp }
\newcommand{\uutaunut}{ \uutaunu\trp }

\newcommand{\cfo}{ \delta_f }
\newcommand{\bias}{ \delta_\tau }
\newcommand{\cfoc}{ \delta'_f }

\newcommand{\deltaf}{ \Delta f }
\newcommand{\deltafc}{ \Delta \fc }
\newcommand{\fc}{ f_c }
\newcommand{\Ntx}{ N_\txrm }
\newcommand{\Nrx}{ N_\rxrm }
\newcommand{\atx}{ \aaa_{\rm{T}} }
\newcommand{\arx}{ \aaa_{\rm{R}} }
\newcommand{\arxc}{ \aaa_{\rm{C}} }

\newcommand{\Ctaub}{ \widebar{\Ctau} }
\newcommand{\Cnub}{ \widebar{\Cnu} }
\newcommand{\Ctaunub}{ \widebar{\Ctaunu} }
\newcommand{\deltataub}{ \widebar{\deltattau} }
\newcommand{\deltanub}{ \widebar{\deltanu} }

\newcommand{\Xcal}{\mathcal{X}}

\newcommand{\Ctau}{ C_{\tau} }
\newcommand{\Cnu}{ C_{\nu} }
\newcommand{\Ctaunu}{ C_{\tau \nu} }

\newcommand{\Nt}{ N_{\rm{T}} }
\newcommand{\Nss}{ N_{\rm{S}}  }
\newcommand{\Nc}{ N_{\rm{C}}  }

\newcommand{\ynm}{ y_{n,m} }
\newcommand{\xnm}{ x_{n,m} }
\newcommand{\pnm}{ P_{n,m} }

\newcommand{\XX}{ \mathbf{X} }
\newcommand{\PP}{ \mathbf{P} }
\newcommand{\RR}{ \mathbf{R} }
\newcommand{\WW}{ \mathbf{W} }
\newcommand{\VV}{ \mathbf{V} }
\newcommand{\DD}{ \mathbf{D} }
\newcommand{\YY}{ \mathbf{Y} }
\newcommand{\ZZ}{ \mathbf{Z} }
\newcommand{\JJ}{ \mathbf{J} }
\newcommand{\HH}{ \mathbf{H} }
\newcommand{\FF}{ \mathbf{F} }

\newcommand{\YYr}{ \YY_{\rm{r}} }
\newcommand{\YYc}{ \YY_{\rm{c}} }

\newcommand{\HHc}{ \HH_{\rm{c}} }
\newcommand{\ZZc}{ \ZZ_{\rm{c}} }
\newcommand{\ZZr}{ \ZZ_{\rm{r}} }
\newcommand{\sigmac}{ \sigma_{\rm{C}} }
\newcommand{\sigmas}{ \sigma_{\rm{S}} }
\newcommand{\sigmarcs}{ \sigma_{\rm{RCS}} }

\newcommand{\Nsensor}{ N_{\rm{sensor}} }
\newcommand{\Ntar}{ N_{\rm{target}} }
\newcommand{\Ns}{N_s}
\newcommand{\Nb}{N_{{\rm{bit}}}}
\newcommand{\Nf}{N_{{\rm{F}}}}
\newcommand{\Lt}{L_\txrm}
\newcommand{\Lr}{L_\rxrm}
\newcommand{\bF}{\mathbf{F}}
\newcommand{\Frf}{\bF_{\rm{RF}}}
\newcommand{\Fbb}{\bF_{\rm{BB}}}
\newcommand{\bW}{\mathbf{W}}
\newcommand{\Wrf}{\bW_{\rm{RF}}}
\newcommand{\Wbb}{\bW_{\rm{BB}}}

\newcommand{\bz}{\mathbf{z}}
\newcommand{\by}{\mathbf{y}}
\newcommand{\xx}{\mathbf{x}}
\newcommand{\tildex}{\tilde{\bx}}
\newcommand{\bH}{\mathbf{H}}

\newcommand{\boldUps}{ \mathbf{\Upsilon} }
\newcommand{\boldUpstau}{ \boldUps_{\tau} }
\newcommand{\boldUpsnu}{ \boldUps_{\nu} }
\newcommand{\boldUpsboth}{ \boldUps_{\tau \nu} }

\newcommand{\taubar}{\bar{\tau}}
\newcommand{\nubar}{\bar{\nu}}

\newcommand{\AT}{\mathbf{A}_\txrm }
\newcommand{\AR}{\mathbf{A}_\rxrm }
\newcommand{\btheta}{\boldsymbol \theta}
\newcommand{\bphy}{\boldsymbol \phy}
\newcommand{\Hrnm}{\mathbf{H}_{\text{r},n,m}}

\newcommand{\pd}{ P_{\rm{d}} }
\newcommand{\pfa}{ P_{\rm{fa}} }

\newcommand{\etabhat}{ \widehat{\boldsymbol{\eta}} }
\newcommand{\etab}{ {\boldsymbol{\eta}} }

\newcommand{\sml}{ s_{m,\ell} }
\newcommand{\sm}{ s_{m} }

\newcommand{\rect}[1]{ { \rm{rect} }\left(#1\right) }

\newcommand{\mtCN}{{\mathcal{CN}}}
\newcommand{\ptot}{P_{\mathrm{tot}}}

\newcommand{\Hzero}{\mathcal{H}_0}
\newcommand{\Hone}{\mathcal{H}_1}

\newcommand{\norm}[1]{\left\lVert#1\right\rVert}
\newcommand{\normsmall}[1]{\big\lVert#1\big\rVert}
\newcommand{\normbig}[1]{\Big\lVert#1\Big\rVert}

\newcommand{\normsq}[1]{ \norm{#1}^{2} }
\newcommand{\normf}[1]{ \norm{#1}_{\rm{F}}^{2} }
\newcommand{\normfsmall}[1]{ \normsmall{#1}_{\rm{F}}^{2} }

\newcommand{\Eee}{\mathbb{E}}
\newcommand{\quot}[1]{``{#1}''}

\newcommand{\diag}[1]{ {\rm{diag}}\left(#1\right)  }
\newcommand{\ddiag}[1]{ {\rm{ddiag}}\left(#1\right)  }
\newcommand{\blkdiag}[1]{ {\rm{blkdiag}}\left(#1\right)  }
\newcommand{\ncoll}[1]{N_{\text{C}}\left[ #1 \right] }
\newcommand{\nmiss}[1]{N_{\text{M}}\left[ #1 \right] }

\newcommand{\cqun}{q_{u,n}}
\newcommand{\cqsn}{q_{s,n}}
\newcommand{\cqin}{q_{i,n}}
\newcommand{\cqsnl}{q_{s,n}^{(l)}}
\newcommand{\cqinl}{q_{i,n}^{(l)}}

\newcommand{\snr}{{\rm{SNR}}}
\newcommand{\gammasnr}{\gamma_{\rm{SNR}}}

\newcommand{\tr}[1]{{#1}^\mathrm{T}}
\renewcommand{\vec}[1]{\bm{\mathrm{#1}}}
\newcommand{\E}[1]{\mathbb{E}\left[ #1 \right]}
\newcommand{\Earg}[2]{\mathbb{E}_{#1}\left[ #2 \right]}
\newcommand{\var}[1]{\mathrm{Var}\left[{#1}\right]}
\newcommand{\svec}[1]{\bm{#1}}
\newcommand{\re}[1]{\text{Re}\left\{ #1 \right\}}
\newcommand{\pr}[1]{\text{Pr}\left\{#1\right\}}
\newcommand{\sigmoid}[1]{\sigma\left( #1 \right)}

\newcommand{\ones}{\vec{1}}
\newcommand{\zeros}{\vec{0}}

\newcommand{\indic}[1]{{\ones}_{\{#1\}}}

\newcommand{\s}{ \mathbf{s} }
\newcommand{\z}{ \mathbf{z} }
\newcommand{\p}{ \mathbf{p} }
\newcommand{\y}{ \mathbf{y} }

\makeatletter
\newcommand*\rel@kern[1]{\kern#1\dimexpr\macc@kerna}
\newcommand*\widebar[1]{%
  \begingroup
  \def\mathaccent##1##2{%
    \rel@kern{0.8}%
    \overline{\rel@kern{-0.8}\macc@nucleus\rel@kern{0.2}}%
    \rel@kern{-0.2}%
  }%
  \macc@depth\@ne
  \let\math@bgroup\@empty \let\math@egroup\macc@set@skewchar
  \mathsurround\z@ \frozen@everymath{\mathgroup\macc@group\relax}%
  \macc@set@skewchar\relax
  \let\mathaccentV\macc@nested@a
  \macc@nested@a\relax111{#1}%
  \endgroup
}
\makeatother



\title{Cross-layer Integrated Sensing and Communication: \\ A Joint Industrial and Academic Perspective}
\author{
Henk Wymeersch\IEEEauthorrefmark{1}, 
Nuutti Tervo\IEEEauthorrefmark{2}, 
Stefan Wänstedt\IEEEauthorrefmark{3}, 
Sharief Saleh\IEEEauthorrefmark{1},
	Joerg Ahlendorf\IEEEauthorrefmark{17},
	Ozgur Akgul\IEEEauthorrefmark{14},
        Vasileios Tsekenis\IEEEauthorrefmark{6},
	Sokratis Barmpounakis\IEEEauthorrefmark{6},
	Liping Bai\IEEEauthorrefmark{1},
	Martin Beale\IEEEauthorrefmark{8},
	Rafael Berkvens\IEEEauthorrefmark{10,11},
        Nabeel Nisar Bhat\IEEEauthorrefmark{10,11},
        Hui Chen\IEEEauthorrefmark{1},
    Shrayan Das\IEEEauthorrefmark{2},
	Claude Desset\IEEEauthorrefmark{11},
	Antonio de la Oliva\IEEEauthorrefmark{9},
	Prajnamaya Dass\IEEEauthorrefmark{4},
	Jeroen Famaey\IEEEauthorrefmark{10,11},
	Hamed Farhadi\IEEEauthorrefmark{3},
	Gerhard P. Fettweis\IEEEauthorrefmark{15},
	Yu Ge\IEEEauthorrefmark{1},
	Hao Guo\IEEEauthorrefmark{1,7},
	Rreze Halili\IEEEauthorrefmark{10,11},
    Katsuyuki Haneda\IEEEauthorrefmark{20},
	Abdur Rahman Mohamed Ismail\IEEEauthorrefmark{11,12},
    Akshay Jain \IEEEauthorrefmark{14},
    Sylvaine Kerboeuf \IEEEauthorrefmark{18},
	Musa Furkan Keskin\IEEEauthorrefmark{1},
    Emad Ibrahim\IEEEauthorrefmark{3},
	Bilal Khan\IEEEauthorrefmark{2},
	Siddhartha Kumar\IEEEauthorrefmark{5},
	Stefan Köpsell\IEEEauthorrefmark{4},
	Apostolos Kousaridas\IEEEauthorrefmark{13},
    Pekka Kyösti\IEEEauthorrefmark{2,19},
	Simon Lindberg\IEEEauthorrefmark{5},
	Mohammad Hossein Moghaddam\IEEEauthorrefmark{5},
	Ahmad Nimr\IEEEauthorrefmark{15},
	Victor Pettersson\IEEEauthorrefmark{1},
	Aarno Pärssinen\IEEEauthorrefmark{2},
	Basuki Priyanto\IEEEauthorrefmark{8},
	Athanasios Stavridis\IEEEauthorrefmark{3},
	Tommy Svensson\IEEEauthorrefmark{1},
	Sonika Ujjwal\IEEEauthorrefmark{16}\\

\IEEEauthorrefmark{1}{Department of Electrical Engineering, Chalmers University of Technology, Gothenburg, Sweden}, 
\IEEEauthorrefmark{2}{Centre for Wireless Communications - Radio Technologies (CWC-RT), University of Oulu, Oulu, Finland}, 
\IEEEauthorrefmark{3}{Ericsson, Sweden}, 
\IEEEauthorrefmark{4}{Barkhausen Institut, Dresden, Germany}, 
\IEEEauthorrefmark{5}{Qamcom Research and Technology AB, Gothenburg, Sweden}, 
\IEEEauthorrefmark{6}{WINGS ICT Solutions SA, Greece}, 
\IEEEauthorrefmark{7}{NYU Wireless Research Center, New York University (NYU), Brooklyn, NY, USA}, 
\IEEEauthorrefmark{8}{Sony Europe,  Lund, Sweden}, 
\IEEEauthorrefmark{9}{University Carlos III of Madrid, Spain}, 
\IEEEauthorrefmark{10}{IDLab, University of Antwerp, Belgium}, 
\IEEEauthorrefmark{11}{IMEC, Belgium}, 
\IEEEauthorrefmark{12}{TELIN-Ghent University, Belgium}, 
\IEEEauthorrefmark{13}{Nokia, Germany}, 
\IEEEauthorrefmark{14}{Nokia, Finland}, 
\IEEEauthorrefmark{15}{TU Dresden, Germany}, 
\IEEEauthorrefmark{16}{Ericsson, Finland}, 
\IEEEauthorrefmark{17}{NXP, Germany}, 
\IEEEauthorrefmark{18}{Nokia, France}, 
\IEEEauthorrefmark{19}{Keysight Technologies, Finland}, 
\IEEEauthorrefmark{20}{Aalto University, Finland}

\thanks{This work is supported by the European Commission through the Horizon Europe/JU SNS project Hexa-X-II (Grant Agreement no. 101095759).}}

\maketitle

\begin{abstract}
Integrated sensing and communication (ISAC) enables radio systems to simultaneously sense and communicate with their environment. This paper, developed within the Hexa-X-II project funded by the European Union, presents a comprehensive cross-layer vision for ISAC in 6G networks, integrating insights from physical-layer design, hardware architectures, AI-driven intelligence, and protocol-level innovations. We begin by revisiting the foundational principles of ISAC, highlighting synergies and trade-offs between sensing and communication across different integration levels. Enabling technologies (such as multiband operation, massive and distributed MIMO, non-terrestrial networks, reconfigurable intelligent surfaces, and machine learning) are analyzed in conjunction with hardware considerations including waveform design, synchronization, and full-duplex operation. To bridge implementation and system-level evaluation, we introduce a quantitative cross-layer framework linking design parameters to key performance and value indicators. 
By synthesizing perspectives from both academia and industry, this paper outlines how deeply integrated ISAC can transform 6G into a programmable and context-aware platform supporting applications from reliable wireless access to autonomous mobility and digital twinning.
\end{abstract}

\begin{IEEEkeywords}
    6G, integrated sensing and communication, joint sensing and communication, cross-layer. 
\end{IEEEkeywords}
\acresetall 

\maketitle

\section{Introduction}\label{introduction}

The evolution toward 6G wireless technology is expected to redefine the landscape of connectivity by merging communication and sensing into an integrated framework known as  \ac{ISAC}. This paradigm envisions a future where communication networks are not only conduits for data transfer, but also provide real-time environmental awareness and situational intelligence (a recent deliverable can be found here \cite{hex24_d43}). Through the integration of sensing capabilities within communication infrastructures, \ac{ISAC} holds the promise of substantial advances in functionality and adaptability, empowering a wide array of applications across sectors such as autonomous systems, industrial automation, transportation, and smart infrastructure.

\rev{\ac{ISAC} in the context of 6G has been extensively treated in the scientific literature \cite{Fan:22,gonzalez2024integrated}, covering various aspects, such as signal design in time, frequency, and space,  as well as signal processing for channel parameter estimation, detection, localization, and tracking. \ac{ISAC} has also been studied in terms of  various candidate 6G enablers, including various frequency bands \cite{katwe2024cmwave}, \ac{NTN} \cite{dureppagari2023ntn}, \ac{MIMO} \cite{gao2022integrated}, \ac{D-MIMO} \cite{guo2024integrated},  \ac{AI} \cite{demirhan2023integrated}, and \acp{RIS} \cite{zhong2023joint}.}

\rev{
In recent years, ISAC has attracted significant research interest. 
For example, at the physical layer, active RISs have been employed to enhance the physical-layer security in  \ac{ISAC} systems, dynamically controlling the propagation environment to improve performance \cite{active_ris}. At the access/protocol layer, \ac{NOMA} schemes have been leveraged to enhance spectral efficiency and support joint sensing and communication operations \cite{noma_isac}, while \ac{RSMA} techniques enable efficient multi-user communication and sensing integration \cite{rsma_isac}. \Ac{NGMA} methods have been proposed to support a massive number of communication users and sensing targets, offering robust access solutions for \ac{ISAC} systems \cite{ngma_isac}. Going higher at the network layer, cooperative multi-cell \ac{ISAC} systems  have been explored for their ability to manage complex multi-user and multi-target scenarios, optimizing communication and sensing trade-offs \cite{ris_multicell}. Resource allocation and target assignment strategies in multi-\ac{BS} cooperative \ac{ISAC} systems have improved performance in tasks such as \ac{UAV} detection \cite{uav_isac}.}

\rev{These above-mentioned examples demonstrate the diverse methodologies and innovative approaches used to optimize \ac{ISAC} systems across different layers. However, an overall assessment from a practical viewpoint, considering not just one layer, but also hardware requirements, functions, interfaces, and protocols, as well as security and privacy has been left unaddressed. A holistic understanding of these different facets is crucial to design, analyze, and compare \ac{ISAC} solutions in a fair way considering not only the physical layer, but also lower and higher layers. \textit{The understanding we describe in this paper aims to show how physical layer properties influence those lower and higher layers. }}

This paper aims to make progress towards filling this gap and provides a cross-layer vision for \ac{ISAC} in 6G, by the Hexa-X-II project (see \url{https://hexa-x-ii.eu/}). Hexa-X-II is a European 6G flagship project, which developed the 6G vision and fundamental concepts, including key technology enablers. Hexa-X-II collects inputs and contributions from its partners, which include academia and \acp{RTO}, \acp{SME}, and large industry, spanning the 6G ecosystem, including chipsets and hardware, systems and technology, platforms, applications and software, and service providers. This paper thus presents the broad vision across both layers and across academia and industry. 

\begin{figure}
    \centering
    \includegraphics[width=0.99\linewidth]{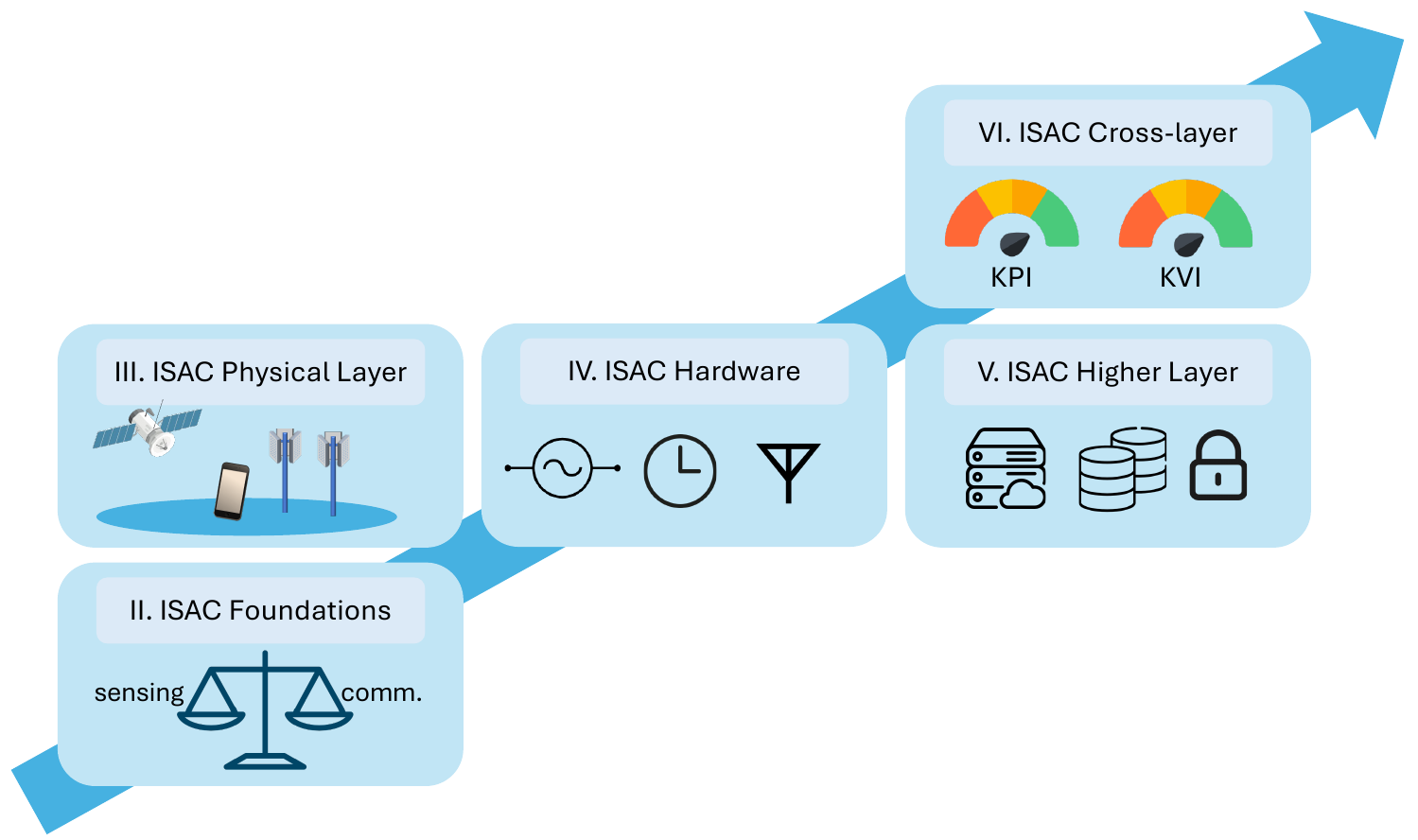}
    \caption{Overview of this paper, starting from the \ac{ISAC} foundations (motivation and use cases) over the \ac{ISAC}  physical layer (relation to the various 6G enablers and sensing configurations), \ac{ISAC} hardware requirements, higher layer aspects of \ac{ISAC}, and cross-layer study of \ac{ISAC}. This holistic analysis shows how radio-level trade-offs enable higher-layer synergies.}
    \label{fig:overview}
\end{figure}

\subsection*{\rev{Methodology}}
This paper considers \textit{candidate physical layer enablers for \ac{ISAC}}  (e.g., frequency bands, \ac{NTN}, massive \ac{MIMO}, \ac{D-MIMO},  \ac{AI}, and \ac{RIS}), addressing essential factors like propagation characteristics, synchronization challenges, and hardware constraints, as well as \textit{\ac{ISAC} hardware requirements}, including full-duplex communication, phase coherence, and array structures, which are crucial for the effective deployment of \ac{ISAC} systems, and \textit{higher layers of \ac{ISAC}}, encompassing protocols, sensing functions, privacy, and security, are explored to address the broader integration challenges and ensure a seamless user experience. 
Additionally, a cross-layer evaluation offers insights into the interdependencies between system design parameters, \acp{KPI}, and \acp{KVI} \cite{wymeersch2025wcm} critical to optimizing \ac{ISAC} operations. This paper builds on the earlier works \cite{wymeersch2021integration,Wymeersch24Joint}, by providing  more detail and depth. 

It is important to notify the reader that this paper is \textit{neither a survey nor a tutorial} and that more comprehensive reviews are available in the technical literature (e.g., \cite{WOS:001377058300001, WOS:001150230200001, WOS:001276356800001, WOS:000934534800001, WOS:001030583300001, WOS:001004192500001, WOS:000848033700014}). Moreover, while this paper cuts across different enablers and layers, there are gaps (e.g., we do not cover \ac{NTN} hardware or higher-layer protocols). These gaps are due to the interests and priorities of the partners within the Hexa-X-II project. \rev{Finally, note that the literature review is spread throughout this paper, rather than restricted to the introduction, due to the large number of topics that are covered. }

\subsection*{\rev{Contributions}}
\rev{The specific contributions of this work are the following: 
  \begin{itemize}
    	\item 	\textbf{Cross-layer perspective on ISAC:}
This paper provides a comprehensive cross-layer vision for ISAC in 6G, combining physical-layer enablers, hardware considerations, protocol-level aspects, and application-layer functions. Unlike previous surveys that focus primarily on isolated layers or components, this work outlines how design trade-offs and synergies propagate across the full system stack, from waveform and synchronization design to sensing \acp{API} and service exposure. Moreover, aspects such as power amplified nonlinearity, phase noise and anchor calibration are treated in some depth, unlike the previous ISAC literature.
	\item 	\textbf{Integration of industrial and academic insights:}
Drawing from the diverse partners in the Hexa-X-II project, the paper synthesizes perspectives from academia, \acp{SME}, and major industry stakeholders to define realistic implementation challenges and enablers for ISAC. This dual-perspective approach bridges theory and practice, highlighting both cutting-edge research ideas and industrial feasibility, which is rarely combined in the existing literature.
	\item 	\textbf{Quantitative cross-layer evaluation:}
A third contribution is the introduction of a cross-layer evaluation framework that links low-level design parameters to both KPIs and high-level \acp{KVI}. This framework enables a holistic understanding of ISAC trade-offs, particularly in resource allocation, sensing accuracy and resolution, latency, and efficiency, supporting system-level decision-making. The framework is applied to generate results from simulation and \ac{PoC} setups for various environments (indoor, urban, and rural), and using different frequency ranges (FR2 and FR3).
\end{itemize}}

\subsection*{Structure of the Paper}

The remainder of this paper is structured as follows (see also Fig.~\ref{fig:overview}): Section \ref{foundations} provides the definition of \ac{ISAC}, considering different levels of integration, its motivation in 6G, and the fundamental trade-offs and synergies. Section \ref{sec:JCASPHY} goes into the physical layer of \ac{ISAC}, relating it to the main 6G enablers and considering the various sensing configurations from a practical perspective. Section \ref{hardware-considerations-of-jcas} considers the hardware aspects of \ac{ISAC}, going deeper into the challenges and opportunities. The higher layer aspects of \ac{ISAC} are considered in Section \ref{protocols-and-functions-for-jcas}, including functions, protocols, exposure, as well as security and privacy. Section \ref{a-quantitative-cross-layer-evaluation} provides a cross-layer study of \ac{ISAC}, yielding quantitative insights into how the different use cases may be supported. Finally, Section \ref{conclusion} provides our conclusions, which we hope provides inputs to the ongoing standardization activities in various standardization organizations, such as \ac{ETSI} and 3GPP.

\section{ISAC Foundations}\label{foundations}

In this section, we review the foundations of ISAC, covering the different ways sensing and communication can be integrated, some envisioned applications for ISAC, and the trade-offs between communication and sensing. We also discuss how ISAC performance can be measured, different channel models, and how ISAC can work in an end-to-end 6G systems.

\subsection{What is ISAC?}\label{what-is-jcas}
\begin{figure}
    \centering
    \includegraphics[width=0.90\linewidth]{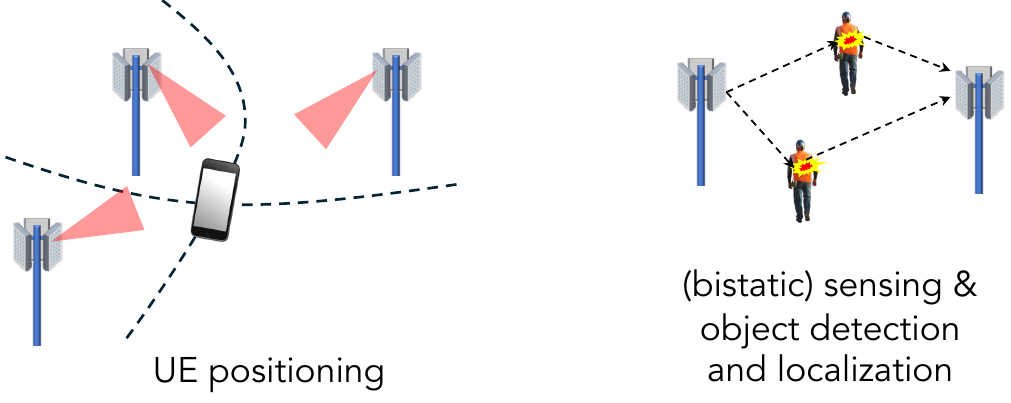}
    \caption{User positioning, sensing, and object localization.}
    \label{fig:localizationAndSensing}
\end{figure}

ISAC represents an advanced new paradigm in wireless
systems, integrating communication and sensing functionalities \cite{Zhang_Survey_ST_2022}. This
integration allows for the re-use of hardware components and conservation
of resources, and provides cross-functional benefits, transcending the
constraints of traditional systems that treat these functions
separately \cite{LiuSPM2023}. The core of ISAC lies in its ability to combine multiple
operational roles into a single framework. The `communication function' in ISAC is primarily responsible for ensuring reliable wireless signal transmissions. It leverages techniques such as coding, modulation, and multi-antenna systems to enable efficient information exchange, even under challenging environments such as deep fading and interference-rich environments. The communication function works closely with the sensing function by sharing spectrum and hardware resources to achieve a seamless integration. In this context, `sensing
functionalities' include three primary aspects (See Fig.~\ref{fig:localizationAndSensing}): sensing, positioning,
and localization  \cite{hex24_d43}. \emph{Sensing}~involves the
transmission, reception, and processing of radio signals to gather information
pertinent to a specific service related to the physical awareness of the surrounding environment. This process can utilize information (including the geometry)
of the \ac{Tx}, \ac{Rx}, and the environment, as typically seen in
radar applications. However, some sensing tasks, like pollution
monitoring, may not depend on such spatial information.
\emph{Positioning}~determines the geometric state (typically the location coordinates, but possibly also the orientation and velocity) of a connected
device in a certain coordinate system using signal metrics, so it
relies on sensing. Finally, in the context of this paper,
\emph{localization}~extends beyond positioning by also estimating the
location {and velocity} of targets in a certain frame of
reference, enabling what is known as device-free localization. Localization is a part of \textit{radar-like sensing}, which includes the detection of targets, localizing and tracking them, and characterizing them in terms of their properties (e.g., material or type).

There are
different levels of integration in ISAC (see Fig.~\ref{fig:reuse}) \cite{EAB24}:
\begin{figure}
    \centering
    \includegraphics[width=0.99\linewidth]{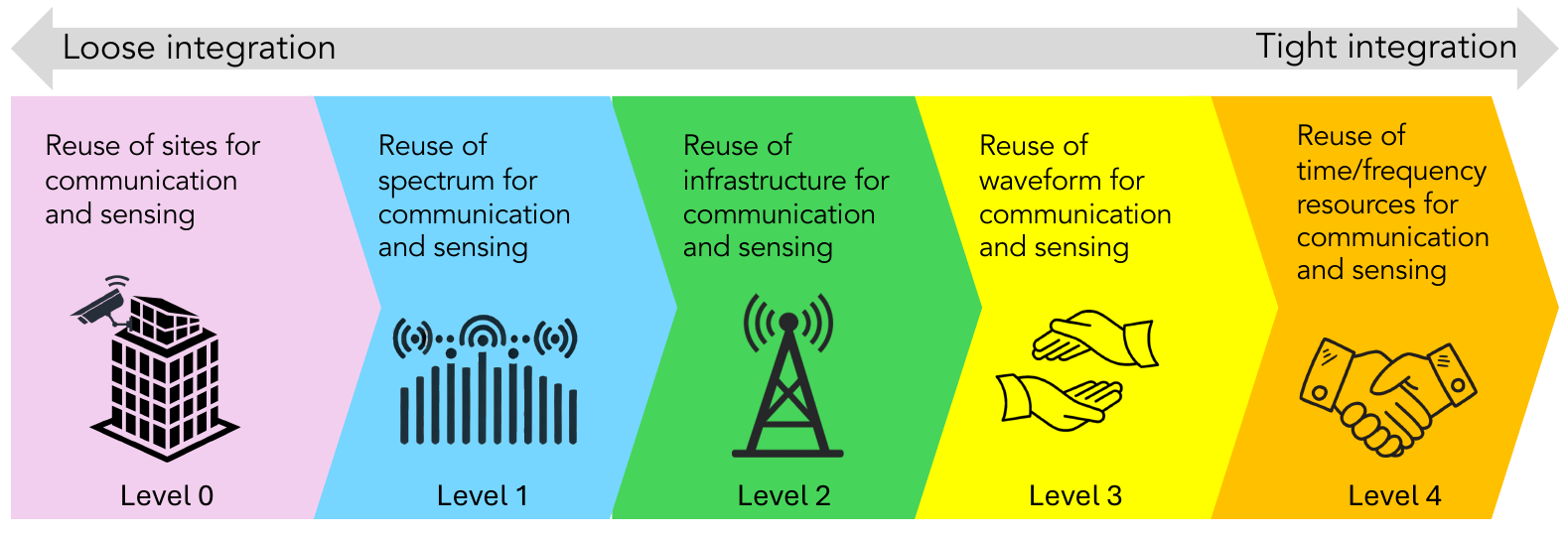}
    \caption{Ericsson proposed several levels of ISAC integration, which can each have their own target applications and benefits. The Hexa-X-II project primarily focuses on levels 3 and 4. }
    \label{fig:reuse}
\end{figure}
\begin{itemize}
\item
  \emph{Level 0 - Integration of Sites:} At this level, external sensors
  (such as cameras or radar systems) are added to \ac{BS} sites or
  user devices, providing auxiliary data to enhance communication
  functions, such as beamforming. This level of integration will not be further considered in this paper. 
\item
  \emph{Level 1 - Integration of Spectrum:} This level ensures that
  sensing and communication services share the same spectral resources,
  emphasizing efficient coexistence.
\item
  \emph{Level 2 - Integration of Hardware:} Common radio hardware
  components, including oscillators, antennas, \acp{DAC}, and \acp{ADC}, analog-to-digital are utilized for both sensing and communication
  tasks, optimizing hardware use.
\item
  \emph{Level 3 - Integration of Waveforms:} Advanced integration is
  achieved when the same waveform, such as \ac{OFDM}, is employed for both sensing and communication
  purposes \cite{ISAC_survey_IoT_2023}. At Level 3, dedicated optimized pilots are used for sensing, similar to positioning pilots in 5G.
\item
  \emph{Level 4 - Integration of Radio Resources:} The most integrated
  level, where the same time and frequency resources are allocated to
  both sensing and communication, ensuring maximal resource utilization efficiency and
  synergy. Such integration is inherent to monostatic sensing,\footnote{The different sensing configurations are covered in detail in Section \ref{analysis-of-jcas-modalities}.} since the
  sensing transmitter and receiver are collocated and share knowledge of
  the data. It is also possible for bistatic or multi-static sensing,
  where the data is either shared via a side-channel or detected by the
  communication receiver \cite{SPM_MC_ISAC_2024}.
\end{itemize}

The Hexa-X-II project primarily focuses on Levels 3 and 4, aiming to
fully exploit the potential of waveform and resource integration to
achieve unprecedented sensing and communication efficiency and performance using communication wireless systems.
This deep integration not only enhances operational capabilities but
also paves the way for innovative applications across various sectors.

\subsection{Why ISAC?}\label{why-jcas}

\begin{figure}
    \centering
    \includegraphics[width=0.99\linewidth]{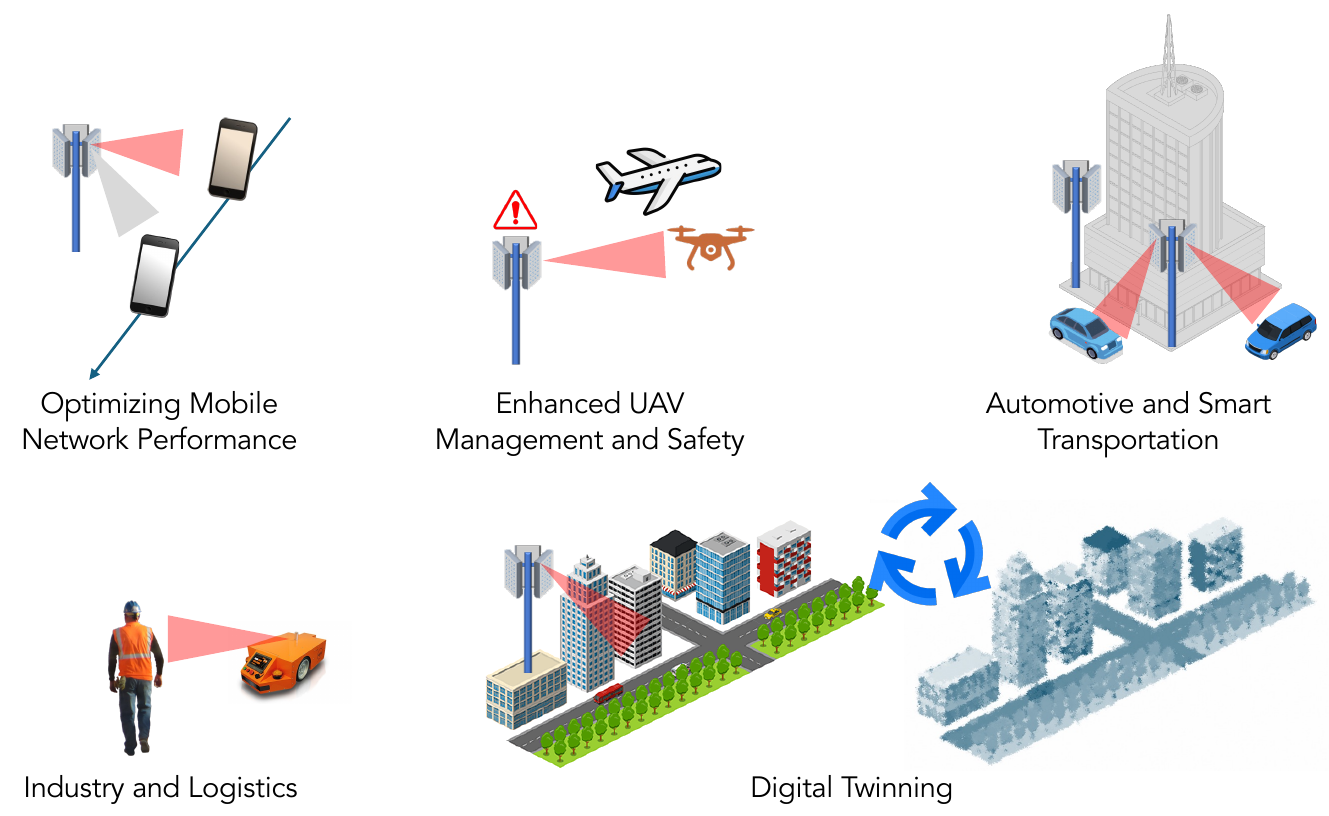}
    \caption{Examples of ISAC use cases.}
    \label{fig:use-cases}
\end{figure}

The driver for adopting ISAC emerges from its versatile utility across
various sectors, including the following \cite{hex23_d12}, visualized in Fig.~\ref{fig:use-cases}. 
\begin{enumerate}[label=(\Alph*)]
\item
  \emph{Optimizing Mobile Network Performance:} In telecommunications,
  ISAC can  contribute to optimizing network performance,  particularly with \ac{mmWave} communications that require  unobstructed \ac{LoS} or communications with predictable movements of network nodes\cite{Bayraktar2023ArXiV}. By preemptively identifying potential interruptions in \ac{LoS}, such as approaching vehicles, ISAC allows networks to anticipate changes in the communication channel, either by adjusting the beam  direction or by triggering a handover,  thereby maintaining seamless service continuity \cite{guo2022highrate}. Similarly, by predicting future locations and orientations of devices, time-consuming beam training can be avoided, increasing communication efficiency \cite{guo2021predictor}. ISAC can also be used to detect changes in environmental conditions (e.g., rain) and adapt transmission strategies accordingly. 
\item
  \emph{Enhanced \ac{UAV} Management and Safety:} ISAC is expected to play a
  pivotal role in the detection and tracking of \acp{UAV}, particularly in safeguarding critical infrastructure \cite{uav_isac}.
  Utilizing existing mobile communication networks to monitor \ac{UAV}
  activity in restricted airspace ensures security while supporting \ac{UAV}
  traffic management \cite{JRC_UseCases_2024}. This integration aids in navigation and collision
  avoidance, enhancing the functionality of \acp{UAV} beyond their
  traditional capabilities.

\item
  \emph{Automotive and Smart Transportation:} In the automotive
  industry, while onboard sensors like radar, \ac{LiDAR}, and cameras provide
  immediate situational awareness, their scope is limited to direct
  \ac{LoS}. {Note that also the performance of \ac{LiDAR} and camera can significantly deteriorate in severe weather conditions. Due to the plethora of nodes and connectivity,} ISAC technology can extend this capability by offering
  broader, beyond the direct \ac{LoS} information, such as detecting pedestrians
  or animals on highways, or ``seeing around the corner" \cite{xu2024seearoundthecorner,rastorgueva2024millimeter}, thus
  significantly improving road safety and traffic management \cite{veh_ISAC_resources_2024}. Similarly,
  in rail transportation, ISAC can be integral in enhancing safety
  measures \cite{HSR_PSO_ISAC_2024}. Dedicated mobile networks can detect and communicate hazards
  like obstacles or trespassers on the tracks to train control systems,
  enabling timely preventive actions and reducing the risk of accidents.
\item
  \emph{Industry Automation:} In the context of industry automation, ISAC can support the navigation and operational efficiency
  of \acp{AGV} and \acp{AMR} \cite{MM_ISAC_2024}. By leveraging  wireless networks tailored for
  industrial applications, ISAC facilitates real-time collision
  avoidance and enhances safety protocols where human workers interact
  with automated machines \cite{MIMO_ISAC_Models_2023}. This may include both private (e.g., inside a factory) or public (e.g., across outdoor facilities) networks. 
\item
  \emph{Digital Twinning:} ISAC  will be instrumental in creating digital twins—dynamic digital representations of physical entities or environments that evolve over time with real-world data \cite{DT_ISAC_2024}. This capability is crucial for applications ranging from network performance simulations in telecommunications to predictive fault monitoring in industrial systems. A particular class of these models, known as \ac{RF} digital twins, virtualizes radio-frequency environments and enables real-time emulation of wireless propagation conditions, antenna configurations, and device interactions. ISAC-enabled \ac{RF} digital twins support adaptive multi-domain optimization by fusing real-time \ac{RF} and non-\ac{RF} data for reliable, low-latency decision-making across diverse applications \cite{yang2024joint, ghaderi2024site, goudy2024network, salehihikouei2024leveraging}. 
\end{enumerate}

These use cases are summarized in terms of exemplifying requirements in
Table \ref{tab:use-cases}.

\begin{table}
\centering
\caption{ISAC use cases and example requirements. The use case are (A)Optimizing Mobile Network Performance, (B) Enhanced \ac{UAV} Management and Safety, (C) Automotive and Smart Transportation, (D) Industry Automation, (E) Digital Twinning.}
\begin{tabular}{|>{\raggedright}p{0.05\linewidth}|p{0.15\linewidth}|p{0.15\linewidth}|>{\raggedright}p{0.15\linewidth}|p{0.15\linewidth}|}
\hline
\textbf{Use case} & \textbf{Sensing accuracy} & \textbf{Sensing latency} & \textbf{Mobility} & \textbf{Sensing coverage} \\
\hline
(A) & 1 -- 10 m & 10 ms & \textless{} 10 m/s & 100 -- 1000 m \\
\hline
(B) & 1 -- 10 m & 100 ms & \textless{} 50 m/s & 1000 -- 10000 m \\
\hline
(C) & 0.1 -- 1 m & 10 ms & \textless{} 40 m/s & 10 -- 100 m \\
\hline
(D) & 0.1 -- 1 m & 1 ms & \textless{} 5 m/s & 10 -- 100 m \\
\hline
(E) & 0.1 -- 10 m & 1 -- 100 ms & \textless{} 10 m/s & 1 -- 1000 m \\
\hline
\end{tabular}
\label{tab:use-cases}
\end{table}

\subsection{ISAC Synergies and
Trade-offs}\label{sensing-and-communication-synergies-and-trade-offs}

At all levels of integration, except level 0 (see Section~\ref{what-is-jcas}), there are fundamental
trade-offs between communication and sensing, since any resource (time,
frequency, computation, storage) provided to sensing is taken from
communication resources. In the case sensing is performed opportunistically in a communication-optimized setting (see Level 4 discussion later in this section), the sensing performance cannot be controlled or optimized. 
These trade-offs can take a variety of forms:

At \textit{Level 1} ISAC, where sensing and communication systems share the same spectrum but operate on separate hardware, trade-offs emerge primarily from the need to manage interference and optimize resource allocation. While this shared spectrum allows efficient utilization of scarce bandwidth, it necessitates strict coordination to avoid mutual interference. This can be done by  allocating separate time slots for sensing and communication operations, which ensures reliable performance for both functions \cite{IT_JCAS_2022}. However, this temporal separation introduces a rate loss during radar operation, and the radar system must forgo sensing during communication slots \cite{holisticISAC_2025}. Furthermore, the need for robust interference management protocols and synchronization mechanisms adds to system complexity. 

At \textit{Level 2}, 
the integration of communication and sensing functions within a single hardware platform in 6G ISAC offers several synergies, particularly in terms of cost-effectiveness and efficient use of spectrum resources \cite{JRC_mod_hardware_2022}. By utilizing the same hardware components, such as antennas and transceivers, for both communication and sensing tasks, the need for dedicated hardware is reduced, which can lower overall costs and streamline device design. Additionally, shared spectrum usage between sensing and communication tasks enhances spectral efficiency, which is crucial given the scarcity of spectrum in low-frequency bands. However, these synergies come with trade-offs. The dual-use hardware must meet stringent design and calibration requirements to support both tasks effectively, which can complicate the design process and increase development costs. Performance optimization for one function (e.g., high-resolution sensing) might limit the efficiency of the other (e.g., reliable communication). For instance, the use of wideband antennas for high-resolution sensing can introduce phase noise and non-idealities that degrade communication performance, especially at sub-THz frequencies. Another example is the use of directional antennas or not. In certain communication scenarios, non-directional antennas create multi-path propagation which can be useful in the form of diversity or multiplexing. However, in general, multi-path propagation degrades the performance of radar sensing.  Furthermore, achieving the necessary synchronization and precision in hardware components, such as oscillators and phase shifters, is critical but challenging, as it impacts both functions simultaneously. In general, common hardware will need to support the most stringent requirements, which may increase the cost per component. Hence,  the  net gain should come from the need of fewer components compared to separate hardware for communications and sensing, rather than the cost per component. 
These design challenges underscore the need for innovative hardware solutions that balance performance, reliability, and cost in future ISAC systems.

At \emph{Level 3}, sensing requires dedicated pilot signals. The
  frequency of these pilots, their duration and bandwidth as well as
  their power are fundamental design parameters that have a direct
  impact on the sensing update rate, latency, resolution, and accuracy \cite{mura2024optimized}.
  For multiple-antenna (MIMO) systems, the spatial design of the pilots
  affects the ability to illuminate part or an entire surveillance
  region \cite{holisticISAC_2025}. Hence, pilots may be optimized for scanning or tracking
  purposes, where in the latter case, prior information of the users or
  targets can be leveraged for improved performance or increased
  efficiency. When there are multiple active transmitters, pilots must
  be coordinated to avoid mutual interference. In some cases, the
  sensing pilots may be reused for communication purposes (e.g.,
  synchronization and channel estimation), but generally sensing pilots
  take away radio resources from communication \cite{dedicatedSensingISAC_2024}.

At  \emph{Level 4}, sensing is based on existing
  communication signals. In this case, the same radio resources are used
  for both sensing and communication, so the trade-off manifests itself
  in the optimization of these radio resources. Frequency resources can
  be optimized for communication or for sensing. For example, optimizing for communication
  may happen by maximizing rate through water-filling, based on information of the communication
  channel. Optimizing for sensing may happen by maximizing the sensing information through
  the \ac{CRB}, based on prior information about the
  sensing targets or users \cite{JRC_CRB_Beamforming}. 
  Combination of these objectives are also possible \cite{OFDM_DFRC_TSP_2021}.
  Similarly, spatial resources can be optimized for communication (e.g.,
  directional beamforming towards users) or sensing (e.g., a grid of
  beams towards possible target locations) or for a combination of these
  services \cite{Puc22,outage_6G_ISAC}.

Hence, at all levels, there are trade-offs between communication and sensing. At best, the sensing does not reduce the communication performance. However, this is not the end of the story, as there are higher-order effects, as already mentioned in Section \ref{why-jcas}:
the knowledge of user locations, blockages, or other contextual information can improve the operation of the communication system \cite{guo2023potential}. This means that the radio-level trade-offs can turn into higher-level synergies and benefits.

\subsection{Measuring ISAC Performance}\label{KPIs}

This section defines the \acp{KPI} for radio sensing in the \ac{PHY}, which complement traditional communication \acp{KPI} such as data rate, coverage, latency, reliability, and energy efficiency. In ISAC systems, these \acp{KPI} can be combined in various ways during resource allocation and signal optimization, including optimizing a linear combination of \acp{KPI}, prioritizing communication \acp{KPI} while meeting sensing constraints, or prioritizing sensing \acp{KPI} while meeting communication constraints. 
\begin{itemize}
    \item 
\textit{Accuracy}, measured in meters, represents the location error norm at specific percentiles, calculated as the distance between true and estimated positions in either 3D or 2D coordinates. This metric applies to both connected devices and targets, where it is referred to as sensing accuracy. Similarly, orientation accuracy, measured in degrees, quantifies the orientation error norm at certain percentiles, calculated by  determining the geodesic distance between rotation matrices or quaternions. Accuracy can be improved by increasing the \ac{SNR}. 
\item \textit{Resolution} measures the smallest detectable difference in position or other dimensions (e.g., range, angle, Doppler) between targets, expressed in meters, radians, or meters per second. In contrast to accuracy, resolution generally does not increase with \ac{SNR}, but depends on the physical properties of the signal (bandwidth, time, and antenna aperture).
\item \textit{Latency}, measured in seconds, defines the time between initiating a sensing or localization procedure and obtaining the estimate for the requested service. 
\item  \textit{Coverage} specifies the area, volume, or percentage of space within which localization error remains below a set threshold. 
\item \textit{Detection probability} reflects the probability of successfully detecting a target when it is present. 
\item \textit{Classification accuracy} is relevant for machine learning-based applications such as gesture or posture classification. It is the ratio between correctly classified items versus all items. 
\item \textit{Efficiency} captures the ability to achieve accurate sensing while minimizing energy consumption, encompassing energy used during sensing operations as well as processing, data transmission, and communication.
\end{itemize}
These \acp{KPI} collectively characterize the performance trade-offs and optimization goals in ISAC systems (see also Section \ref{a-quantitative-cross-layer-evaluation}).

\subsection{ISAC Channel Models}\label{sec:channel-models}
Compared to conventional communication channels, ISAC channels require precise modeling of additional sensing characteristics. The sensing channels include propagation paths influenced by the sensing target, i.e., {\it the target channels}, and the impact of background targets on the sensing channel, i.e., {\it the background channels}~\cite{10292797}. Most existing statistical geometric channel models, including 3GPP \ac{TR} 38.901, in their current form, do not sufficiently represent the target channels~\cite{3rd20183gpp} though background channels have been supported well. Site-specific models, on the other hand, already support both the target and background channels implicitly and, in principle, can be further extended to encompass nearly all of the needed features of the channel model for ISAC use cases. Recognizing this limitation of the statistical geometric models, 
the 3GPP has recently begun to address these issues by discussing suitable modifications to the \ac{TR} 38.901. The present subsection outlines the needed features of channel models for ISAC use cases, popular legacy channel models, their recent extensions for ISAC use cases and finally open issues for further extensions.

\subsubsection{ISAC Channel Model Features}
\begin{figure}
    \centering
    \includegraphics[width=0.99\linewidth]{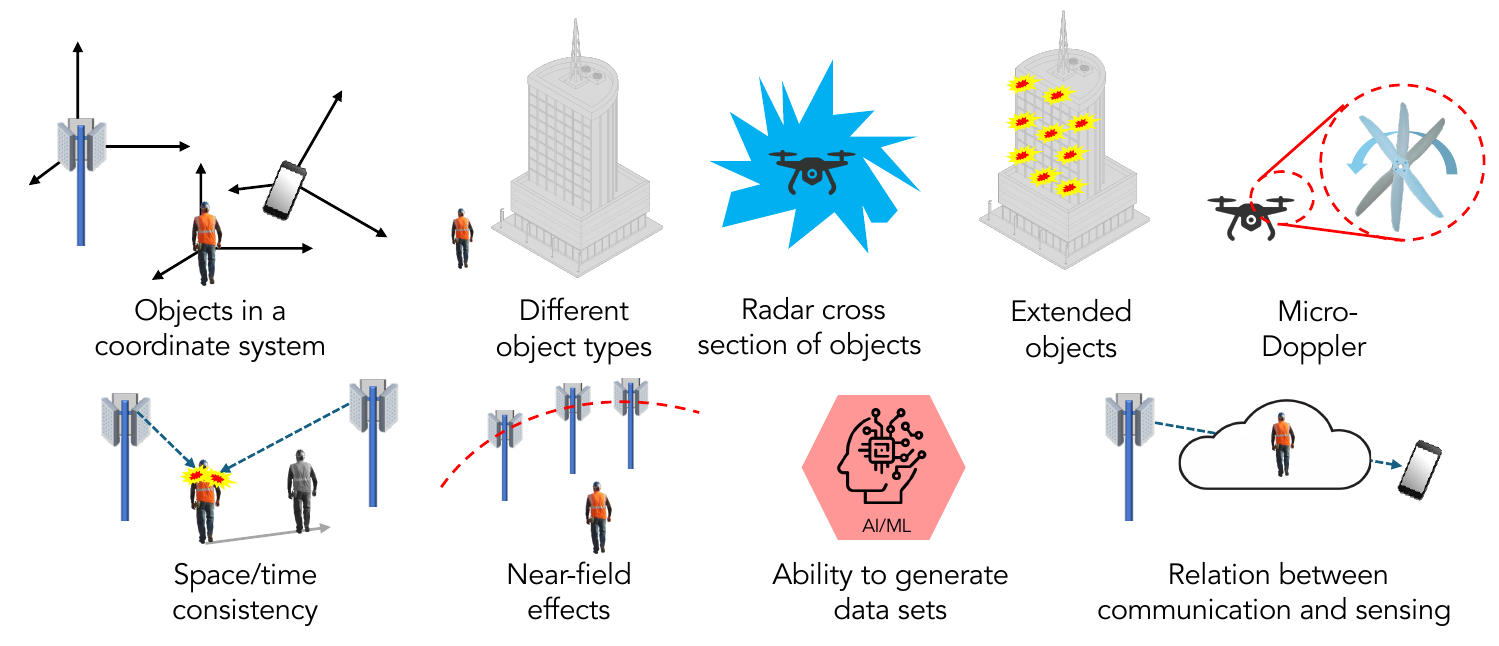}
    \caption{A non-exhaustive list of desired ISAC channel model features.}
    \label{fig:ISACchannelFeatures}
\end{figure}

From the use cases defined in Section~\ref{why-jcas}, it becomes apparent that several features must be supported by channel models to evaluate ISAC performance. In \cite[Section~3.6]{hex24_d43}, several features for ISAC channel models were identified, described below and  depicted in Fig.~\ref{fig:ISACchannelFeatures}. Not all features must be supported simultaneously since they are not all relevant for each use case.
\begin{itemize}
\item {\it Objects in a coordinate system:} Objects (e.g., buildings, cars, people, \acp{UAV}) should be defined in a globally consistent coordinate system with associated 3D position, 3D orientation, and 3D velocity vectors. Objects refers both to objects of interest (e.g., to be detected and tracked) and to targets that are part of the background and that generate clutter.
 \rev{Clutter refers to the scattered echoes caused by objects other than the target and forms a key component of the background channel. Clutter can degrade sensing performance by masking weak targets. 
 Clutter can be modeled stochastically or geometrically.
Commonly used stochastic models include the Rayleigh distribution, log-normal distribution, Weibull distribution, and K-distribution \cite{shnidman2002generalized}, depending on the type of radar and type of clutter. }
 
\item {\it Different object types:} the objects should include large flat objects (e.g., walls, floors), which have wave interaction points that depend on locations of transmitter and receiver and small objects (e.g., pillars, tables), which have interaction points that are independent on the location of transmitter and receiver. The notions of large and small are relative to the signal wavelength. The object classes should include objects of interest to the use cases, such as people (hands), animals, cars, \acp{UAV}, as they lead to different channel responses and can affect the polarization of the signals.
\item {\it \Ac{RCS} of objects:} Both \ac{RCS} for monostatic and bistatic sensing, and \ac{RCS} for deterministic and stochastic object modeling should be supported. For large bandwidths, frequency dependence of the \ac{RCS} may become important.
\item {\it Extended objects:} For large models with several resolved interaction points, each interaction point should be described with an \ac{RCS} or reflection coefficient, accounting for aspect angles.
\item {\it Micro-Doppler:} In addition to Doppler shifts induced by objects, mechanical vibration or rotation of parts of these objects leads to micro-Dopplers (sidebands around the main Doppler shift). Such micro-Dopplers of objects should be supported.
\item {\it Space/time consistency:} Channel models should allow generating a set of channels from a transmitter to one or more objects (in the global coordinate system) to several spatially separated receivers. Similarly, there might be a need to generate a set of channels from several transmitters to one or more objects (in the global coordinate system) to a receiver. These channels must be spatially consistent and correctly account for the \ac{FoV} of each transmitter and receiver. The correspondence of the time of the transmitters, receivers, or moving objects must be maintained.
\item {\it (Geometric) Near-field effects:} Wave-front curvature, between transmitter and objects, between objects and receiver, and between transmitter and receiver should be supported. 
Moreover, channel non-stationarity across larger arrays (i.e., variations of amplitude and phase due to differences in distance and blockages) is an additional effect due to the wave-front curvature and radiative near-field. Note that plane-wave far-field propagation is the baseline assumption in legacy channel models.  \rev{The phase variation across receiving antennas is determined by the propagation distance from the transmitter, allowing a received to directly infer the location of a user or object from this phase variation \cite{chen20246g,GueGuiDarDju:J21}. In the classical far-field regime, the phase variation depends only on the \ac{AoA}, so that distance estimation requires complementary \ac{ToA} measurements for locating users or objects. }
\item {\it Ability to generate data sets for classification use cases:} Such use cases require labeled training data for classification (e.g., gesture recognition, pose classification), when these are not provided as inputs.
\item {\it Relation between communication and sensing channel:} The communication and sensing channel between a transmitter and a receiver are in principle identical and are reciprocal after compensation for transceiver nonlinearities and calibration of  the transceivers.

\end{itemize}
Some use cases may need to include even more features (e.g., different types of \ac{RIS}), but they are not considered here.

\subsubsection{Legacy Channel Models}
Widely-used and well-established channel models describe the communication channel between a transmitter and a receiver as a wireless propagation environment, where this environment may contain influential objects for communications and sensing. Legacy models can be roughly grouped into two categories~\cite{hex24_d43}:
\begin{itemize}
\item {\textit{Site-specific models:}} Communicating devices in a site-specific model are defined on a map. The map can vary from simple and featureless ones with only flat walls and grounds to complex ones including walls with detailed structures, cluttering and time-varying objects. Background and target channels can both be considered implicitly. Since most site-specific simulations of wave propagation have so far been intended for \ac{BS} deployment and its coverage simulations, small clutters and time-varying objects are usually neglected. Material properties and an \ac{RCS} of objects are necessary for deriving polarimetric gains of multipath in the background and target channels. Material properties of popular objects in our living environment are available in~\cite{ITU2040}, while the dielectric and conductive parameters of water~\cite{Ellison07_JPC} are applicable to model a human body. While experimental verification is crucial for any channel model, site-specific models have primarily focused on validating received signal strength. Moreover, extensive validation of small-scale channel characteristics, such as power delay profiles and angular distributions observed at the link end points, is scarce for these models. A site-specific model is in principle a deterministic channel model, and often includes stochastic modeling of diffuse scattering, e.g.,~\cite{Lu19_TAP}. In addition, it can also have random scattering and shadowing objects, leading to a hybrid model~\cite{Kyosti17_TAP, Li22_Globecom, Chen23_PIMRC}. 
\item {\textit{Statistical geometrical models:}} Communicating devices in statistical geometrical models are defined on a coordinate system. Clusters or objects contributing to clusters are statistically defined either through multipath characteristics such as angle-of-arrivals, angle-of-departure, and time-of-arrival as was defined in the 3GPP 38.901~\cite{3rd20183gpp}, or through their locations on the same coordinate system as the communicating device like the COST 2100 model~\cite{6393523} and the QuaDRiGa model~\cite{jaeckel2014quadriga}. The former sub-category, with random multipath angles and delays without any coordinate based locations of clusters, is much more widely used but also less functional for ISAC evaluations. On the other hand, the latter sub-category with randomly drawn coordinates of interacting objects, is suitable for ISAC with minor updates. \rev{Stochastic clutters can be generated via random cluster generation as in TR 38.901~\cite{3rd20183gpp}.}
\end{itemize}

In Table ~\ref{tab:features}, the relation between the ISAC channel model features and the legacy channel models is evaluated. This reveals that legacy channel model must be extended to accommodate ISAC use cases.

\begin{table*}
\centering
\caption{
Relation between the desired ISAC features and the existing channel models;
(A) = is the concerned feature currently present? (B) = is it possible to include the concerned feature? If (A) is answered Yes, then (B) is shown “N/A”.
}
\renewcommand{\arraystretch}{1} 
\begin{tabular}{|>{\raggedright}p{0.1\linewidth}|p{0.40\linewidth}|p{0.40\linewidth}|}
\hline
\textbf{ISAC feature} & \textbf{Statistical geometric models, e.g., 3GPP \ac{TR} 38.901} & \textbf{Site-specific models, e.g., ray-launching} \\ \hline
Objects in a coordinate system & (A) No, except for some models e.g., COST2100 and QuaDRiGa; (B) Yes, e.g., in case of 3GPP 38.901, this is provided by the alternative map-based model & (A) Yes, they are implicitly considered on a map of the simulation site; (B) N/A. \\ \hline
Different object types & (A) No; (B) Yes, can be possibly modelled using the radar cross-section. & (A) No, only large planar surfaces are supported with different material characteristics and roughness; (B) Yes, using the radar cross-section. \\ \hline
\ac{RCS} & (A) No; (B) Yes, the gain of a propagation path can be determined with a specific \ac{RCS} of the cluster or sub-path. However, the \ac{RCS} value can vary depending on the distance between the transmitter/receiver and the target~\cite{10404988}. Additionally, the \ac{RCS} value for a given target may differ based on whether mono-static or bi-static sensing is used. In bi-static sensing, when the angle between the departing and incident waves is small, the \ac{RCS} value closely resembles that of mono-static sensing~\cite{zhang2024latest}. However, as this angle increases, variations in the target channel's propagation conditions~\cite{zhang2024latest} can lead to random fluctuations in the \ac{RCS} value. 
& (A) Yes, but usually neglected in coverage simulations due to overall complexity of the physical model of the propagation environment. Building walls are much more influencing object than small scatterers in outdoor environments; (B) N/A, but \ac{RCS} estimates are not always available for the object and frequency of interests. \\ \hline
Extended objects & (A) No; (B) No. It is quite difficult to associate different clusters with objects.  & (A) Yes, but experimental verification of the method to simulate waves and extended objects would be necessary; (B) N/A. \\ \hline
Doppler and Micro-Doppler due to motions of wave scattering objects & {(A) No; (B) Yes, if a linear direction and speed of a wave scattering object are defined, the Doppler frequency is determined uniquely for a propagation path emanating from it. The Micro-Doppler effects can also be considered by considering probability distribution of the direction and speed of scattering objects.} & Same as statistical models\\ \hline
Space/time consistency & (A) No, except for COST2100 and QuaDRiGa models; (B) Yes, similar to the QuaDRiGa model & (A) Yes, implicitly; (B) N/A. \\ \hline
Near-field & (A) No (B) Yes, with the knowledge of objects’ coordinate system, location of the first and last bounce scatterers and the plane wave model replaced by the spherical wave model. & (A) No, the far-field plane wave propagation is assumed; (B) Yes, far-field spherical wave propagation is straightforward while fields in the radiative near-field region cannot be considered based on ray-optics approximation of wave propagation. \\ \hline
Data sets for classification use cases & (A) No; (B) No. & (A) No, ray-optics approximation of wave propagation is invalid for scattering characterization from objects; (B) No, measurements or numerical full-wave simulations must be performed, e.g.,~\cite{Csernyava22_EuCAP, Li22_Globecom, Chen23_PIMRC}. \\ \hline
Communication and sensing channel compatibility, and reciprocity & (A) No, because spatial consistency is not ensured for widely separated devices; (B) Yes, similar to QuaDRiGa. & (A) Yes, but the reciprocity of the simulated channels must be checked by swapping the transmit and receive sides and comparing their channels. Moreover, verification of the simulated channels against measurements is always worthwhile, especially in monostatic scenarios where not many measurements are available; (B) N/A. \\ \hline
\end{tabular}
\renewcommand{\arraystretch}{1.0} 
\label{tab:features}
\end{table*}

\subsubsection{Extensions of the 3GPP Channel Model}
The target channel consists of all propagation paths influenced by the sensing target, while the background channel includes propagation paths that are unrelated to the target at a time instant~\cite{10292797}. Propagation paths related to the target channel can be further classified into \acp{DIP} and \acp{IDP}. \acp{DIP} travel directly from the \ac{Tx} to the target and then from the target to the \ac{Rx}, while \acp{IDP} undergo multiple reflections or diffractions before reaching the receiver~\cite{10292797}. One approach to modeling \acp{IDP} is through statistical methods similar to those used for generating \ac{NLoS} paths in \ac{TR} 38.901~\cite{3rd20183gpp}. However, \acp{EO} (such as street lamps or walls) that significantly affect wave propagation should be incorporated into the target channel to enhance sensing accuracy~\cite{zhang2024latest}. The target channel is modeled as a concatenation of the \ac{Tx}-target and target-\ac{Rx} component channels. Recent measurement studies confirm the modeling approach in that, for large-scale fading, the \ac{Tx} to \ac{Rx} path loss can be expressed as the summation of the \ac{Tx} to target and target to \ac{Rx} path losses in the logarithmic scale. Additionally, for small-scale fading, the \ac{Tx} to \ac{Rx} \ac{CIR} can be represented as the convolution of the \ac{Tx} to target and target to \ac{Rx} CIRs~\cite{zhang2024latest}. 

In ongoing standardization efforts, the prevailing perspective is that the background channel should be modeled using statistical clusters as defined in \ac{TR} 38.901. However, two key aspects require further investigation. The first aspect concerns whether deterministic components, such as EOs, should be incorporated into the background channel in addition to the target channel so that it takes the form of the hybrid channel model finally. EOs are objects that can be characterized by their \ac{RCS}, similar to sensing targets. Including these components into the background channel could be essential for accurately assessing signal levels from the background clutter and their impact on sensing performance. The second aspect involves the potential interaction between the background and target channels. For example, when a sensing target moves dynamically over time, the target may shadow a propagation path of the background channel at one time instant and make the path a new part of the target channel. Most existing channel models generate communication and sensing channels independently under a common framework. However, since both functionalities share hardware resources and operate in the same propagation environment, their signals often interact with a subset of common scatterers. Thus, the communication and sensing channels are related to each other due to these shared scatterers. Channel measurements conducted in both \ac{LoS} and NLoS indoor scenarios further confirm the presence of these shared clusters~\cite{10334037}. To quantify the effect of these shared scatterers on the communication and sensing channel, the \textit{sharing degree} metric has recently been proposed in \cite{10334037}. However, further study is required to determine whether this characteristic should be explicitly incorporated into the channel modeling framework.

Ongoing discussions within 3GPP have resulted in the establishment of a framework for ISAC channel modeling. The main distinction between the existing 3GPP model and its potential modifications for ISAC lies in the modeling of the target channel characterized by the \ac{RCS} of the sensing target, along with the inclusion of EOs. This is in addition to the background channel, which is already well-supported in \ac{TR} 38.901. Modeling approaches for both target and background channels have been recently proposed in \cite{10292797} while \ac{RCS} measurements for mono and bi-static sensing are currently ongoing~\cite{10745726}. 
Finally, additional field measurements and electromagnetic simulations in representative scenarios are necessary to further validate these modeling assumptions.

\subsection{ISAC in the End-to-End 6G System}
\begin{figure*}
    \centering
    \includegraphics[width=0.99\linewidth]{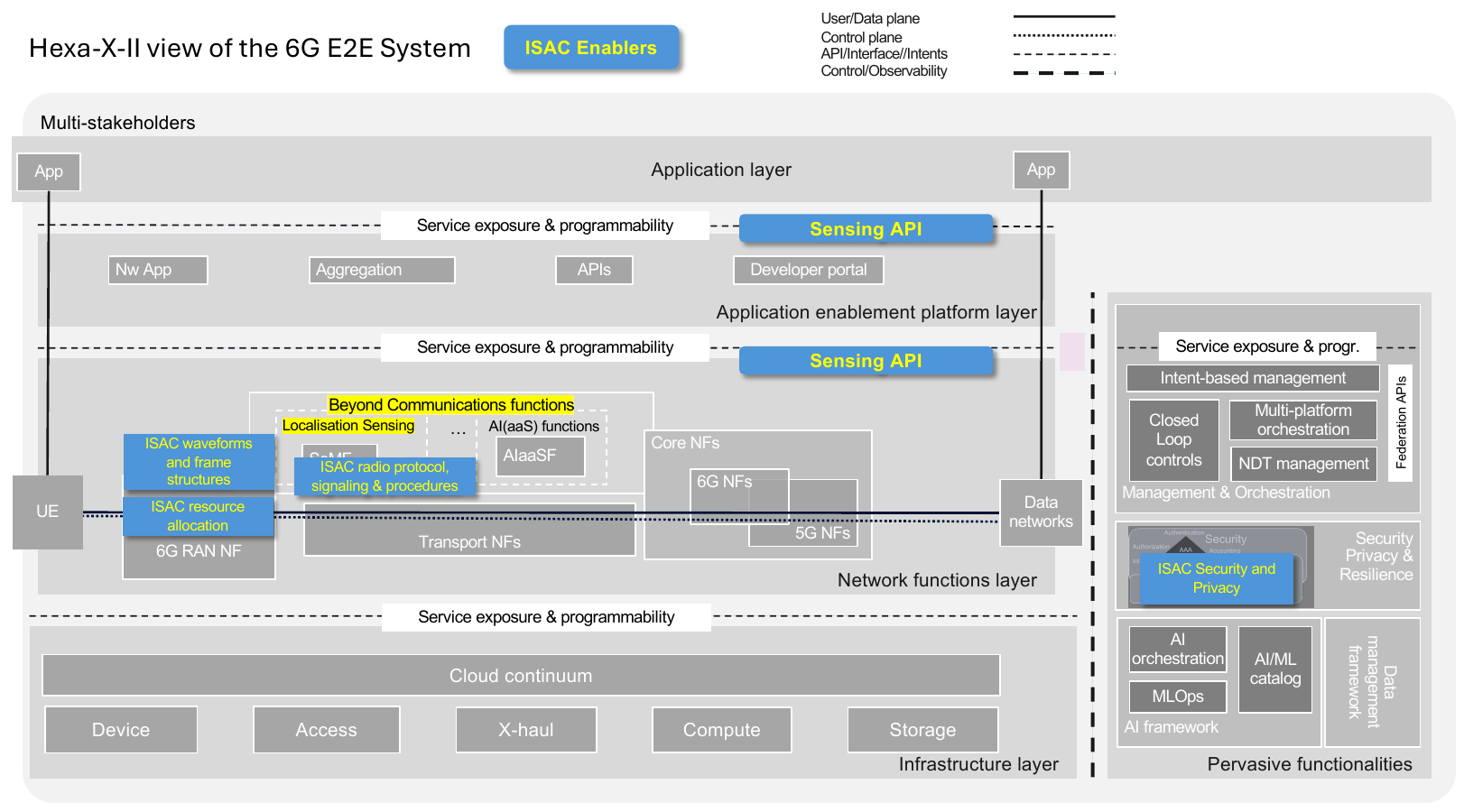}
    \caption{ISAC in the Hexa-X-II 6G E2E system, based on \cite{hex24_d23}.}
    \label{fig:ISACe2e}
\end{figure*}

In the broader picture of the 6G end-to-end system, ISAC impacts not only on the radio \ac{PHY} design, but additional functionalities need to be considered in the core network as well as other pervasive functionalities. To describe this in further detail, Fig.~\ref{fig:ISACe2e} is introduced.

Concretely, Fig.~\ref{fig:ISACe2e} presents the mapping of the ISAC technological enablers onto the Hexa-X-II 6G E2E system \cite{hex24_d23}. The physical and \ac{MAC} layer enablers alongside the radio protocols, signaling and procedures will be implemented mainly at the 6G \ac{RAN} and between the 6G \ac{RAN} and beyond communication functions, i.e., the \ac{SeMF} \cite{hexd23}.  
Specifically, {ISAC protocols, signaling, and procedures} enabler addresses network integration of sensing capabilities \cite{hexd23}. It introduces a \ac{SeMF} for next-gen communication systems to control sensing processes and data processing. \ac{SeMF} coordinates sensing procedures efficiently, considering requirements, capabilities, and constraints. Enhanced procedures and signaling, aligned with existing and evolved communication processes, facilitate secure and efficient data collection, processing, and exposure. This enabler supports ISAC services, enabling precise localization and tracking through advanced sensing protocols. 

Next, {ISAC waveforms and frame structures}, given the 3GPP mandate to continue the use of \ac{OFDM} waveforms, will need to utilize \ac{OFDM}. Concretely, the ISAC algorithms will be based on the underlying fact that all communication and sensing will be done using \ac{OFDM} waveforms and frame structures standardized in 3GPP \cite{hexd23}. Moreover, the {ISAC resource allocation} enabler highlights the need for efficient resource allocation between sensing and communication. This is because sensing resources will not be used for communications and vice versa. Hence, to be able to maintain the \acp{SLA}, i.e., in simpler terms user \ac{QoS}, the resource allocation will need to be done intelligently between sensing and communication tasks \cite{hexd23}. 

Importantly, there is a need for the exposure of the sensing data and service to the applications, e.g., via \acp{API}. Such \ac{API} will include service request detailing the sensing purpose (what, when, where, who), the identity and authorization control, result delivery, and service termination. The application can reside in the \ac{UE} side, in the \ac{RAN} or in the data network. There can be several types of exposure depending on the nature of the request (small or massive data flows), such as querying for object presence or tracking objects over large areas involving exposure of measurement steam. Different levels of \ac{API} exposure are possible for connecting applications to sensing systems, e.g., including direct access from 3GPP \ac{NEF} interfaces \cite{29522}, intermediate \acp{API} like \ac{SEAL} \cite{23434}, and CAMARA \acp{API} at a more aggregated level \cite{CAM}.

Lastly, considering the sensing functionality within ISAC, there will be concerns regarding trust, security and privacy of operations and data management. Subsequently, the {ISAC Security and Privacy enabler(s)} aim to provision authentication, authorization, logging, data control, security and privacy of data transfer and storage, etc., as some of the security and privacy measures. Furthermore, these technological enablers will also consider aspects of data anonymization, pseudonymization and privacy preserving computation, among others as ways of maintaining integrity and security of the user and network identity and the sensed data \cite{hexd23}.

\section{ISAC at the Physical Layer} \label{sec:JCASPHY}

In this section, we go deeper into the 6G \ac{PHY} (or radio) enablers for ISAC as well as the various sensing configurations and the use of \ac{AI}. One topic intentionally omitted is the aspect of waveforms, as \ac{OFDM}  and its precoded version (such as \ac{DFTS-OFDM}) are likely to become the 6G standardized waveform. The support for \ac{OFDM} combines a large popularity in most recent high-throughput wireless standards, well-understood characteristics, low-complexity \ac{FFT}-based equalization, a natural combination with \ac{MIMO}, especially for precoded schemes, and an excellent capture of frequency-domain diversity when properly combined with channel coding. Despite its higher \ac{PAPR}, there is no sufficient justification to replace \ac{OFDM} by single-carrier options based on end-to-end performance analysis including various hardware non-idealities and related compensation techniques. Another argument comes from more limited multi-path in sub-THz channels, but removing the \ac{FFT}-based equalization would still require some time-domain replacement with other challenges in estimation, tracking, and combination with remaining frequency-domain features.

\subsection{Machine Learning and Artificial Intelligence for ISAC}\label{ai}

The use of \ac{AI}/\ac{ML} has been envisioned to have an impact on the overall 6G networks \cite{HexaXAIML23, Ryd23}, and in particular redesign several aspects of the \ac{PHY} \cite{Far25}. The \ac{AI}-driven radio air interface can be treated as a radio technology with one/multiple learnable functionalities for communication and/or sensing across the \ac{Tx} and/or \ac{Rx}. 

\subsubsection{Motivations for \ac{AI}-driven ISAC}\label{ai:motivations}
The motivation for the adoption of \ac{AI}-driven methods for ISAC are outlined in the following: \textit{(i) \ac{AI}/\ac{ML} accelerators:} \ac{AI}/\ac{ML}-dedicated hardware advancements reduce training costs and enable the real-time operation of \ac{AI}-driven methods for the \ac{PHY}. Training chipsets provide high memory bandwidth and precision for complex data processing, enhancing model training efficiency, while inference chipsets are optimized for real-time, low-power processing. These advancements facilitate the adoption of \ac{AI}-driven approaches for ISAC methods.
\textit{(ii) Performance enhancements:}
\ac{AI}-driven methods enhance ISAC by improving reliability, throughput, and sensing accuracy through holistic optimization, while mitigating the impact of hardware impairments beyond what classical models can address \cite{HexaXAIML23}. Unlike traditional approaches that optimize individual components separately, \ac{AI} enables end-to-end optimization across both communication and sensing functions resulting in more efficient and robust ISAC systems.
\textit{(iii) Flexibility:}
\Ac{AI} enhances the flexibility of ISAC systems by enabling real-time adaptation to dynamic and heterogeneous environments, including variations in traffic patterns, user mobility, and radio propagation. Through data-driven policy learning and predictive modeling, \ac{AI} supports context-aware resource allocation and cross-layer coordination. This allows ISAC systems to autonomously reconfigure sensing and communication strategies in response to evolving conditions and diverse application requirements, ensuring robust and efficient operation across scenarios.
\textit{(iv) Scalability:}
The ability to re-train models with new data as scenarios evolve enables scalable \ac{AI}-driven ISAC solutions. This capability ensures the network remains adaptable and scalable, regardless of the growing number and diversity of devices and services it must support. To maintain this scalability, model \ac{LCM}—covering the entire process of model development, deployment, and management—must be supported.

\subsubsection{\ac{AI}-driven Use Cases for ISAC}\label{ai:use cases}
The \ac{AI}-driven radio air interface
for ISAC may impact \ac{Tx}-side optimization and/or \ac{Rx}-side optimization. Example use cases of \ac{AI} for ISAC are illustrated in Fig.~\ref{fig: AI for ISAC use cases}) and are outlined in the following: 
\begin{figure}
    \centering
\includegraphics[width=0.99\linewidth]{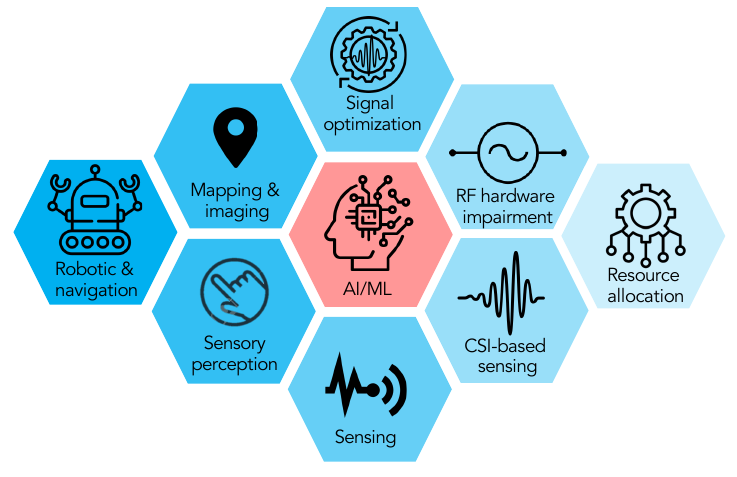}
    \caption{Representative \ac{AI}/\ac{ML} for ISAC use cases.}
    \label{fig: AI for ISAC use cases}
\end{figure}
(i)
  \textit{calibration and operation under model mismatch:} data-driven
  methods can be used to cope with various impairments and model
  mismatches, such as hardware impairments (e.g., power amplifier nonlinearity \cite{AIfornonlinearPA}, and oscillator's phase noise \cite{AIforphasenoise}); 
(ii)
  \textit{sensing and imaging:} One of \ac{AI}'s main success stories has  been in image processing. Since sensing data relates to real-world  targets, image processing tools applied to ISAC radio signals can be used to de-noise data, to
  classify different object types, and to track targets over time. In  particular, \ac{AI} can be used to infer information from the data where  traditional methods fail, such as gesture or pose recognition in
  cluttered environments;
(iii)
  \textit{discovery of new signals:} \ac{AI} can be used to learn new signals
  (in time, frequency, and spatial domains) that are optimized for ISAC, considering e.g., multi-objective optimization.
  Such problems are hard to formulate and solve using model-based signal
  processing.
(iv)
  \textit{Representation:} \ac{AI} can learn compact representations of observations (e.g., novel source coding schemes), avoiding sending massive amounts of data between devices or over the backhaul.
(v)
  \textit{robotics and navigation:} \ac{AI}/\ac{ML} have greatly enhanced the functionality of robotics, navigation, and ISAC. In robotics, these techniques enable more accurate path planning and obstacle avoidance by interpreting complex environments in real time. For navigation, \ac{AI}/\ac{ML} models integrate multimodal data, such as \ac{LiDAR} and \ac{mmWave} signals, to improve localization and mapping, even in challenging scenarios. For instance, the study in \cite{yin2022millimeter} demonstrates how \ac{ML}-based link state classification in \ac{mmWave}-assisted navigation allows robots to effectively distinguish between \ac{LoS} and \ac{NLoS} paths, improving target localization and navigation in dynamic environments.

\subsubsection{Challenges of \ac{AI}-driven ISAC}\label{ai:challenges}
\ac{AI}-driven methods in ISAC for 6G face several challenges:
(i) \textit{Complexity:} 
\ac{AI} models for both communication and sensing introduce high computational demands, where efficient models and hardware accelerators are needed to handle real-time processing without overwhelming energy-constrained devices.
(ii) \textit{Training Data:} 
A crucial aspect of any \ac{AI}/\ac{ML} system is its training. In contrast to communication, training data for ISAC is more involved, as it requires knowledge of sensing targets, their type, locations, etc. Obtaining such  information would need external sensors and human labelling  {or advanced and well-structured radio training procedures. An alternative solution is the use of knowledge transfer methods which can significantly reduce the training overhead.}
High-quality, standardized datasets for diverse ISAC scenarios are lacking, making model development challenging. Reference datasets from measurement campaigns together with synthetic data from simulation or generative \ac{AI} can help mitigate this issue.
(iii) \textit{Real-Time Processing:} The need for low-latency processing in ISAC requires optimized \ac{AI} models and hardware, leveraging parallel processing and accelerators like \acp{GPU} or \acp{ASIC} to meet real-time demands.
(iv) \textit{New Design Paradigm:}
Integrating \ac{AI} in ISAC requires new processes for model \ac{LCM}, particularly in distributed models. Standardized \ac{LCM} approaches are necessary for scalability and multi-vendor coordination.
(v) \textit{Generalization:}
\ac{AI} models are desired to generalize across diverse environments. Techniques like self-supervised learning can enable rapid adaptation without extensive retraining. However, generalization may not be always feasible or lead to extended \ac{AI}/\ac{ML} model complexity. Hence, \ac{AI}/\ac{ML} model performance monitoring is required as part of \ac{AI}/\ac{ML} model \ac{LCM} to ensure the \ac{AI}/\ac{ML} model inference meets the expected quality for the desired use case. 
(vi) \textit{Contributions to 6G Values:} \ac{AI}-driven ISAC systems must align with 6G's core values—sustainability, trustworthiness, and inclusiveness. While \ac{AI} can enhance network efficiency and sensing accuracy, training \ac{AI} models can be energy-intensive. To meet sustainability goals, model compression and energy-efficient inference are essential. Privacy and security concerns, particularly regarding sensitive data collection for training, must be addressed through methods like federated learning and differential privacy. 

Addressing these challenges is essential for enabling scalable, efficient, and secure \ac{AI}-driven ISAC systems in 6G networks.

\subsubsection{\ac{AI} and \ac{RF} Digital Twins}
\ac{AI}-driven \ac{RF} digital twins are emerging as powerful tools for enabling and enhancing ISAC functionality across physical, protocol, and application layers. Building on the motivations, use cases, and challenges outlined above, \ac{RF} digital twins provide a virtualized, continuously updated representation of the radio environment, informed by real-time measurements, historical data, and model-based predictions. These digital replicas enable the training, evaluation, and deployment of \ac{AI} models for ISAC in a controlled, risk-free environment—effectively addressing key issues such as data scarcity, generalization, and real-time adaptability.

Specifically, \ac{RF} digital twins support multi-domain learning and optimization, allowing \ac{AI} models to co-optimize communication and sensing under complex and dynamic conditions. For instance, in vehicular edge computing scenarios, a digital twin-aided framework integrating visible light communication and sensing was shown to enhance task offloading and resource allocation \cite{yang2024joint}. Similarly, site-specific outdoor propagation studies have demonstrated how \ac{RF} digital twins, combined with ray-tracing analysis, can inform wireless deployment strategies in dense urban environments by modeling building layouts and antenna configurations \cite{ghaderi2024site}. At the network level, \ac{RF} digital twins enable multi-domain planning by simulating \ac{RF} transmissions, device behavior, and control policies \cite{goudy2024network}. Furthermore, the integration of non-\ac{RF} modalities—such as GPS and \ac{LiDAR}—within \ac{RF} digital twin frameworks has been proposed to support beam management and sector selection via deep learning, showcasing the versatility of digital twins in \ac{AI}-assisted ISAC \cite{salehihikouei2024leveraging}.

Through this tight coupling of \ac{AI} and \ac{RF} digital twins, ISAC systems can become more data-efficient, adaptive, and resilient—unlocking a new design paradigm that leverages digital replicas to train, fine-tune, and validate \ac{AI} models before real-world deployment.

\subsection{Candidate 6G Radio
Enablers for ISAC}\label{jcas-and-the-6g-radio-enablers}

\begin{figure}[t]
    \centering
    \includegraphics[width=0.99\linewidth]{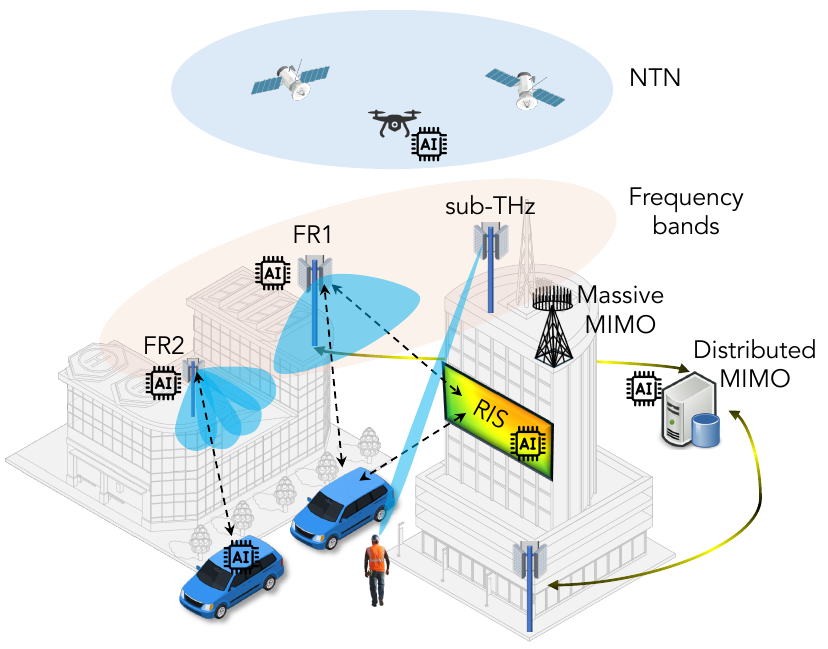}
    \caption{A selection of candidate 6G radio enablers that complement ISAC. }
    \label{fig:enablers}
\end{figure}

Various candidate 6G radio enablers studied in Hexa-X-II \cite{hex24_d43} interact with ISAC (namely the different frequency bands, \ac{RIS}, \ac{NTN}, massive \ac{MIMO}, \ac{D-MIMO}), as visualized in Fig.~\ref{fig:enablers}. We briefly recap these technical enablers, which are presented in an arbitrary order, without implying any preference or priority.   
First of all, in terms of frequency bands, 4 different cases are considered. \ac{FR} 1, between 410 MHz and 7.125 GHz, with bandwidths between 20 and 100 MHz; FR2, between 24 and 71 GHz, with bandwidths between 50 and 400 MHz;  FR3 (not standard naming but adopted here for convenience), between 7 and 15 GHz, with bandwidths from 20 to 400 MHz; and Sub-THz, between 100 and 300 GHz, with bandwidths from 1 to 10 GHz.
Second, \ac{RIS} refers to controllable intelligent surfaces, which provide control of the electromagnetic waves, e.g., to create non-natural reflections. RISs are a special case of network-controlled repeaters (NCR) with negative amplification gain \cite{carvalho2024network}. 
  A   \ac{RIS} can be passive (no amplification) or active (with amplification and noise). A \ac{RIS} can be part of the infrastructure or a personal   \ac{RIS}, collocated with the user, and may be locally controlled (autonomous \ac{RIS}) or remotely controlled by the network.
Third, \ac{NTN} refers to infrastructure not physically on the ground, including
  \acp{UAV}, \acp{HAPS}, and \ac{LEO} satellites. \ac{NTN} is here
  considered to be transparent to the \ac{UE}, so the \ac{UE} would not need
  dedicated hardware to communicate with the \ac{NTN} infrastructure.
Fourth, massive \ac{MIMO} refers to a type of \ac{BS} that is equipped with a large
  (tens, hundreds or even thousands) number of phase-coherent antenna elements.
Fifth,   D-MIMO refers to \ac{MIMO} operation with \acp{DU}, each with a small number of antenna
  elements, connected with a central  unit (which may be an
  existing \ac{BS}). The DUs may be phase or time synchronized, by use of a common oscillator, or may require regular synchronization in case they have individual oscillators. 
Sixth and final, an \ac{AI}-driven radio air interface refers to a radio air interface for
  communication and/or sensing, wherein one or multiple functionalities
  in the lower layers across the \ac{Tx} and/or \ac{Rx} are
  replaced or enhanced by \ac{AI}-based methods. These methods learn functionalities to
  enhance performance and adaptability in wireless networks.
\begin{table*}[t]
\centering
\caption{ISAC aspects in relation to the different frequency bands.}
\label{tab:bands}
\begin{tabular}{
|>{\raggedright}p{0.04\linewidth}
|>{\raggedright}p{0.15\linewidth}
|>{\raggedright}p{0.15\linewidth}
|>{\raggedright}p{0.15\linewidth}
|>{\raggedright}p{0.15\linewidth}
|p{0.15\linewidth}
|
}
\hline
\textbf{Band} & \textbf{Coverage} & \textbf{Resolution} & \textbf{Mobility} & \textbf{Channel} & \textbf{Hardware} \\
\hline
FR1 &  High (several km) & {Moderate delay resolution. High spatial resolution under phase coherent operation (D-MIMO), and high angular resolution under massive \ac{MIMO} operation.} & High (100 m/s) & Rich, reflection, limited shadowing & Limited impairments. Near-ideal operation \\
\hline
FR3 & High (though limited outdoor-to-indoor penetration) &{High  delay resolution. High spatial resolution under phase coherent operation (D-MIMO), and high angular resolution under massive \ac{MIMO} operation.} & High (20 m/s) & Less rich due to increased shadowing & Probably limited impairments. \\
\hline
FR2 & Medium (several 100 m) & High delay resolution, medium angle resolution & Medium (10 m/s) & Sparse with few multipath clusters, related to geometry & Mature technology, but more sensitive to impairments \\
\hline
Sub-THz & Short (10s of m) & High delay resolution,  high angle resolution &  slow (\textless{} 1 m/s) & sparse with  few dominant propagation clusters; High signal attenuation due to atmospheric absorption & Severe impairments. Immature technology. \\
\hline
\end{tabular}
\end{table*}

In the next few sections, these six enablers are considered from the
ISAC perspective in terms of \emph{coverage}, \emph{resolution},
\emph{mobility}, \emph{channels}, and \emph{hardware aspects} (e.g., impairments,
calibration, and synchronization).

\subsubsection{Frequency Bands}\label{frequency-bands}
\begin{figure}
    \centering
    \includegraphics[width=0.99\linewidth]{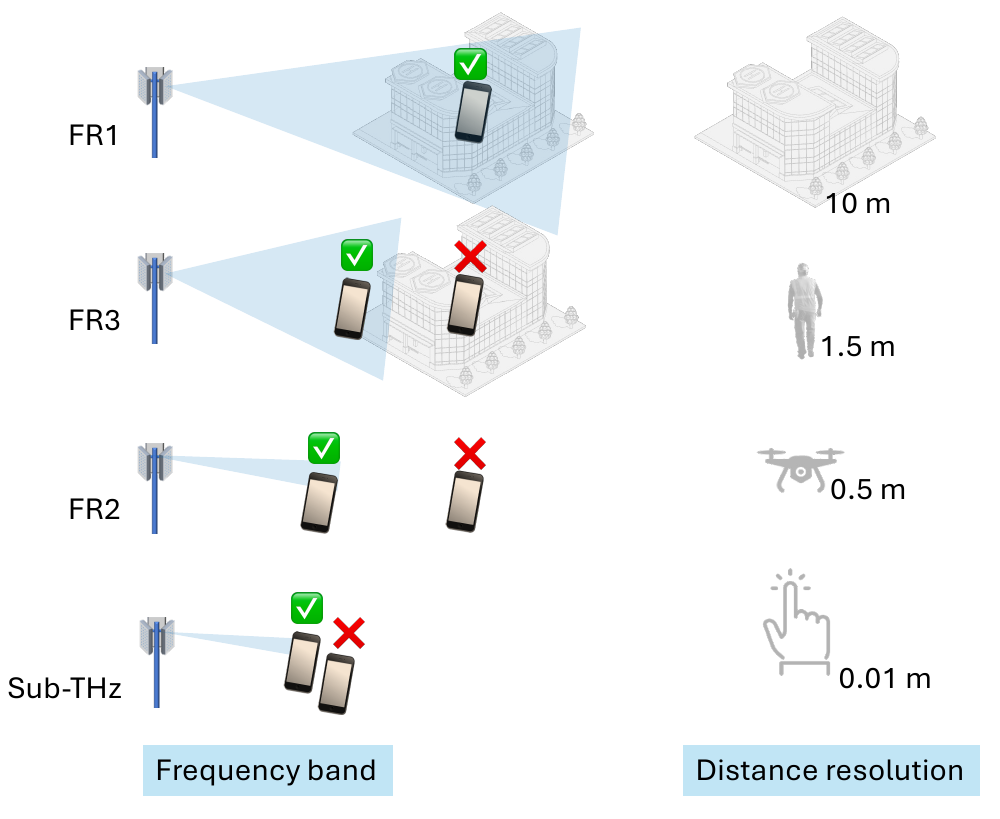}
    \caption{Different 6G frequency bands and their relation to ISAC, in terms of coverage, obstacle penetration, and resolution.}
    \label{fig:frequency-bands}
\end{figure}

The selection of frequency band has severe implications on ISAC, as depicted in Fig.~\ref{fig:frequency-bands}. Higher bands refer to FR2 and sub-THz, while lower bands cover FR1 and FR3.

First, \textit{coverage} in ISAC varies significantly across different frequency bands due to their unique propagation characteristics \cite{cai2022towards}. FR1  offers large coverage areas, robust obstacle penetration, and support for high mobility but is limited in resolution due to narrow bandwidth. FR2  provides higher data rates and temporal resolution, suitable for precise sensing, but has limited coverage and poor obstacle penetration. The upper mid-band FR3 balances coverage and resolution, excelling in urban environments. Sub-THz bands deliver exceptional resolution for static \ac{LoS} applications but suffer from severe coverage constraints and blockage issues.

Second, in terms of
{\textit{resolution}},  ISAC relies on the ability to detect separate
  targets and estimate their state. At higher frequencies, more
  bandwidth is available (typical bandwidth ranges from around 1\% to
  around 10\% of the carrier frequency). Hence, at sub-THz, distance
  resolution (computed from the speed of light divided by the bandwidth) will be at least 30 cm and possibly down to 1 cm, which is
  sufficient for many use cases. At FR2, distance resolution degrades to
  around 0.5 m, and at FR1 to around 10 m under low (e.g., 20 MHz) bandwidths \cite{ge2024v2x}. Moreover, the larger
  number of antennas at higher frequencies provides also angular
  resolution.  On the other hand, when phase coherent
  multistatic operation is considered (see D-MIMO in Section \ref{d-mimo}),
  resolution can be provided by the large aperture. Similarly, massive \ac{MIMO} in combination with moderate bandwidths (e.g.,100 MHz) provides high delay and angle resolution. 
    FR3 bands will likely provide good delay resolution under either large contiguous bandwidth or carrier aggregation, as well as good angle resolution.
  %
%

  A third aspect relates to \textit{mobility}.  At higher frequencies, the use of narrow beams to
  compensate path loss implies more beam training overhead and
  sensitivity to beam pointing errors. Hence, higher bands are
  mainly suitable for low mobility \cite{wang20206g}. 
  In terms of supporting high object mobility, the use of
  analog beamforming limits the time each object can be illuminated when illuminating an environment.
  Hence, only tracking of a small number of targets can supported at
  higher frequencies. 

 Fourth, in terms of the \textit{propagation channel},  at higher frequencies, the channel is
  subject to more severe shadowing, limited scattering, and limited
  dense multipath. This leads to a more geometric channel, determined by
  few multipath clusters, which are themselves due to few-bounce
  interactions with the environment (single or double bounce). At lower
  frequencies, obstacle penetration, dense multipath, and multi-bounce
  interaction lead to a channel that is harder to relate to the
  geometry, unless a  large bandwidth is available (as in ultra-wide
  bandwidth systems). From the point of view of geometric positioning
  and radar-like sensing, operation at higher frequencies is preferred for ISAC.
  When considering large bandwidths, sensing must account for
  frequency-dependent effects to properly estimate channel parameters \cite{10108050}. 
  Finally, as discussed in Section \ref{sec:channel-models}, 
    ISAC channel models are needed to evaluate ISAC signals and methods in a fair and comprehensive way. 
 %
 
  Finally, in terms of \textit{hardware impairments, calibration, and synchronization}, we note that  ISAC
  relies on the ability to extract angles, delays, and Dopplers from
  received signals \cite{chen2023modeling}. All received signals are subject to a variety of
  hardware impairments and hardware limitations, including quantization,
  power amplifier nonlinearity, and phase noise. These impairments tend to be more severe at higher carrier frequencies, making it harder to
extract geometric information. At higher frequencies, different
  forms of calibration are also more challenging (e.g., location
  calibration for phase coherent processing requires knowledge of
  antenna elements' locations within a fraction of the wavelength). 

Table \ref{tab:bands} summarizes the ISAC aspects in relation to the different frequency bands.

\begin{remark}[Deeper insights about FR3 and ISAC from recent studies] The FR3 band (7–24 GHz) is increasingly seen as a key resource for 5G-Advanced and 6G networks \cite{chafii2023twelve, kang2024cellular}, and is critical for advanced wireless communication systems and ISAC applications. However, the unique propagation characteristics of FR3 present both challenges and opportunities in the design and optimization of these systems.
In a study conducted in an indoor factory environment, researchers investigated the propagation characteristics of FR3 at 16.95 GHz. The study highlighted that FR3 exhibits frequency-dependent behaviors, such as reduced path loss in \ac{LoS} scenarios and higher path loss in \ac{NLoS} scenarios due to the dense scattering environment \cite{ying2024upper}. The angular and delay spreads observed were crucial for understanding multipath effects, which directly influence the performance of ISAC systems by affecting radar resolution and communication reliability. Similarly, another comprehensive study on FR1 and FR3 in indoor environments measured material penetration losses and developed channel models for these frequencies. The findings revealed that FR3 bands are more susceptible to attenuation through walls and targets, which poses challenges for NLOS scenarios in ISAC systems. However, these challenges can be mitigated by advanced beamforming and channel estimation techniques \cite{shakya2024comprehensive}.
In another experimental study \cite{Bomfin2024experimental}, FR3 frequencies at 6.5 GHz and 8.75 GHz were analyzed specifically for ISAC applications. The results indicated that lower FR3 frequencies have more distinguishable multipath components in the presence of targets, while higher frequencies experience more signal blockage. This highlights the need for careful frequency selection in ISAC systems to balance resolution and coverage requirements.
\end{remark}

\begin{table*}
\centering
\caption{ISAC aspects in relation to \ac{RIS}.}
\label{tab:RIS}
\begin{tabular}
{|>{\raggedright}p{0.04\linewidth}
|>{\raggedright}p{0.15\linewidth}
|>{\raggedright}p{0.15\linewidth}
|>{\raggedright}p{0.15\linewidth}
|>{\raggedright}p{0.15\linewidth}
|p{0.15\linewidth}
|}
\hline
\textbf{Aspect} & \textbf{Coverage} & \textbf{Resolution} & \textbf{Mobility} & \textbf{Propagation channel} & \textbf{Hardware/Calibration} \\
\hline
FR1 & Short (10s of meters), nearby the \ac{RIS}. & Improved angle resolution & Limited by \ac{RIS} update rate and synchronization with network. & Complex stochastic channel. Limited location dependence. & Mature technology, but limited use for \ac{RIS}. \\
\hline
FR3 & Short (10s of meters), nearby the \ac{RIS}. & Improved angle resolution & Limited by \ac{RIS} update rate and synchronization with network. & Geometric in delay dimensions & Unknown \\
\hline
FR2 &  short (\textless{} 10 m), nearby the \ac{RIS}. Mainly when LOS is blocked. & Somewhat improved angle resolution & Limited by \ac{RIS} update rate and synchronization with network. & High loss, near-field channel effects. Controllable geometric paths. & Susceptible to impairments. Location and orientation need to be calibrated. \\
\hline
Sub-THz &  short (\textless{} 1 m), nearby the \ac{RIS}. Mainly when \ac{LoS} is blocked. & Limited added value in terms of resolution & Limited by beamforming & High loss, near-field channel effects. Controllable geometric paths. & Susceptible to impairments. Location and orientation need to be calibrated. \\
\hline
\end{tabular}
\end{table*}

\subsubsection{Reconfigurable Intelligent Surfaces}\label{ris}

\begin{figure}
    \centering
    \includegraphics[width=0.99\linewidth]{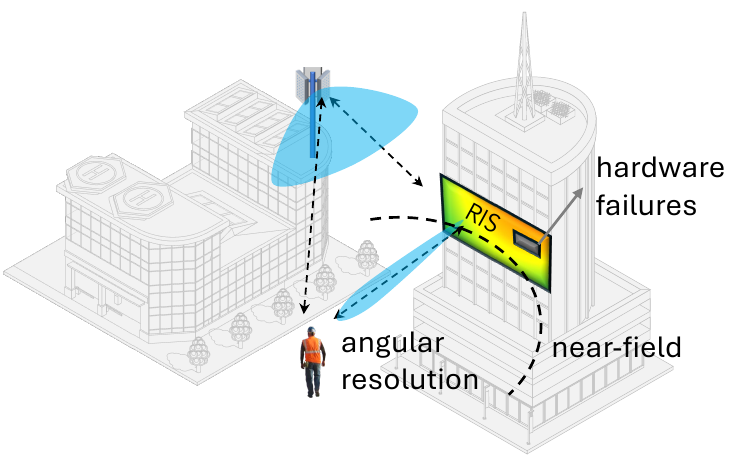}
    \caption{\ac{RIS} for ISAC: the monostatic \ac{BS} with few antennas relies on the \ac{RIS} to provide high sensing angular resolution. If the \ac{RIS} is large, the object may be in the geometric near-field. }
    \label{fig:RIS}
\end{figure}

RISs provide additional controlled propagation paths (see Fig.~\ref{fig:RIS}). While for
communication these paths are mainly to overcome blockages between a
\ac{Tx} and an \ac{Rx} (especially at higher frequencies) for
positioning and sensing, they provide additional geometric information.
The idea is that a path via a \ac{RIS} contains geometric information (via
delays, angles, Dopplers) of the \ac{Tx}, \ac{Rx}, and surrounding
targets. While there are many variations of \ac{RIS}, the basic one is
passive and reflecting, which will be the focus of this analysis.

There are several properties of \ac{RIS} that are important to understand
from an ISAC perspective \cite{9999292}.
%

 In terms of \textit{coverage}, while paths provided by \ac{RIS} are useful to overcome
  blockages, these paths are much weaker than direct paths, depending on
  the size of the \ac{RIS} and the carrier frequency (for a given frequency,
  a larger \ac{RIS} leads to a higher \ac{SNR}; for a given electrical size of the
  \ac{RIS}, a lower carrier frequency leads to higher \ac{SNR}) \cite{bjornson2022reconfigurable}. This makes the
  extraction of the geometric information more difficult and the
  estimated \ac{RIS} channel parameters noisier, when compared to direct
  paths, which act as a  strong interferer. For positioning, this
  implies that \ac{RIS} paths will provide relatively little information in
  terms of the positioning accuracy. For sensing, this implies that \ac{RIS}
  paths are mainly useful when the \ac{RIS} is positioned near a \ac{Tx}, \ac{Rx}, or a target object. 
  This story changes when active RISs are deployed, which can amplify the
  signal in interference-limited deployments. \rev{Distributed RIS deployment to cover large areas require interference coordination and synchronization to avoid unwanted signals \cite{gan2025modeling}. }

In terms of \textit{resolution}, the large aperture brought by the \ac{RIS} provides an
  opportunity for narrow beamforming, even when the \ac{Tx} or
  \ac{Rx} have few antennas. This leads to better multipath resolution
  in the angular domain, which can translate to improved positioning and
  sensing \cite{xu2023joint}. \rev{At the same time, to harness this spatial resolution, careful calibration of the RIS is required. }

In terms of \textit{mobility}, when \ac{Tx}, \ac{Rx}, targets, or \ac{RIS} are mobile, angle
  measurements will be affected by Doppler, which leads to a coupling
  effect between angles and Dopplers (since both depend on the variation
  of the signal phase over time) that is challenging to process \cite{ercan2024ris}. Sensing
  methods that rely on a sequence of \ac{RIS} configurations, during which the target or \ac{UE} are assumed to be quasi-static,  are limited by
  the time it takes to refresh the \ac{RIS} and the level of synchronization
  to other network elements.

In terms of the \textit{propagation channel},  RISs bring several aspects to the
  channel from an ISAC perspective.
\rev{Careful modeling of the RIS propagation is needed to accurately capture the RIS-induced multiplicative multi-hop path loss and multi-bounce propagation \cite{kim2023ris}.}
    The large aperture of the \ac{RIS}  leads to \textit{near-field
    effects}, such as wavefront curvature (i.e., the signal phase
    depends on both angle and distance) and channel non-stationarity
    (i.e., the amplitude and phase of the signal from different \ac{RIS}
    elements varies across the \ac{RIS}). These effects provide additional
    geometric information and can thus be exploited for positioning and
    sensing but come with additional training and processing overheads \cite{bjornson2022reconfigurable}.
    {A \ac{RIS} provides \textit{additional propagation paths}, which in
    turn provides} additional geometric information. 
    Separating \ac{RIS} paths from the uncontrolled multipath requires careful \ac{RIS} profile design. 
    These effects are also to
    be more pronounced when more than one \ac{RIS} is deployed or when mobility is considered.
   
Finally, in terms of  \textit{hardware impairments, calibration, and synchronization},  for a
  \ac{RIS} to serve as a geometric anchor for positioning and sensing, the
  position and orientations needs to be known. Any offsets in position
  and orientation leads to corresponding errors in positioning and
  sensing \cite{ghazalian2024calibration}. If \ac{RIS} is to be a low-cost technology, it will be unlikely
  that each individual \ac{RIS} will be perfectly calibrated. This means that
  limited knowledge will be available about individual \ac{RIS}
  configurations (e.g., main direction, location of nulls), preventing
  the use of high-resolution processing methods that rely on complete
  knowledge of \ac{RIS} weights and steering vectors. This is an important
  limitation that will affect the accuracy of positioning and sensing.
  Additional hardware impairments, such as nonlinearities, losses, and
  \ac{RIS} element failures compound this limitation \cite{ghaseminajm2022ris}.

Table \ref{tab:RIS} summarizes the ISAC aspects in relation to \ac{RIS}.

\subsubsection{Non-Terrestrial Networks}\label{ntn}

\begin{figure}
    \centering
    \includegraphics[width=0.99\linewidth]{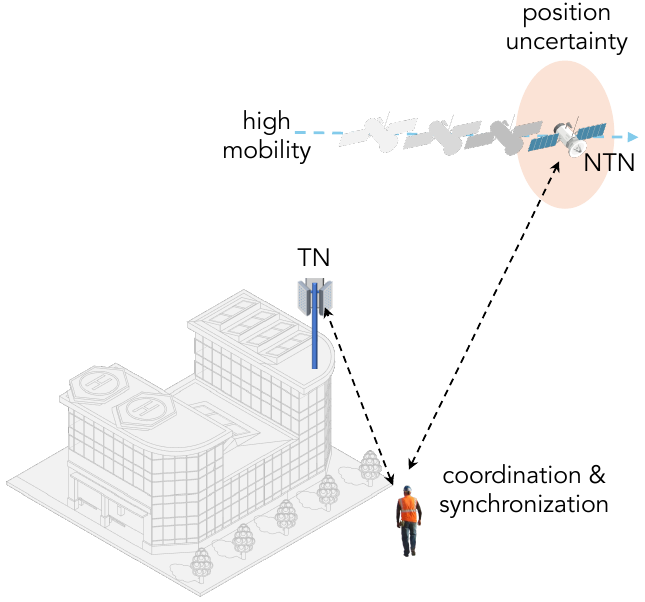}
    \caption{\ac{NTN} for ISAC, where a user is localized based on fusing information from the terrestrial network and the fast-moving \ac{LEO} satellite.}
    \label{fig:NTN}
\end{figure}

\ac{NTN} enables wider communication coverage than conventional terrestrial
infrastructure, while also providing high-rate downlink transmission to \acp{UE}. The use of \ac{NTN} will be considered here in the form of \ac{LEO} satellites (see Fig.~\ref{fig:NTN}) in conjunction with conventional \ac{UE} devices without highly directional arrays (e.g., as a standard GPS receiver), in a way that is transparent to the user (i.e., with \ac{LEO} integrated in the 6G network) \cite{10878492}. Other forms of \ac{NTN} (not covered in this paper) include \acp{UAV} and \ac{HAPS}, as well as satellites with higher orbits. \acp{UAV} are similar to terrestrial nodes, while \ac{HAPS} is a more futuristic technology that will likely not be part of 6G \cite{9275613}.
\ac{LEO} \ac{NTN} with known location provides additional anchors for
positioning and sensing.

In terms of  \textit{coverage},  the large distances between the \ac{NTN} and \ac{UE} (600 --
  2000 km) in combination with limited array gains at the \ac{UE} side leads
  to low \ac{SNR} due to the propagation losses. Hence, while lower bands
  operation is feasible without directional antennas, operation at FR2
  will require continuous adaptive beamforming. Sub-THz operation seems
  unlikely because of high signal attenuation due to atmospheric absorption. Due to \ac{LEO} satellites high mobility, static users are only connected for short periods
  on the order of minutes \cite{gu2024isac}, providing both challenges and opportunities.

In terms of 
  \textit{resolution},  due to preferred operation or lower bands with standard \acp{UE}, resolution in delay will be limited  \cite{dureppagari2023ntn}. In downlink,   angle-of-departure (AOD) is unlikely to support resolvability, due to large distance between the \ac{UE} and satellite. This is in contrast to   terrestrial \acp{BS}, where AOD is available and  informative. On the   other hand, high mobility of the \ac{LEO} satellites  provides opportunities for increased aperture and, thus, resolution \cite{dureppagari2023ntn}.

 High \textit{mobility} of satellites is the main
  distinguishing feature of \ac{LEO} \ac{NTN}, directly affecting the \textit{propagation channel} as well as handover procedures. With carrier frequency offsets, larger Doppler spreads will occur, resulting in shorter coherence intervals. \ac{SNR} for positioning and sensing can be boosted
  with longer integration times \cite{saleh2025integrated}. However, the high satellite mobility
  means that the satellite will move during the observation time, which
  requires careful modeling in terms of delays and Doppler. Under such
  correct models, the mobility will be a benefit, due to the synthetic  aperture effect. This also means that the integration of \ac{NTN} with \ac{RIS} requires careful design to ensure coherent operation. A separate propagation consideration is related to the atmospheric effects, such as scintillation, which requires  dedicated correction procedures \cite{de2025integrated}. 

  Finally, in terms of \textit{hardware impairments, calibration, and synchronization}, the
  location and velocity of the satellite must be known for the
  positioning and sensing to be performed. This information can be
  obtained from GNSS, complemented with terrestrial observations at
  fixed stations. In case several satellites or terrestrial stations are
  used, time and frequency synchronization effects must be considered to
  ensure that signals end up at the \ac{Rx} without inter-symbol or
  intercarrier interference. Hence, the quality of the \ac{LO} at the satellite will play an important role.
  The synchronization depends on the \ac{UE}
  position and velocity, which makes this a closed-loop problem.

Table \ref{tab:NTN} summarizes the ISAC aspects in relation to \ac{NTN}.

\begin{table*}[ht]
\centering
\caption{ISAC aspects in relation to \ac{NTN}-\ac{LEO}.}
\label{tab:NTN}
\begin{tabular}{
|>{\raggedright}p{0.04\linewidth}
|>{\raggedright}p{0.15\linewidth}
|>{\raggedright}p{0.15\linewidth}
|>{\raggedright}p{0.15\linewidth}
|>{\raggedright}p{0.15\linewidth}
|p{0.15\linewidth}
|}
\hline
\textbf{Aspect} & \textbf{Coverage} & \textbf{Resolution} & \textbf{Mobility} & \textbf{Propagation channel} & \textbf{Hardware/Calibration} \\
\hline
FR1 &  Large (several km footprint), time-varying. & Only in delay and Doppler, provided there is a clean \ac{LoS} channel. & Can support highly mobile \acp{UE}  and targets. & High Doppler and varying \ac{LEO} location during transmission interval. Ionospheric effects. & \ac{LEO} location and velocity must be known. Time and frequency sync with network needed for interference coordination. \\
\hline
FR3 &  Large (several km footprint), time-varying, not indoors. & High delay resolution. & Can support highly mobile \acp{UE}  and targets. & High Doppler and varying \ac{LEO} location during transmission interval. Tropospheric impacts. & \ac{LEO} location and velocity must be known. Time and frequency sync with the network needed for interference coordination. \\
\hline
FR2 &  Large (several km footprint), time-varying, not indoors. & High delay resolution, good angle resolution at \ac{UE}. & Lower mobility with \ac{UE}-side beamforming. & High Doppler and varying \ac{LEO} location during transmission interval. Tropospheric impacts. & \ac{LEO} location and velocity must be known. Time and frequency sync with network needed for interference coordination. \\
\hline
Sub-THz & N/A & N/A & N/A & N/A & N/A \\
\hline
\end{tabular}
\end{table*}

\subsubsection{Massive \ac{MIMO}}\label{massive-mimo}

\begin{figure}
    \centering
    \includegraphics[width=0.99\linewidth]{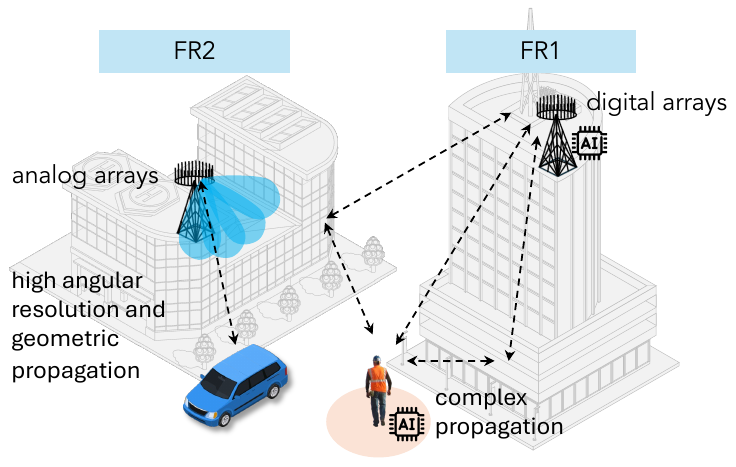}
    \caption{Massive \ac{MIMO} for ISAC, in lower and higher bands.}
    \label{fig:mMIMO}
\end{figure}

Massive \ac{MIMO} (see Fig.~\ref{fig:mMIMO}) relies on a large number of antenna elements at the base
station, in order to spatially multiplex many users over the same
time-frequency resources or to support the transmission to individual users. Due to the large aperture, narrow beams can be
generated, and conversely high-resolution angle estimation becomes
possible. Massive \ac{MIMO} at lower bands  commonly relies on digital arrays,
with dedicated \ac{RF} chains per antenna element, while at higher bands,
analog and hybrid arrays are more practical.
From the ISAC perspective, massive \ac{MIMO} leads to the following
considerations \cite{9971740}.

In terms of \textit{coverage},  at higher bands, massive \ac{MIMO} large arrays provide high-gain pencil beams, in order to boost \ac{SNR} and thus coverage. These directional beams rely on dedicated beam training procedures, which are time-consuming. Beam training can be avoided by harnessing contextual information (e.g., locations of users and targets), provided this information varies slowly. At lower bands, under relatively narrowband transmission (say, 20 MHz), massive \ac{MIMO} beams have little to no geometric information, due to the rich and unresolved multipath in the channel, but provide extended coverage \cite{lu2014overview}. Hence, larger bandwidths (in excess of 100 MHz) to provide meaningful information .

The \textit{resolution} in the angular domain is proportional to the antenna aperture, which for half wavelength spacing yields an angle resolution of at best \(2/N\) radians, assuming $N$ antennas.
Hence, angle resolution alone is generally not sufficient for sensing applications and should be complemented with delay resolution, preferably in the order of 100s of MHz \cite{gao2022integrated}. 
  Massive \ac{MIMO} technology offers varying degrees of \textit{mobility support} depending on the frequency band used. In lower frequency bands massive \ac{MIMO} can effectively handle user mobility thanks to wider beam patterns, robust channel conditions, and manageable Doppler shifts, allowing efficient tracking of moving users with fewer beam updates. In contrast, higher frequency bands  present more significant challenges due to narrower beamwidths, higher Doppler sensitivity, and faster channel variations. Mobility management in these bands demands advanced beam tracking methods, ultra-fast beam alignment procedures, and predictive beamforming techniques to maintain reliable connectivity \cite{lu2014overview}.

The \textit{propagation channel} is characterized by {near-field  effects}. When arrays
  (or RISs) are large, various near-field effects become apparent \cite{wei2021channel}. These include wavefront curvature and channel non-stationarity. Wavefront curvature refers to the variation in delay of waves that impinge
  across an array. In near-field, these delays are a function of the   positions of the source and the position and orientation of the array,   while in far-field, these delays are a function of the direction of
  the source. This difference in delay is due to differences in distance   between the source and each array element and leads to three effects:   location-dependent phase variation across the array (effect of the   delay on the carrier), and varying delay of the signal across the   array (effect of the delay in the time domain), and variable path loss   across the array (effect of path loss on distance). This latter effect   is also known as channel non-stationarity, which can be further   exacerbated by variations in the array element patterns and blockages
  in the environment. Combined, this makes the near-field model more  complicated and computationally harder to extract sensing information from.
  On the other hand, the near-field model is much richer in terms of geometry, providing both higher resolution and increased
  identifiability.

  In terms of \textit{hardware impairments, calibration, and synchronization}, in higher bands, massive arrays will be analog   for cost and complexity reasons. Analog arrays complicate sensing for   several reasons: as mentioned earlier, they require beam alignment or   beam scanning, which is time-consuming. In addition, since the signals
  at the individual array elements are not available, angle estimation   relies on so-called beam-space observations, which require more   complex processing. In lower bands, arrays will likely be digital,   supporting lower latency (and thus higher mobility) and simpler  sensing methods. Another aspect is quantization. In higher bands,
  signal streams can be quantized with a large number of bits, since the   number of streams is limited (one stream in the case of an analog   array). At sub-THz, the massive data rates will likely lead to   few-bits quantization. At lower bands, digital arrays generate massive   amounts of data and will be quantized with few bits per sample. This   leads to nonlinear distortions, which affect the ability to perform   sensing \cite{gustavsson2014impact}. Finally, like \ac{NTN} and \ac{RIS}, to serve as a geometric anchor for  positioning and sensing, the position and orientation of the massive  \ac{MIMO} \ac{BS} needs to be known. Any offsets in \ac{BS} position and  orientation leads to corresponding errors in positioning and sensing.

Table \ref{tab:mMIMO} summarizes the ISAC aspects in relation to massive \ac{MIMO}.

\begin{table*}[ht]
\centering
\caption{ISAC aspects in relation to massive \ac{MIMO}.}
\label{tab:mMIMO}
\begin{tabular}{|>{\raggedright}p{0.04\linewidth}
|>{\raggedright}p{0.15\linewidth}
|>{\raggedright}p{0.15\linewidth}
|>{\raggedright}p{0.15\linewidth}
|>{\raggedright}p{0.15\linewidth}
|p{0.15\linewidth}
|}
\hline
\textbf{Aspect} & \textbf{Coverage} & \textbf{Resolution} & \textbf{Mobility} & \textbf{Propagation channel} & \textbf{Hardware/Calibration} \\
\hline
FR1 & Large (\textgreater{} 1 km) & High angle resolution & Can support highly mobile \acp{UE} and targets. & Stochastic channel & Digital arrays with few-bit quantization. Requires specialized sensing methods, e.g., based on \ac{AI}. \\
\hline
FR3 & Large, but not indoors & High angle and resolution. & Can support highly mobile \acp{UE} and targets. & Geometric & Unknown \\
\hline
FR2 & Limited (\textless{} 300 m) & High delay and angle resolution & Limited mobility due to analog beamforming & Complex channel with wideband effects and near-field effects & Analog arrays with beamspace observations. \\
\hline
Sub-THz & Limited (\textless{} 10 m) & Extreme delay and angle resolution & Limited mobility due to analog beamforming & Complex channel with wideband effects and near-field effects & Analog arrays with beamspace observations. \\
\hline
\end{tabular}
\end{table*}

\subsubsection{Distributed \ac{MIMO}}\label{d-mimo}

\begin{figure}
    \centering
    \includegraphics[width=0.99\linewidth]{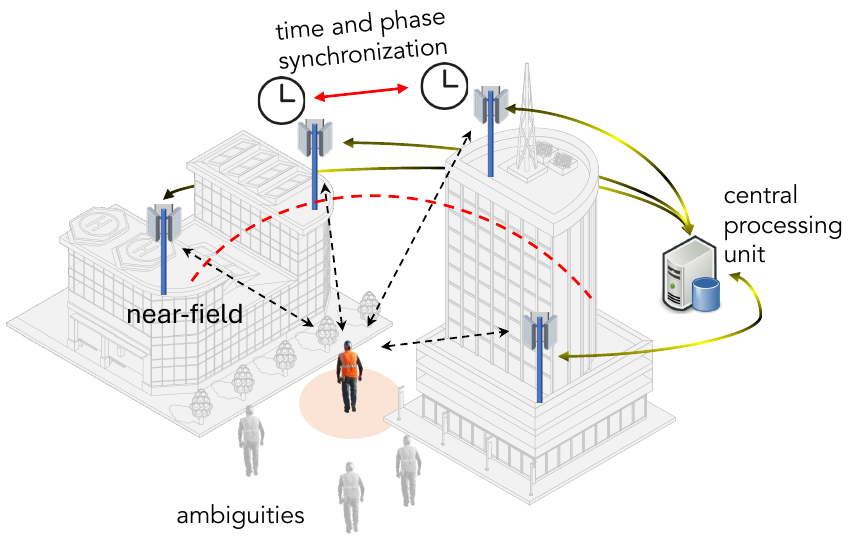}
    \caption{D-MIMO for ISAC relies on cooperation between widely \acp{DU} via central processing. The level of synchronicity between the units is critical for both communication and sensing.}
    \label{fig:dMIMO}
\end{figure}

Distributed MIMO can be treated similarly to massive MIMO, with the difference that \ac{BS} elements (referred to as \acp{AP} or \acp{DU} are widely distributed in space (see Fig.~\ref{fig:dMIMO}). These bring some specific ISAC aspects \cite{guo2024integrated}.

In terms of \textit{coverage}, D-MIMO (whether phase-coherent in lower bands or
  time-coherent at higher bands) provides a widely distributed  perspective, and thus a way to localize and track \acp{UE} and targets beyond the \ac{FoV} of a single DU \cite{li2023toward}.

  In terms of \textit{resolution},  at lower bands, phase-coherent operation of different D-MIMO DUs extends the high angular resolution into the spatial domain since the distributed array sees targets and \acp{UE} from different directions. Due to gaps in the array (with respect to  half-wavelength spacing), the high spatial resolution is coupled with ambiguities, which may mask weak targets in the presence of stronger (more reflective) targets \cite{fascista2025joint}. Nevertheless, in combination with delay resolution, high sensing performance can be expected. At higher bands, resolution will instead be provided in terms of delay and angle, thanks to large available bandwidths and large arrays at each DU.

  In lower bands, high \textit{mobility} can be supported at the expense of regular updating of beamforming weights. At higher bands, beam tracking will limit mobility \cite{elhoushy2021cell}.

The
  \textit{propagation channel},  under phase-coherent operation, leads to near-field propagation effects such as wavefront curvature and channel non-stationarity. In contrast to \ac{RIS}, each channel is directly observable, making it easier to estimate the channels and use them for ISAC \cite{guo2024integrated}.

  Finally, \textit{hardware impairments, calibration, and synchronization} include calibrating the position and orientation of each
  individual element and also synchronization among the DUs. For phase coherent operation,
  this knowledge should be correct within a fraction of a wavelength \cite{guo2024integrated}.

In D-MIMO systems, achieving effective ISAC requires coordination among DUs, involving either centralized or distributed processing strategies. Centralized processing demands extensive backhaul bandwidth to aggregate raw \ac{IQ} samples from DUs, particularly in lower bands, leveraging coherent processing for spatial resolution. Conversely, in higher bands, local estimation at DUs followed by centralized fusion reduces backhaul load but increases computational complexity at the fusion center. Additionally, efficient ISAC operation places stringent throughput, latency, and synchronization demands on fronthaul networks, driving the adoption of advanced technologies such as optical fronthaul and \ac{mmWave} backhaul to achieve the necessary bandwidth, latency, and reliability trade-offs \cite{guo2024integrated}.

Table \ref{tab:DMIMO} summarizes the ISAC aspects in relation to D-MIMO.
\begin{table*}[ht]
\centering
\caption{ISAC aspects in relation to D-MIMO.}
\label{tab:DMIMO}
\begin{tabular}{|>{\raggedright}p{0.04\linewidth}
|>{\raggedright}p{0.15\linewidth}
|>{\raggedright}p{0.15\linewidth}
|>{\raggedright}p{0.15\linewidth}
|>{\raggedright}p{0.15\linewidth}
|p{0.15\linewidth}
|}
\hline
\textbf{Aspect} & \textbf{Coverage} & \textbf{Resolution} & \textbf{Mobility} & \textbf{Propagation channel} & \textbf{Hardware/Calibration} \\
\hline
FR1 & Large (\textgreater{} 1 km) & High spatial resolution under phase coherent operation. & High, but high overhead. & Near-field channel model & Phase sync needed. \\
\hline
 FR3 & Large (\textgreater{} 1 km) thanks to node cooperation. & High spatial resolution under phase coherent operation. High delay resolution per DU. & High, but high overhead & Near-field channel model. & Phase sync needed. \\
\hline
 FR2 & Large (\textgreater{} 1 km) thanks to node cooperation. & High delay resolution and angle resolution per \ac{DU}. & Medium & Far-field channel model. & Time sync needed. \\
\hline
Sub-THz & Large (\textgreater{} 1 km) thanks to node cooperation.& Extreme delay resolution and angle resolution per \ac{DU}. & Medium. & Far-field channel model & Time sync needed. \\
\hline
\end{tabular}
\end{table*}

\subsection{Analysis of ISAC
Configurations}\label{analysis-of-jcas-modalities}

ISAC will be considered in terms of different sensing configurations, including monostatic, bistatic, and multistatic sensing. In each configuration, a distinction will be made in terms of transmitters and
receivers being \acp{UE} or BSs.

\begin{remark}[Location knowledge regarding \acp{UE} and BSs]
    BSs
 are typically considered as static entities with known
location and orientation. In the case of \ac{NTN}, however, BSs may be mobile
(e.g., \acp{UAV} or \ac{LEO} satellites). Their location and orientation may still
be considered to be known (from a combination of GNSS and internal
measurements). \acp{UE} are typically considered to be mobile with unknown
location and orientation. Nevertheless, 3GPP also considers static \acp{UE}
with known location and orientation, known as positioning reference
units (PRUs). {From the geometric perspective,
these PRUs fulfill a similar role as BSs, though functionally they are \acp{UE}}, with the distinction that they
have no wired connection to the network and are not synchronized to the
network (to the extent needed for positioning purposes). These same PRUs
can thus also fulfill the role of sensing reference units (SRUs).
\end{remark}

In the following, we consider each sensing configuration  in light of the applications
from Section \ref{why-jcas} and the enablers from Section \ref{jcas-and-the-6g-radio-enablers}. The analysis does
not extend to higher-level processing (e.g., cooperation among different
entities, fusion of different sensing configurations) or the comparison with
existing sensing technologies (e.g., radar, camera, UWB), which will be treated in Section \ref{protocols-and-functions-for-jcas}.

\subsubsection{Participation of \acp{UE} in ISAC}

Depending on the capabilities and resources of \acp{UE}, they may participate in ISAC in different manners. They could actively perform sensing, rely on network assistance, collaborate with other \acp{UE} or BSs, or fully rely on network-based sensing. Four different approaches are identified. 
\begin{itemize}
    \item 
\textit{Individual (Monostatic) Sensing:} {Mainly relevant for unlicensed spectrum.} The \ac{UE} actively transmits sensing signals and processes the reflected signals directly on-device. This approach offers the lowest possible latency but requires advanced on-board sensing and processing capabilities (hardware and software), and results in significant resource consumption. It is most useful to ensure situational awareness (e.g., collision detection, self-localization, navigation) in critical scenarios and \acp{UE} with powerful capabilities (e.g., mobile robots, autonomous vehicles, \acp{UAV}) . 
\item  \textit{Assisted Sensing:} 
In the assisted sensing scenario, the \ac{UE} relies on network functions for sensing. The network provides fully or partially processed sensing information, such as a point cloud, a complete map, object detection and tracking, or derived information like collision warnings. This processing can be assisted by \ac{AI}. The network may deliver sensing data to a single dedicated \ac{UE} (unicast), a group of \acp{UE} (multicast), or all nearby \acp{UE} (broadcast).
The \ac{UE} may or may not have its own sensing capability, as the network can integrate data from the \ac{UE} itself, other \acp{UE}, or BSs via the network function. Data fusion is based on available data and sensing information, enabling enhanced sensing accuracy, such as improved positioning. In contrast to monostatic sensing, assisted sensing offloads data processing to the network, making it accessible for less powerful \acp{UE}. However, this results in higher latencies, which may make it less suitable for time-critical applications.
%
Network functions (see Section \ref{protocols-and-functions-for-jcas}) like “sensing request,” “sensing control,” and “sensing processing” may be required to support this approach. For power efficiency, the \ac{BS} might implement power management mechanisms, such as initiating sensing and processing only when requested by a \ac{UE}.

\item \textit{Collaborative Sensing:} 
In the collaborative sensing scenario, a group of \acp{UE} and/or BSs work together to enhance sensing capabilities\rev{~\cite{peng2024trajectory}}. The \ac{UE} sends its own raw or pre-processed sensing data to one or multiple nearby sink nodes, such as BSs or peers, which have sufficient processing power to act on or forward sensing requests and fuse the received data. While transmitting raw data provides greater flexibility and can yield more accurate sensing outputs, it requires significantly higher data rates and precise time synchronization across sources.
Data fusion is performed based on the requested sensing service and required accuracy. The \ac{UE} may or may not use its own sensing data, and this approach can be combined with assisted sensing, where the \ac{UE} either provides only its sensing data or also consumes sensing data provided by the network. The network can enhance sensing accuracy, such as positioning, by incorporating data from other localization methods like GPS or additional sensor inputs. Time information is often critical to ensure proper data fusion.
This scenario requires network functions such as “sensing request,” “sensing control,” and “sensing processing.” 
\item \textit{Sensing Network Function:} 
In the sensing network function scenario, the network implements sensing or radar as a service, enabled by BSs and/or dedicated network sensing devices that collect and process sensing data. \acp{UE} can request the use of these sensing functions and receive outputs in formats similar to those used in assisted sensing. The sensing function may also be initiated by the network, where the \ac{BS} might request sensing data from an attached \ac{UE}, if the \ac{UE} agrees.
The network can broadcast, multicast, or unicast its sensing capabilities, specifying available options such as local or extended map information across several BSs (e.g., in Distributed MIMO), object tracking based on object detection via a point cloud, or motion estimations, which can be assisted by \ac{AI}. The network may also employ data fusion from other localization sources, such as GPS or additional sensor data, to enhance sensing accuracy.
To support this scenario, the \ac{UE} may require an initial attach procedure to request and provide sensing capabilities. This scenario requires network functions like “sensing request,” “sensing control,” and “sensing processing.”

\end{itemize}
\subsubsection{Monostatic Sensing}\label{monostatic-sensing}

\begin{figure}
    \centering
    \includegraphics[width=0.99\linewidth]{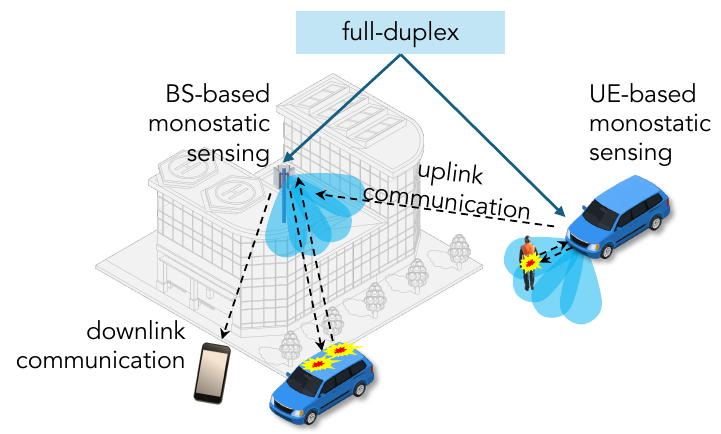}
    \caption{Monostatic sensing may be performed by \acp{UE} (in uplink) or BSs (in downlink), and require full-duplex capabilities, to suppress self-interference. }
    \label{fig:monostatic}
\end{figure}

In monostatic sensing \cite{lu2022degrees,holisticISAC_2025}, the \ac{Tx}
and \ac{Rx} are co-located (see Fig.~\ref{fig:monostatic}). This generally means that they share a clock and knowledge of the \ac{Tx} signals. Importantly, full-duplex
capabilities are needed to transmit and receive at the same time. In turn, this demands powerful interference cancellation that still preserves the information in the back-scattered channel.
Monostatic sensing also has specific
relations to some of the radio enablers from Section \ref{jcas-and-the-6g-radio-enablers}, in particular,
to \emph{RIS} and \emph{massive MIMO}. When installed on a wall, a \ac{RIS}
can serve as a geometric reference (e.g., to help a \ac{UE} to localize
itself in a global coordinate system) and provide around-the-corner
sensing capabilities. A massive MIMO \ac{BS} acting as a monostatic sensor
can provide 3D or 4D imaging capabilities in the vicinity of the \ac{BS}.

Monostatic sensing can be performed by either a \ac{UE} or a \ac{BS}.  When monostatic
sensing is performed by the \ac{UE}, sensing data collected by the \ac{UE} is in
the coordinate system of the \ac{UE} \cite{guidi2015personal}.  When monostatic sensing is performed by the \ac{BS}, sensing data
collected is provided in an absolute coordinate system.  These functionalities can support some of the use cases from Section \ref{why-jcas}  (see Tables \ref{tab:monoUE}--\ref{tab:monoBS}).

\begin{table}
\centering
\caption{ISAC use cases matched to \ac{UE}-based monostatic sensing.}
\label{tab:monoUE}
\begin{tabular}{|p{0.25\linewidth}|p{0.6\linewidth}|}
\hline
\textbf{Use case} & \textbf{Example} \\
\hline
Optimizing Mobile Network Performance & \ac{UE} can sense targets in its vicinity and inform the network about foreseen obstacles. \\
\hline
Enhanced \ac{UAV} Management and Safety & Collision avoidance. \\
\hline
Automotive and Smart Transportation & \ac{UE} can operate like a vehicle radar and provide relative location information. \\
\hline
Industry and Logistics & \ac{UE} can operate like a vehicle radar and provide relative location information. \\
\hline
Digital Twinning & \ac{UE} can sense a survey in a specific area to provide update/refine inputs about targets or events to the digital twin. \\
\hline
\end{tabular}
\end{table}

\begin{table}
\centering
\caption{ISAC use cases matched to \ac{BS}-based monostatic sensing.}
\label{tab:monoBS}
\begin{tabular}{|p{0.25\linewidth}|p{0.6\linewidth}|}
\hline
\textbf{Use case} & \textbf{Example} \\
\hline
Optimizing Mobile Network Performance & The \ac{BS} can sense targets (possibly including users), allowing it to allocate radio resources (e.g., beamforming) with reduced overhead. \\
\hline
Enhanced \ac{UAV} Management and Safety & Good match for this application, especially since the channel to/from the \ac{UAV} is relatively uncluttered. Benefits from high Doppler resolution. \\
\hline
Automotive and Smart Transportation & A \ac{BS} can sense and survey a certain area and track vehicles and other road users. \\
\hline
Industry and Logistics & A BS can sense and survey a certain area, including automated vehicles and personnel. \\
\hline
Digital Twinning & A BS can sense and survey a certain area to provide inputs about targets or events to the digital twin. \\
\hline
\end{tabular}
\end{table}

\subsubsection{Bistatic Sensing}\label{bistatic-sensing}

\begin{figure}
    \centering
    \includegraphics[width=0.8\linewidth]{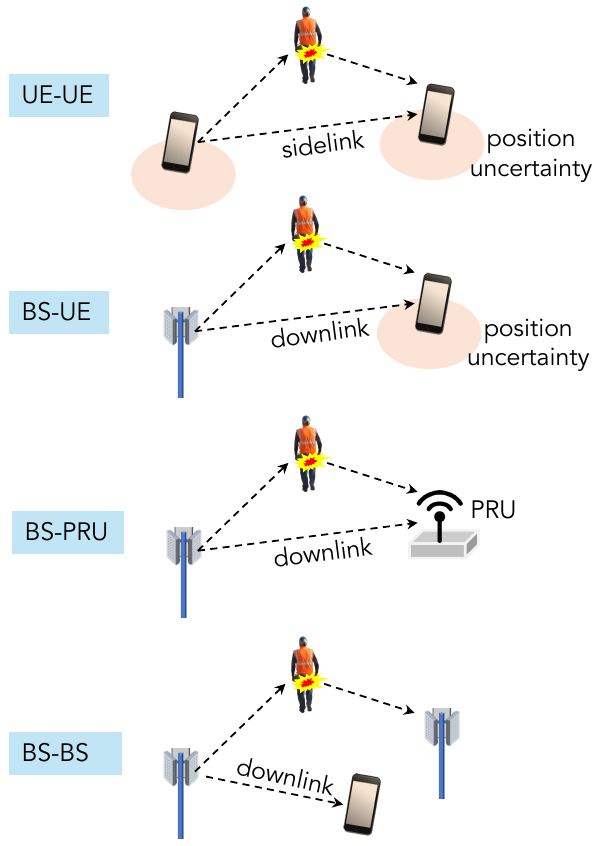}
    \caption{Bistatic sensing. The focus of the figure is on downlink, though uplink is also feasible.}
    \label{fig:bistatic}
\end{figure}

In bistatic sensing, the \ac{Tx}
and \ac{Rx} are physically separate. This implies generally that the \ac{Rx} is not perfectly synchronized with the \ac{Tx} and that the \ac{Rx} may not have full knowledge of the transmitted signal (unless dedicated pilots are used) \cite{wu2024sensing}. In addition, \ac{Tx} and \ac{Rx} must agree on a common reference frame in which to place the detected targets.
Bistatic sensing also has specific
relations to some of the radio enablers from Section \ref{jcas-and-the-6g-radio-enablers}. When the BS is
a \ac{LEO} satellite, the BS-\ac{UE} signal over an extended observation period can be used to localize a \ac{UE} and also to sense targets near the \ac{UE}. Care must be taken to synchronize BSs in time and frequency to avoid interference and to account for location uncertainties of the satellite. 
Depending on whether the \ac{Tx} or \ac{Rx} are a \ac{UE} or BS, there are four cases to consider (see Fig.~\ref{fig:bistatic}): \ac{UE}-\ac{UE} (first node is \ac{Tx}, second \ac{Rx}), BS-\ac{UE}, \ac{UE}-BS, BS-BS. 
\begin{enumerate}
    \item \textit{\ac{UE}-\ac{UE} bistatic sensing: } 
    {Mainly relevant for  operation under unlicensed spectrum.}
    \ac{UE}-\ac{UE} bistatic sensing operates via sidelink transmission \cite{chen2023riss}. To avoid
problems related to synchronization and limited knowledge regarding the
transmitted signal, a round-trip-time (RTT) protocol and the
transmissions of pilots can be considered. From that, the \acp{UE} have
knowledge of their distance, as well as differential distances to
surrounding targets, in their local coordinate system. The different
radio enablers can provide specific benefits. Finally, the use of \ac{UE}-\ac{UE}
bistatic sensing can be related to the use cases from Section \ref{why-jcas}  as
shown in Table \ref{tab:biUEUE}.
\item \textit{BS-\ac{UE} and \ac{UE}-BS bistatic
sensing:}
The two cases where the two nodes are a BS and mobile \ac{UE} are discussed jointly, given that the channel is reciprocal, so that the same information is available in uplink as in downlink. Information from the
corresponding received signals can be used for sensing, provided the position and orientation of the UE are known (either from being fixed,
 or inferred from the bistatic sensing
measurements\footnote{Also referred to as radio \ac{SLAM} \cite{ge2022mmwave} or channel \ac{SLAM}.}, or from
other sensors) \cite{ge2022mmwave}. The use of BS-UE and UE-BS bistatic sensing relates to
the use cases from Section \ref{why-jcas}  is detailed in Table \ref{tab:biUEBS}.
\item \textit{BS-BS bistatic sensing:}
The last bistatic sensing configuration is BS-BS \cite{fodor2025bistatic}. The sensing signals can be of different forms, including (i)
  Downlink signals from a \ac{Tx} BS, received by the receiving BS.  {This mode is expected to be practically most likely approach for bistatic sensing}; (ii)
  Dedicated BS-BS signals in the form of new pilots;
BS-BS bistatic sensing can rely on knowledge of the BSs location and orientation, so that all sensing information is automatically provided
in a global and coordinate system. As BSs are typically static, they can be tightly synchronized via GPS or wired links.  The use of BS-BS bistatic sensing relates to the use cases from Section \ref{why-jcas}  is detailed in Table \ref{tab:biBSBS}.
\end{enumerate}

\begin{table}
\centering
\caption{ISAC use cases matched to UE-UE bistatic sensing.}
\label{tab:biUEUE}
\begin{tabular}{|p{0.25\linewidth}|p{0.6\linewidth}|}
\hline
\textbf{Use case} & \textbf{Example} \\
\hline
Optimizing Mobile Network Performance & UE position information can be used to reduce beam alignment overheads in the case of sidelink communications.  \\
\hline
Enhanced \ac{UAV} Management and Safety &  Swarm coordination. \\
\hline
Automotive and Smart Transportation & Relative positioning between \acp{UE} (e.g., vehicles) can support platooning. \\
\hline
Industry and Logistics & Relative positioning between \acp{UE} (e.g., robots) can support inter-robot coordination. \\
\hline
Digital Twinning & Predictive collision avoidance for collaborative warehouse robots. \\
\hline
\end{tabular}
\end{table}

\begin{table}
\centering
\caption{ISAC use cases matched to UE-BS and BS-UE bistatic sensing.}
\label{tab:biUEBS}
\begin{tabular}{|p{0.25\linewidth}|p{0.6\linewidth}|}
\hline
\textbf{Use case} & \textbf{Example} \\
\hline
Optimizing Mobile Network Performance & UE position information can be used to reduce beam alignment overheads. Knowledge of future blockages can be used to trigger handovers. \\
\hline
Enhanced \ac{UAV} Management and Safety & Swarm coordination. \\
\hline
Automotive and Smart Transportation & UE position information and sensing of nearby targets in an absolute coordinate system is important in this use case. \\
\hline
Industry and Logistics & UE position information and sensing of nearby targets in an absolute coordinate system is important in this use case. \\
\hline
Digital Twinning & Tracking of \acp{UE} provides a real-time picture of the network status. This information can be augmented with bistatic sensing information in an absolute coordinate system. \\
\hline
\end{tabular}
\end{table}

\begin{table}
\centering
\caption{ISAC use cases matched to BS-BS bistatic sensing.}
\label{tab:biBSBS}
\begin{tabular}{|p{0.25\linewidth}|p{0.6\linewidth}|}
\hline
\textbf{Use case} & \textbf{Example} \\
\hline
Optimizing Mobile Network Performance & Information about presence and location of (moving) targets enables advance warning of blockages. \\
\hline
Enhanced \ac{UAV} Management and Safety & Sensing of moving targets between two base stations can detect and localize \acp{UAV}. \\
\hline
Automotive and Smart Transportation & Information about presence and location of (moving) targets enables advanced collision warning. \\
\hline
Industry and Logistics & Information about presence and location of (moving) targets enables advanced collision warning. \\
\hline
Digital Twinning & Tracking of targets provides a real-time picture of their absolute location in an environment, suitable for real-time digital twin. \\
\hline
\end{tabular}
\end{table}

\subsubsection{Multistatic Sensing}\label{multistatic-sensing}

\begin{figure*}
    \centering
    \includegraphics[width=0.89\linewidth]{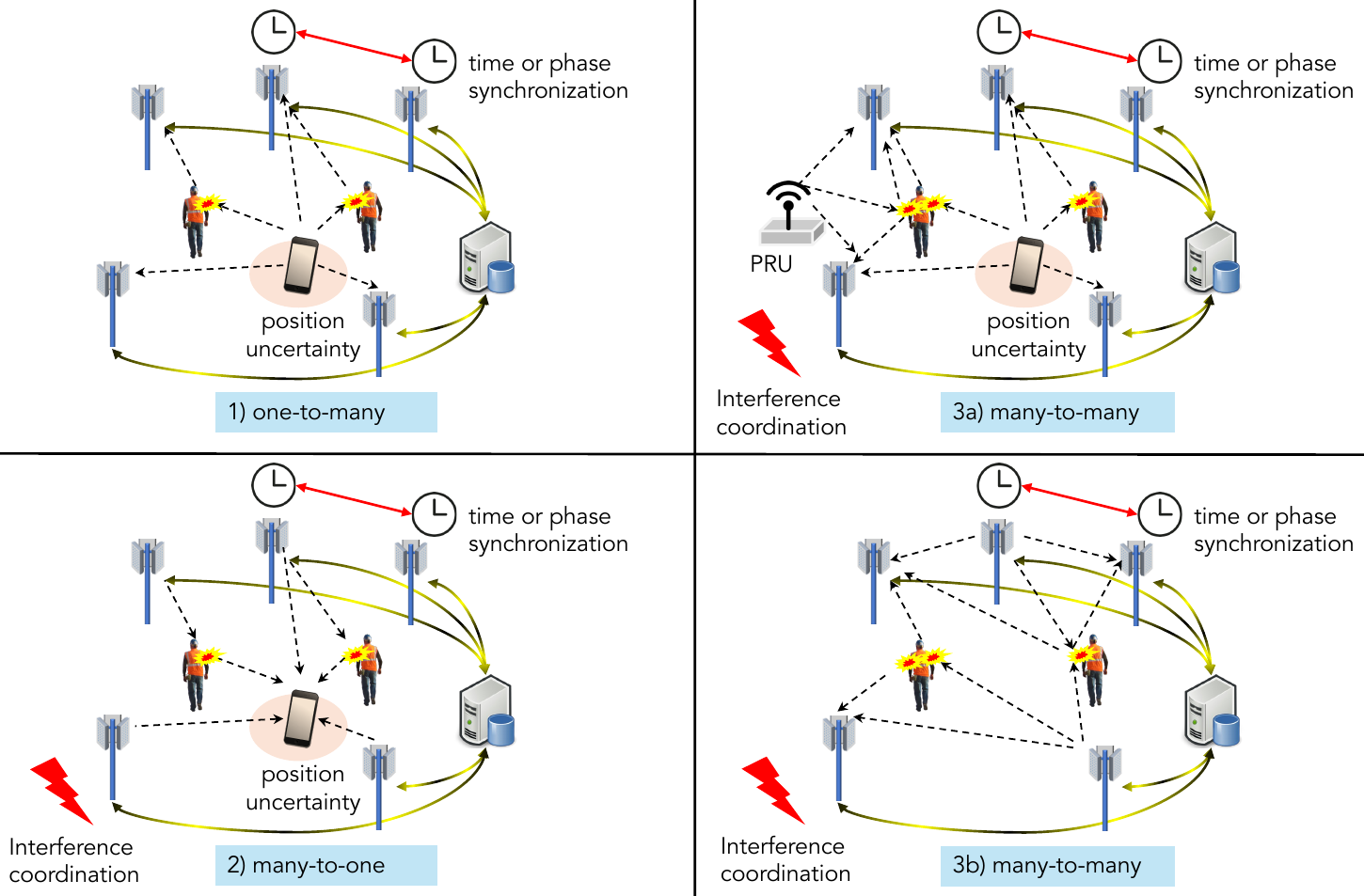}
    \caption{Multistatic sensing exemplifying scenarios: 1) uplink UE positioning and target sensing; 2) downlink UE positioning and target sensing; 3a) uplink transmission from several \acp{UE} (one is a PRU), which requires coordination to avoid interference; 3b) downlink transmission from several DUs, where other DUs receive and process the signal scattered from the targets. Again the transmitters need to coordinate to avoid or control interference. }
    \label{fig:multistatic}
\end{figure*}

 The final sensing configuration is
multistatic sensing, where there are multiple transmitters and/or
multiple receivers \cite{liu2024cooperative,hong2024cooperative}. The transmitters and receivers could be BSs, \acp{UE}, or
PRUs. Multistatic sensing can rely on \emph{phase-coherence} between the
different nodes (which typically requires wired connections),
\emph{time-coherence} between the different nodes (which can be achieved
via over-the-air synchronization or from GNSS), or \emph{non-coherent}
nodes (neither time nor phase synchronized). In the last case, RTT
signals can be used to provide useful sensing information or to
time-synchronize the different nodes. Hence, the discussion below will
be limited to the two synchronized cases. Under phase-coherent
operation,  precise location calibration is needed (at the
sub-wavelength level), which is challenging even at lower frequencies
and applicable to FR2 and sub-THz for narrow niche applications. Also at
phase coherent operation, near-field channel models  should be used due
to the large aperture, which also provides high spatial resolution.
Phase-coherent operation is limited
to lower bands, while time-coherent operation is possible for higher bands as well. D-MIMO is a key technology to support multistatic
sensing.
 Both time- and phase coherent
multistatic sensing can be used inside the use cases from Section \ref{why-jcas} 
as detailed in Table \ref{tab:multi}.

\begin{table}
\centering
\caption{ISAC use cases matched to multistatic sensing.}
\label{tab:multi}
\begin{tabular}{|p{0.25\linewidth}|p{0.6\linewidth}|}
\hline
\textbf{Use case} & \textbf{Example} \\
\hline
Optimizing Mobile Network Performance & Detecting and localizing targets or blockages can be used to improve communication beamforming. \\
\hline
Enhanced \ac{UAV} Management and Safety & Sensing of moving targets between base stations or between BS and UE/PRU can detect and localize \acp{UAV}. \\
\hline
Automotive and Smart Transportation & Information about presence and location of (moving) targets enables advanced collision warning. \\
\hline
Industry and Logistics & Information about presence and location of (moving) targets enables advanced collision warning. \\
\hline
Digital Twinning & Tracking of targets provides a real-time picture of their absolute location in an environment, suitable for real-time digital twin. \\
\hline
\end{tabular}
\end{table}

Multistatic sensing is broken down into three cases:  one
\ac{Tx} to several (time- or phase-synchronized) receivers,
 several (time- or phase-synchronized) transmitters to one
\ac{Rx};  several (time- or phase-synchronized)
transmitters to several (time- or phase-synchronized) receivers (see Fig.~\ref{fig:multistatic} for examples).
\begin{enumerate}
    \item 
\textit{One-to-many multistatic
sensing:}
One-to-many multistatic sensing mainly relates to uplink sensing (UE to several BSs, PRU to several BSs). The analysis breaks down into two cases, depending on the type of synchronization.  {Under time-synchronized BSs}, this type of sensing provides the means for time- and angle-based UE positioning (e.g., TDoA or RTT-based) and sensing targets. With a PRU, positioning is not needed so only targets are sensed. Since the BSs are not phase coherent, separate
processing of each BSs signal is possible, followed by non-coherent fusion. The different radio enablers can provide specific benefits. On the other hand,
{when the BSs can be  phase-synchronized}, this type of sensing provides the means for carrier phase-based positioning of a UE, supported by a PRU as a reference station. Joint processing across the BSs is needed for full exploitation of the phase information and large aperture in uplink, placing high demands on backhaul to carry all the raw information.

\item 
\textit{Many-to-one multistatic
sensing:}\label{many-to-one-multistatic-sensing}
Many-to-one multistatic sensing mainly relates to downlink sensing (several BSs to one UE, several BSs to one PRU). In the downlink case, the signals from the different base stations should be properly coordinated.
{Under time-synchronized BSs}, this coordination is of the form of orthogonal signals (e.g., using orthogonal time-frequency resources), so that the \ac{Rx} UE or PRU can separate the signals from the different transmitters and extract the (bistatic) sensing information from each link.
On the other hand, {under phase-synchronized BSs}, this coordination can take two
forms: (ii)
  \emph{Transmitting orthogonal signals} (e.g., using orthogonal   time-frequency resources): this leads to the largest spatial   resolution due to the resulting virtual array aperture and allows the
  \ac{Rx} UE/PRU to separate the signals from the different   transmitters; (ii) \emph{Transmitting beamformed signal} (e.g., focusing on a specific   location): this leads to the largest signal power in the targeted   location. From the \ac{Rx} UE/PRU, the transmission appears to come
  from a single node.
\item 
\textit{Many-to-many multistatic
sensing}\label{many-to-many-multistatic-sensing}
The last multistatic sensing case is the many-to-many case. This can pertain to several BSs to/from several \acp{UE}, several BS to/from several
PRUs, and several BSs to several other BSs. As in the many-to-one case, signals from different transmitters should be
coordinated, depending on if the BSs nodes are time-synchronized or
phase-synchronized. In addition, there is a sensor selection problem in
the BSs-to-BSs case in terms of deciding which BSs should act as
transmitters and which as receivers.
\end{enumerate}

\section{\ac{RF} Hardware Considerations of
ISAC}\label{hardware-considerations-of-jcas}

Traditional transceivers used in cellular networks are usually primarily designed for wideband communications. However, when the same HW resources are utilized for sensing, the hardware architecture, calibration, and various impairments such as phase noise and nonlinear distortion must also be addressed from the sensing perspective. Moreover, the impact and requirements are also highly dependent on the sensing configurations discussed in Section \ref{analysis-of-jcas-modalities}. For example, monostatic sensing requires full duplex support highlighting problems with self-interference and nonlinear distortion, while bistatic and MIMO approaches would benefit from phase stability and coherency. Sensing approaches relying on multiple antennas and accurate beamforming make it inevitable also to consider the impact of beamforming/array architectures. Moreover, the device/UE side has different limitations from the \ac{RF} HW employed at the infrastructure/\ac{BS} for example in terms of power and array size, as well as possible HW complexity and cost. 
In general, the HW effect is also highly dependent on the used waveform: When using traditional communication waveforms with rather high \ac{PAPR}, the effect of \ac{PA} nonlinear distortions cannot be ignored. The waveform's ability to handle the impairments is one of the keys when considering the hardware impairments in ISAC as discussed in \cite{Abdur_ISAC}.   

The section discusses first the transceiver array architectures and their utilization for sensing. Then, the full-duplex aspects related to \ac{RF} HW are discussed in the context of monostatic sensing. Thereafter, the section introduces and analyzes the key \ac{RF} impairments in sensing scenarios with a focus on \ac{PA} nonlinear distortion and phase noise. Phase noise mitigation is also discussed. Then, the importance of the anchor HW calibration and deployment for ISAC are discussed. Finally, existing hardware prototypes used for ISAC demonstrations are reviewed and discussed.

\subsection{\ac{RF} Transceiver Array Architectures and Full Duplex}\label{array-structures} 

\ac{RF} \ac{TRx} architecture in general refers to the architecture of how a radio signal is generated from the digital waveform and delivered to the antennas and vice versa. Typical radars, such as \ac{FMCW}, use a different architecture from traditional communication radio. In today's systems, antenna arrays are considered as de-facto. Hence, in this context, the \ac{RF} \ac{TRx} architecture also typically refers to the beamforming/array architecture. In this subsection, beamforming architectures are first discussed in general, followed by enabling architectures specifically for ISAC. Then, full-duplex aspects of ISAC systems are discussed.

\subsubsection{Beamforming Architectures}

Modern base stations are typically equipped with tens, in some cases even hundreds of antennas depending on the used operation frequency. Architectures on how to feed and control signals in \ac{Tx} and \ac{Rx} antennas have been a research topic for a long time \cite{AhmedIrfan_BFSurvey2018}. Generally, these are often referred to as beamforming architectures. In addition to traditional fully digital, fully analog, and hybrid approaches, many other alternatives have been proposed and utilized, including lens-based systems \cite{TTD1, TTD2, Rasilainen_JSAC2023,chen2016handbook}, Butler-matrices \cite{ButlerMatrixReview}, and Rotman-lens-based fixed beam systems \cite{Turalchuk_RotmanLens2018,RotmanLensReview2014}. In general, lower frequencies use more traditional antenna arrays, whereas \ac{mmWave} and sub-THz systems consider more exotic solutions such as lenses. Furthermore, reflect arrays and intelligent reflective surfaces are also used to control the spatial characteristics of transmitted and received signals \cite{Rasilainen_JSAC2023}. Many of these architecture discussions and findings are also highly relevant to ISAC scenarios. From an array point of view, both sensing and communications benefit from high beamforming gain and narrow beams. However, sensing scenarios especially require also fast scanning of the environment. This results in a basic trade-off between beam width and environment scanning time, having an impact on sensing accuracy as well as the latency of the sensing result. For the scanning speed, digital beamforming approaches at the \ac{Rx} have the superior advantage of seeing all directions with "one look", while the requirement to support wide bandwidth to all elements with high dynamic range makes them power-hungry approaches to use for large arrays. On the other hand, in monostatic sensing, digital beamforming approaches require extremely high-resolution and high dynamic range data converters to be able to separate the self-interference from the desired scattered signal bouncing back from the environment.

\subsubsection{Enabling \ac{RF} \ac{TRx} Array Architectures for ISAC}

Many ISAC works rely on array architectures that have simultaneous multi-beam capability \cite{zhang2019multibeam,Baquero2020ICC}. In addition to traditional beam shapes (e.g., directional beams pointing towards a certain angle with the maximum possible energy), arrays used in traditional standalone radars may also employ so-called \textit{derivative/difference beams} to further improve the angular sensitivity \cite{monopulse_review,crb_mimo_radar_2008}. Such beams have also been proven to enhance localization/sensing accuracy in localization or ISAC applications \cite{spatialSignal_TVT_2022}. In order to support these types of beams, accurate array calibration is crucial and the required phase and amplitude control is easier to achieve with digital beamforming. In communications, multi-beam approaches typically lead to either fully digital or hybrid beamforming architectures, with the latter one being mainly used in higher frequencies at FR2 and above. The most used hybrid beamforming architecture is based on multiple subarrays, typically implemented as phased arrays. In this section, we focus mainly on new aspects in the phased array domain, providing benefits for sensing applications. 

\begin{figure}
\centering
\subfloat[]{\includegraphics[width=0.31\columnwidth]{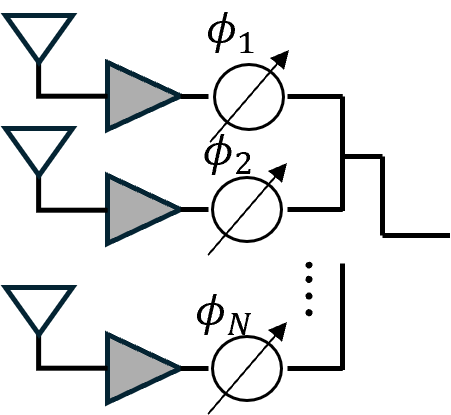}}
\subfloat[]{\includegraphics[width=0.31\columnwidth]{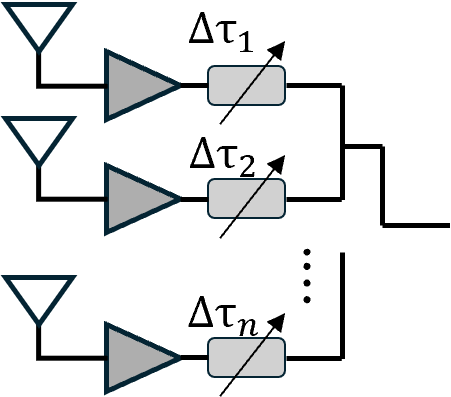}}
\subfloat[]{\includegraphics[width=0.37\columnwidth]{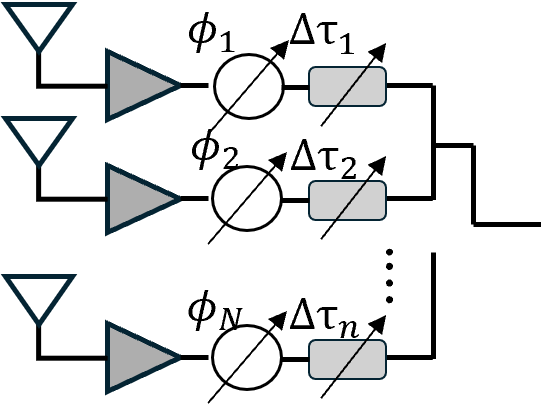}}
\caption{Analog beamforming approaches: (a) typical phased array, (b) true-time-delay based array, and (c) delay-phase controlled array. Architecture (c) can be used to configure beam effectively to frequency scanning (ISAC) and wideband beamforming (communication) modes. 
 \label{fig:PhasedArrayConfigurations}}
\end{figure}

Phased array-based systems that typically employ analog beamforming are often used, especially in higher frequencies at \ac{mmWave} and sub-THz regime \cite{Kolodziej_EURAD2023}. A traditional phased array architecture is shown in Fig.~\ref{fig:PhasedArrayConfigurations}(a). One of the fundamental challenges in large phased arrays is beam squint, where individual frequencies within the signal band are steered to slightly different directions \cite{beamsquintref}. The beam squint problem as such can be mitigated by using true-time-delay (TTD) based beamforming \cite{TTD1}. A TTD-based analog beamforming array is shown in Fig.~\ref{fig:PhasedArrayConfigurations}(b). While beam squinting is typically considered a problem from a communications perspective, it has also been identified as an enabler of fast beam scanning \cite{AoAEstim1, AoAEstim2, AoAEstim3, BoljanovicTCAS2021}. The fast beam scanning feature also enables faster sensing. Especially when used together with controllable delay elements in the individual paths or subarrays, the system can be efficiently tuned for frequency-angular-scanning and wideband communication modes \cite{Ratnam, TTD3, TTD4, TTD5, TTD6, Lin_JSSC2022}. Moreover, beam squint also has similar frequency-dependent behavior in the distance domain when used in the near field \cite{DelfiniArray2024}. An array architecture that utilizes TTDs for beam scanning is shown in Fig.~\ref{fig:PhasedArrayConfigurations}(c). In frequencies above 300 GHz, so-called prism approaches have also been using similar frequency scanning properties to localize multiple transmitters with one shot \cite{Sengupta_Prism2021}.

\begin{figure}
\centering
\subfloat[]{\includegraphics[width=0.3\columnwidth]{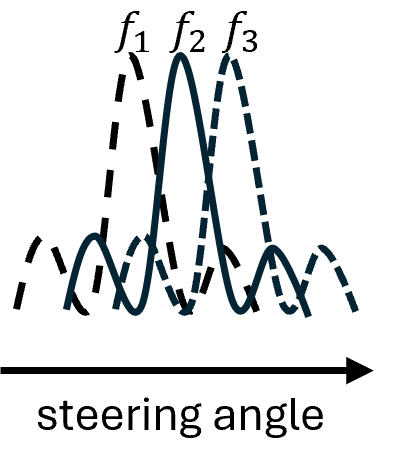}}
\subfloat[]{\includegraphics[width=0.3\columnwidth]{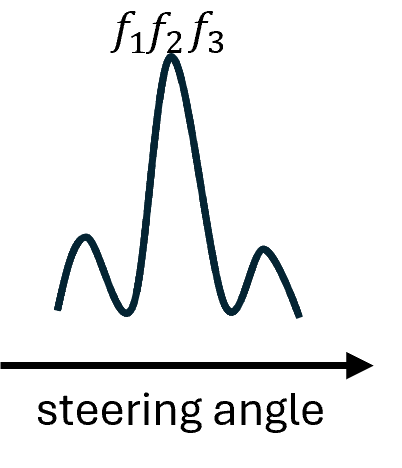}}
\caption{Array configuration modes with phase-TTD array architecture for cases (a) frequency-beam scanning mode for ISAC and (b) wideband mode for communications. In ISAC mode the TTDs are used to spread beams over frequencies while in communication mode wideband beam can be provided to desired direction. 
 \label{fig:PhasedArrayModes}}
\end{figure}

When designing an array transceiver for both communications and sensing, one of the keys is reconfigurability to different modes of operation: Communications require a wideband operation, meaning that all energy across the signal band should be focused on the directions that enhance the link capacity \cite{BoljanovicTCAS2021}. For this mode of operation, the array would like to have the beam squint minimized. Sensing, or in this case scanning the environment requires transmitting and receiving to/from different directions of the array as fast as possible. For the second mode of operation, the array would like to have the beam squint maximized. This kind of mode of operation has been proposed for example in \cite{BoljanovicTCAS2021,Lin_JSSC2022} by using TTD circuits to control the delay of the antenna paths. Different modes of operation can be effectively configured by using the array approach depicted in Fig.~\ref{fig:PhasedArrayConfigurations}(c). The fundamental principle of the modes of operation is shown in Fig.~\ref{fig:PhasedArrayModes}. Reconfigurability for different modes can be also achieved by using asymmetrical signal routing from antennas to the sum node \cite{AlokPaper,EuCAP2023,EuCAP2024}. Proper utilization of beam squint can also potentially enhance scanning resolution in smaller arrays, making it an attractive solution for the device side where the possible antenna area is limited.

\subsubsection{Full-duplex
Operation for Monostatic Sensing}\label{full-duplex-communication}

\begin{figure}
	\centering
	\includegraphics[width=0.9\linewidth]{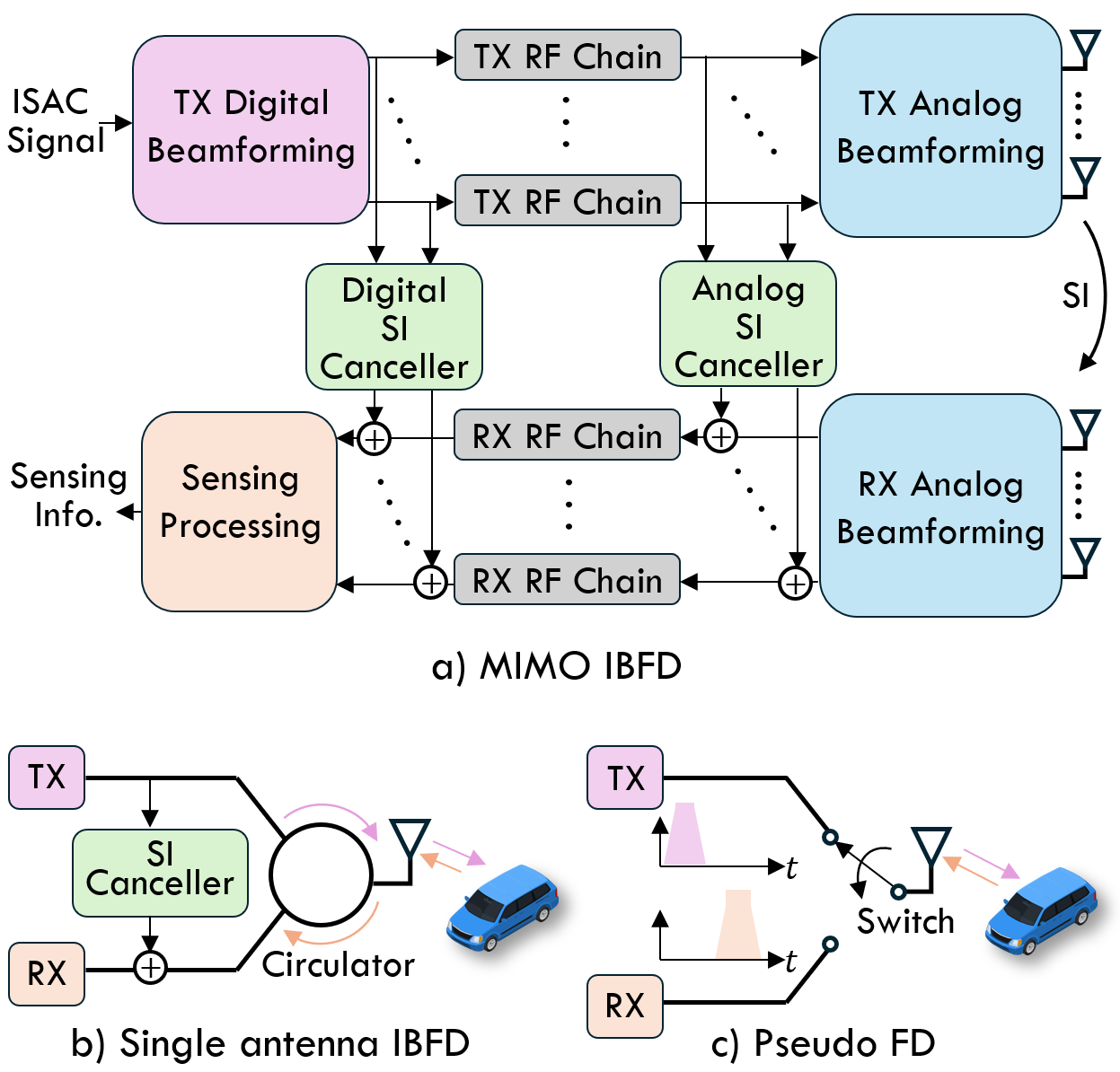}
	\caption{Full duplex architectures for ISAC. (a) MIMO IBFD provides additional isolation over (b) single antenna IBFD due to separated antennas and/or beamforming. (c) Pseudo FD relies on very fast switching between transmit and receive modes.}
\label{fig:IBFD_arch}
\end{figure}

As discussed in Section~\ref{monostatic-sensing}, monostatic sensing is performed by collocating the \ac{Tx} and the \ac{Rx} of the sensing setup of interest at the same point in space. In particular, in such a setup, the \ac{Tx} emits \ac{RF} signals which after being reflected by the surrounding environment are collected by the \ac{Rx}. Provided the knowledge of the transmitted signal and the tight synchronization between the \ac{Tx} and the \ac{Rx}, conclusions can be drawn regarding the surrounding environment via either radar signal processing or \ac{CSI}-based sensing. It is clear from the description in Section~\ref{monostatic-sensing} that the efficient implementation of monostatic sensing requires a full-duplex (FD) or pseudo-FD operation, as highlighted in Fig.~\ref{fig:IBFD_arch}.

Theoretically, without considering any limitation regarding implementation, the in-band full duplex (IBFD) operation (i.e., the ability to transmit and receive simultaneously in the same frequency band) is the preferred operation for a monostatic sensing transceiver. This is because it provides flexibility in terms of pulse design (pulse duration and pulse repetition interval) and consequently design of the coverage area \cite{Richards2014}. In addition, IBFD operation can act as an enabler to concurrent communication and sensing \cite{9099670}. However, the practical implementation of IBFD in the monostatic case is challenging, primarily due to a) self-interference (SI), i.e. the direct interference of the \ac{Rx} from the \ac{Tx}; and secondary, in certain deployments, due to b) cross-link interference (CLI), i.e. interference from neighboring communication or sensing links.

Focusing on a single monostatic transceiver, the main challenge for bringing IBFD into reality is the cancellation of the SI in the sensing transceiver. In general, a suppression of more than 100 dB of SI is required \cite{6832471}. This has to be done as the combination of: a) isolation of the transmit chain from the receive chain, and b) cancellation of the remaining SI \cite{6832471} in the analog and digital domains. Regarding the isolation of the transmit from the receive chain for a single antenna transceiver, this can be achieved via the use of a circulator \cite{ bmks2013}. However, circulators provide a relatively small isolation of about 20 dB at 6 GHz \cite{ 6832471} and in higher frequency systems they are bulky, lossy, and challenging to integrate \cite{Prata_IMS2017}. An alternative approach of isolation is based on the use of multiple antennas, where, the sensing transceiver is potentially decomposed to separate transmit and receive chains, with each chain having its own dedicated antenna array \cite{10856334}. Without considering any form of (analog or digital) cancellation, and mainly for the FR2 bands, this method achieves isolation of up to 40 dB at the cost of physical space and advanced antenna design \cite{6832471}.

An additional tool for suppressing SI, especially for the IBFD multiantenna architectures, is the use of \ac{RF}, digital, or hybrid processing. The objective of this processing, provided the knowledge of the transmitted signal, is to remove/subtract the remaining SI signal or its replicas \cite{6832471}. However, the removal of the remaining SI signal can be a challenging task. This is due to hardware imperfection which can result in a random component to the actual transmitted signal. In general, this kind of cancellation results in values ranging between 30 and 60 dB \cite{6832471}. Recently, the use of machine learning has emerged as a research topic to further improve performance in face of hardware imperfections \cite{10622082}.

From the above, it becomes clear that the practical implementation of IBFD (sensing) transceivers is challenging. In fact, this is clearly reflected in standardization forums, such as 3GPP, and the absence of commercialization of IBFD architectures. However, initial results for an FD scheme, with duplexing in the bands of the same carrier, called sub-band full duplex (SBFD) for the case of \ac{BS} sensing can be found in  \cite{10622082}. Beyond the previous approaches for implementing FD monostatic sensing, pseudo-FD operation can be exploited. For example, by carefully controlling the duration of the transmitted pulses, it is possible to have a serial transmission and reception of signals. This approach closely follows the classical monostatic radar approach where transmission is switched off during reception and vice versa \cite{Richards2014}. More advanced architectures can be designed using a combination of spatial filtering, possibly using two antenna arrays and time duplexing of transmission and reception. However, special care should be given with respect to the switching times, connecting both with beam switching/forming and transition from transmission to reception, along with the treatment of the interference from the secondary lobes.

\subsection{\ac{RF} Impairments and Calibration in ISAC}

Modeling and mitigation of \ac{RF} impairments for wireless communications have been an active research topic for decades. However, studies focusing on the sensing scenarios in the ISAC context are few. In communication systems, the two most studied \ac{RF} impairments are nonlinear distortion of PAs and phase noise of \acp{LO}. Moreover, the calibration of phases across multiple access points and mobile devices is important when utilizing phase information for sensing. This section addresses \textit{nonlinear distortion}, \textit{phase noise}, and  \textit{calibration}.

\subsubsection{\ac{PA} Nonlinear Distortion in ISAC} \label{sec:PAdistortion}
\begin{figure}
\centering
\subfloat[OFDM]{\includegraphics[width=0.5\columnwidth]{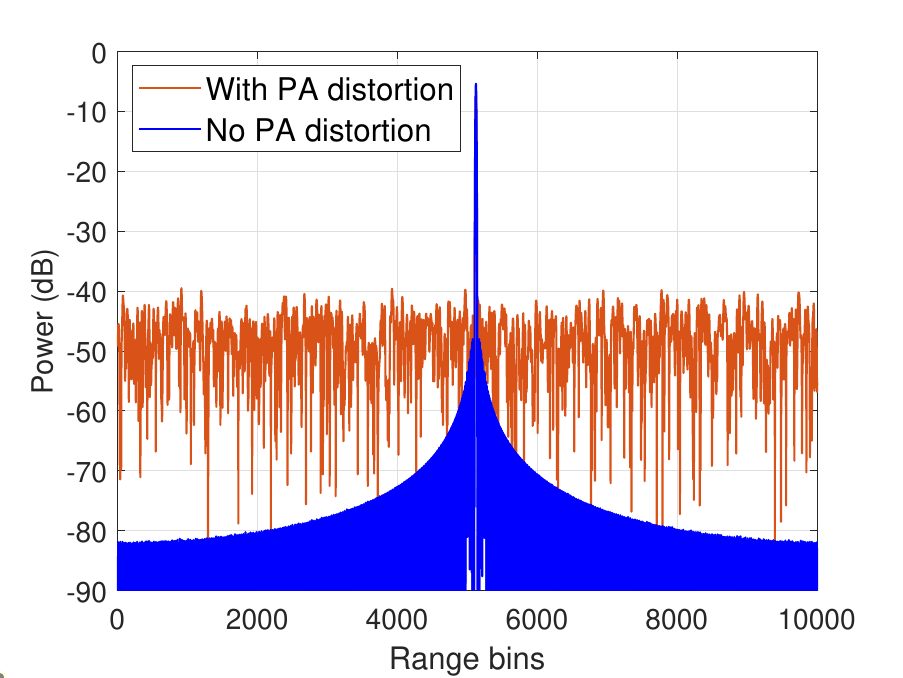}}
\subfloat[Single carrier]{\includegraphics[width=0.5\columnwidth]{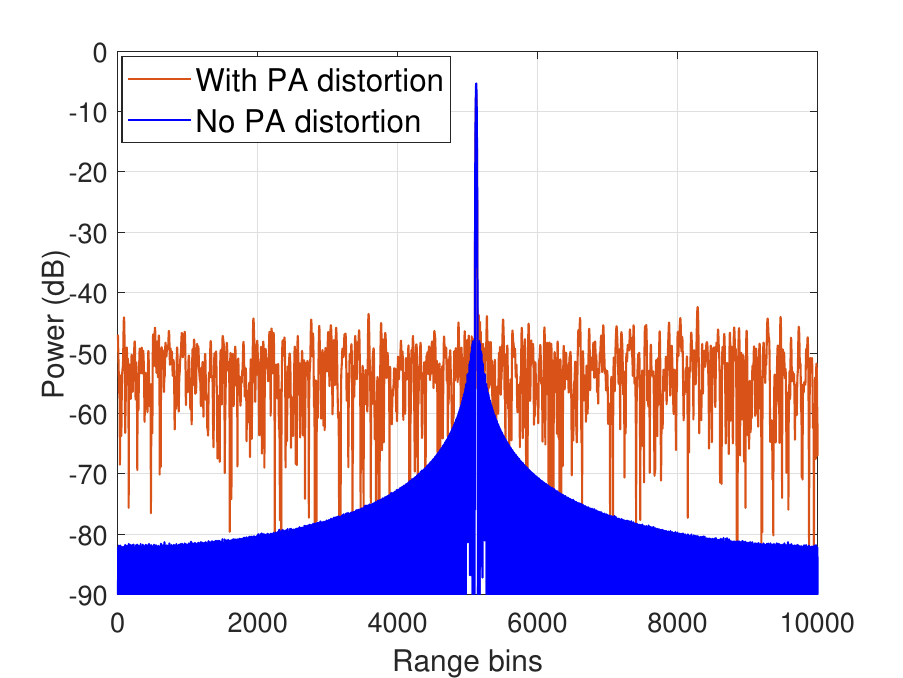}}
\caption{Range ambiguity function with and without \ac{PA} distortion for one realization
of the transmitted information symbols. A system with 64 QAM constellation, block length 1024, root raised cosine filter with roll-off factor 0.3 and a cubic \ac{PA} model at 0~dB back-off in a thermal noise-less setup is assumed \cite{Abdur_ISAC}. 
 \label{fig:RAF_PA_NF}}
\end{figure}

Over the last decades, several studies have shown that non-linear distortion from the \ac{PA} increases the error vector magnitude (EVM) of the received information symbols in a communication system \cite{Mollén_2016_Waveforms}, \cite{Andre_2008_Digital}, \cite{Abdur_SC}, \cite{Abdur_SC_HB}. However, the effect of a non-linear \ac{PA} on the sensing performance of a ISAC system has been relatively less investigated in the literature. \cite{Abdur_ISAC} studied the sensing performance of \ac{OFDM} and \ac{SC} modulation in a monostatic ISAC system with non-linear \ac{PA}. It considers a memory-less \ac{PA}, for which the time-domain \ac{PA} distortion over consecutive samples is independent and identically distributed, irrespective of the waveform. Therefore, due to the central limit theorem, the \ac{PA} distortion in the frequency domain is shown to be complex-normally distributed. Hence, similar to thermal noise, which increases the noise floor of the range-Doppler (RD) ambiguity function, the \ac{PA} distortion also increases the floor of the RD ambiguity function in \ac{OFDM} and \ac{SC}. This is depicted in Fig.~\ref{fig:RAF_PA_NF}. 
Furthermore, \cite{Abdur_ISAC} shows that the variance of the frequency-domain \ac{PA} distortion is modulation-dependent. \ac{SC} has a lower variance compared to \ac{OFDM}, thanks to the low \ac{PAPR} of the \ac{SC} modulation. Interestingly, this does not ensure a better sensing performance using \ac{SC} modulation. This is because the computation of the RD map involves dividing the received signal by the transmitted signal in the frequency domain (fast-time processing). During the division, \ac{SC} causes severe noise enhancement compared to \ac{OFDM}, as the distribution of the frequency-domain transmitted signal in \ac{SC} and \ac{OFDM} is different. Therefore, the beneficial effect of the lower \ac{PAPR}, and thus the lower \ac{PA} distortion in the \ac{SC} system is outweighed by the noise enhancement. Consequently, \ac{SC} modulation has an inferior sensing performance compared to \ac{OFDM} in most operating regions in the presence of a non-linear \ac{PA}.  
Although \cite{Abdur_ISAC} investigated the effect of \ac{PA} distortion in a monostatic ISAC system, the impact of \ac{PA} distortion in other flavors of ISAC requires further investigation. Notably, the impact of \ac{PA} with memory in a multi-antenna setup operating in mono/bi/multi-static ISAC modes remains an open question.

\subsubsection{Phase Noise in ISAC Systems}\label{sec_impact_PN}
Phase noise results from imperfections in the \ac{LO}, causing random phase variations in the transmitted signal \cite{PN_Coh_BW_2022}. In communication systems, phase noise can be effectively compensated by using different techniques. In 5G NR, the \ac{PTRS} is introduced to mitigate \ac{PN} and improve the performance of high-frequency communications, particularly in FR2 bands \cite[Sec.~10.5]{dahlman20205g}. \ac{PTRS} is used primarily to track and compensate for \ac{PN}, which becomes more significant at higher frequencies (e.g., above 24 GHz) due to the instability of oscillators \cite{PN_Coh_BW_2022}. The location and density of \ac{PTRS} in the resource grid are configurable, depending on the subcarrier spacing and modulation scheme. 

\begin{figure}
	\centering
    \vspace{-0.1in}
	\includegraphics[width=0.95\linewidth]{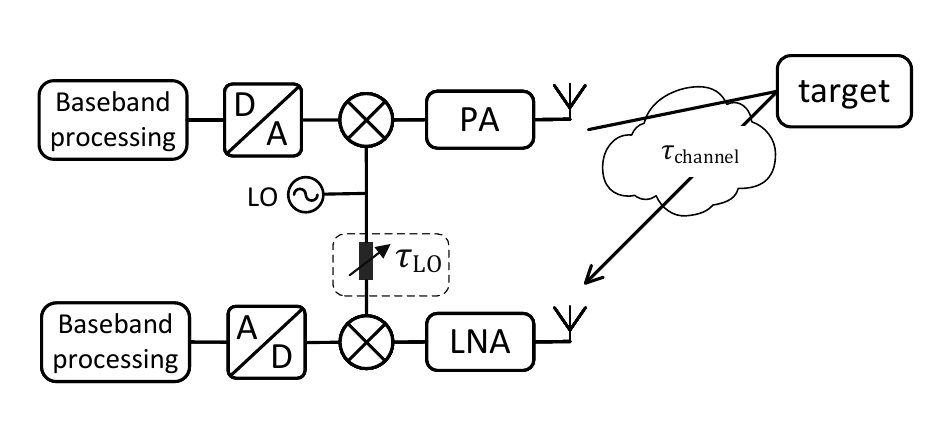}
	\caption{Block diagram of shared \ac{LO} source in monostatic sensing and \ac{PN} mitigation using delayed \ac{Rx} \ac{LO} signal.}
	\label{fig_block_diag_monostatic}
\end{figure}

Under the ISAC framework, \ac{PN} in monostatic sensing exhibits a peculiar characteristic that differs from bistatic/multistatic sensing and communication setups due to the use of a shared oscillator for up-conversion of the transmit ISAC signal and down-conversion of its backscattered echoes in the co-located sensing \ac{Rx}, as depicted in Fig.~\ref{fig_block_diag_monostatic}. More specifically, the \ac{PN} in the monostatic sensing observations corresponds to a \textit{differential/self-referenced \ac{PN} process} that results from taking the difference of the original \ac{PN} process and its time-shifted version delayed by the amount of round-trip delay of the target \cite{OFDM_PN_Sensing_2023}. This leads to the so-called \textit{range correlation} effect, which refers to delay-dependent statistics of \ac{PN} in monostatic sensing \cite{Range_Correlation_93,SPM_PN_2019,OFDM_PN_Sensing_2023}. Such a property does not arise at communications or bistatic sensing receivers since different oscillators are employed at remotely located transmitters and receivers, leading to independent \ac{PN} processes, unlike in a shared-oscillator monostatic ISAC transceiver. An intriguing implication of \textit{delay-dependent \ac{PN} statistics} is the possibility to exploit this characteristic to improve ranging performance, moving beyond the traditional approach of treating impairments purely as detrimental effects to be mitigated, but rather to be exploited \cite{OFDM_PN_Sensing_2023}.

\begin{figure}
	\centering
    \vspace{-0.1in}
    \subfloat[]{
	\includegraphics[width=0.5\columnwidth]{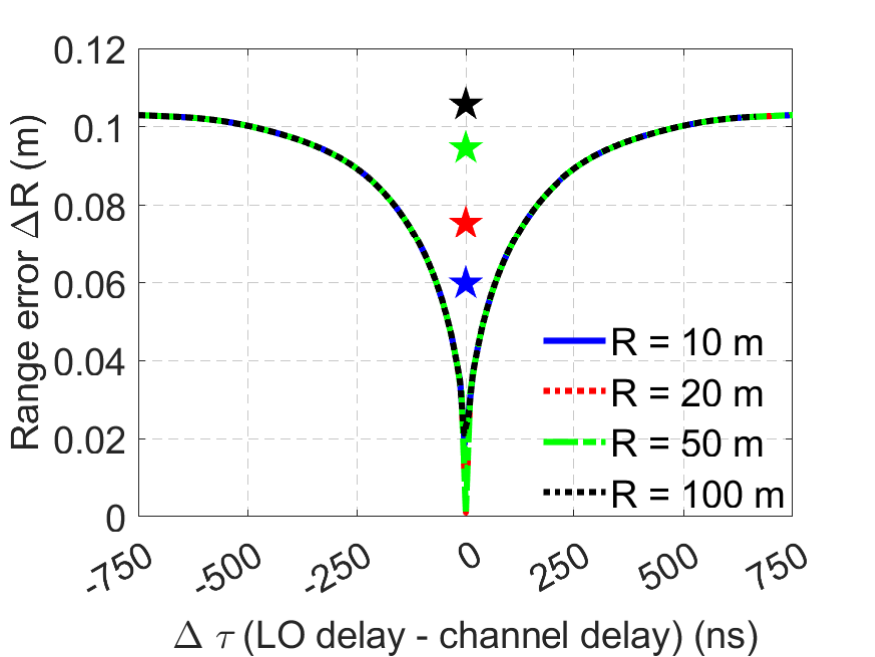}}
    \subfloat[]{\includegraphics[width=0.5\columnwidth]{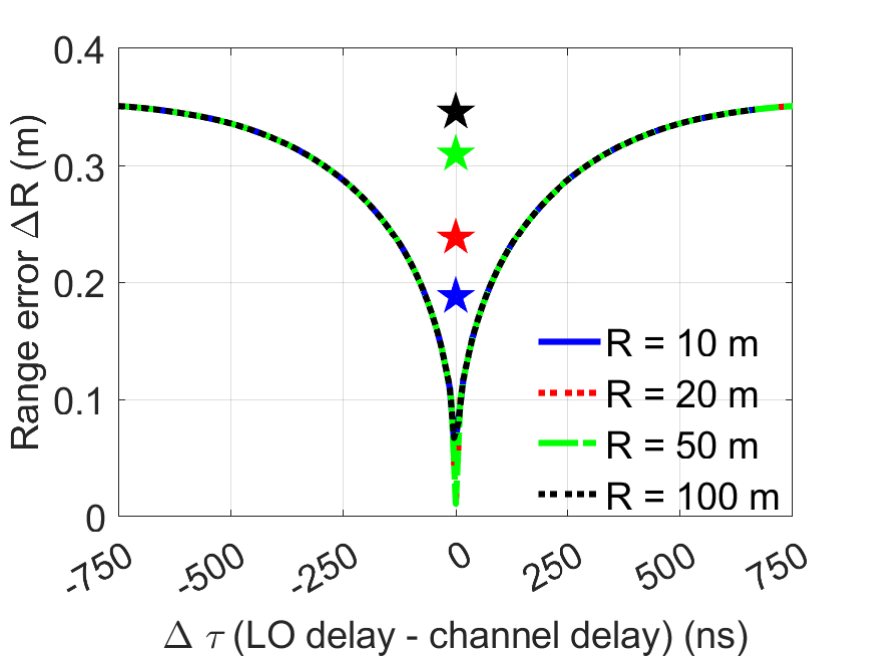}}
	\caption{Range error in monostatic sensing caused by the \ac{LO} \ac{PN} vs the delay difference between the delay used in the \ac{Rx} \ac{LO} ($\tau\textsubscript{LO}$) and the actual channel delay ($\tau\textsubscript{channel}$). The \ac{PN} is measured over $100$~MHz bandwidth and evaluated for different carrier frequencies ($f_c$), i.e., (a) $f_c = 30$~GHz, (b)$f_c = 300$~GHz. The $\star$ sign represents the range error when $\tau\textsubscript{LO}$ is set to $0$.}
	\label{fig_range_error}
\end{figure}

In the single-target scenario, the impact of phase noise can be minimized by delaying the common \ac{LO} signal in the \ac{Rx} path as shown in Fig.~\ref{fig_block_diag_monostatic}, such that when $\tau\textsubscript{LO} = \tau\textsubscript{channel}$, the phase noise can be perfectly eliminated, recovering the \ac{PN}-free sensing performance. 
\ac{PN} mitigation through delayed \ac{Rx} \ac{LO} is tested for different carrier frequencies and target distances (R) and is presented in Fig.~\ref{fig_range_error}. In a typical monostatic T\ac{Rx} architecture, i.e., without the \ac{LO} delay line ($\tau\textsubscript{LO} = 0$), the \ac{PN} of the reflected signal from the target is correlated with the \ac{Rx} phase noise for small target distances. The received \ac{PN} from the channel upon combining with the \ac{PN} of the \ac{Rx} causes a slight reduction in phase noise due to the delay difference caused by channel delay. However, for large target distances, the \ac{PN} in the backscattered and the \ac{Rx} \ac{LO} becomes highly uncorrelated due to large delays in the channel, thereby increasing range error as shown by the star sign in Fig.~\ref{fig_range_error}, especially at higher carrier frequencies.

\smallskip 
\begin{remark}[Generalization to D-MIMO]
In uplink D-MIMO \ac{OFDM} communications, the impact of \ac{PN} can significantly vary depending on whether distributed \acp{AP} use separate \acp{LO} or a shared \ac{LO} \cite{wu2024uplink}. In the case of separate \acp{LO} per \ac{AP}, a distributed \ac{PN}-aware channel estimator can be developed to independently estimate channels per \ac{AP}. A shared \ac{LO} reduces hardware costs and simplifies synchronization but introduces correlated \ac{PN} among \acp{AP}, which degrades centralized combining performance due to correlated interference. However, this correlation can be exploited in channel estimation to mitigate \ac{PN} effects \cite{wu2024uplink}. In particular, one can devise a centralized estimator that iteratively refines common phase error (CPE) and channel estimates to mitigate correlated \ac{PN}. Such an estimator can substantially improve performance in shared \ac{LO} scenarios compared to conventional estimators, reducing the adverse impact of correlated \ac{PN}. Overall, while a shared \ac{LO} setup introduces challenges in D-MIMO setups, properly exploiting \ac{PN} correlation in estimation can recover much of the lost performance due to \ac{PN}.
\end{remark}

\subsubsection{Anchor Calibration and Deployment for ISAC} \label{sec:AnchorCal}

\begin{figure}
	\centering
    \vspace{-0.1in}
	\includegraphics[width=0.99\linewidth]{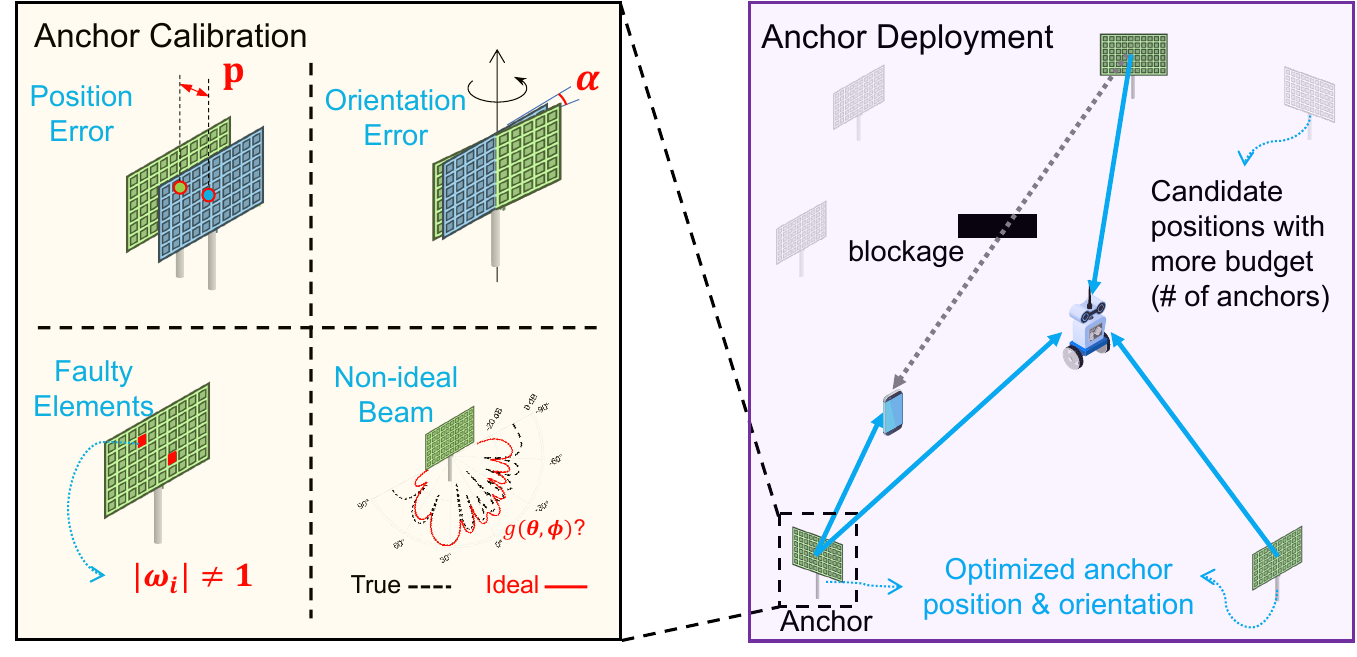}
	\caption{Illustration of anchor calibration and deployment in ISAC systems.}
	\label{fig_calibration_deployment}
\end{figure}

The performance of ISAC systems relies on both effective anchor (BS or PRU) deployment and precise calibration (e.g., in terms of location, orientation, or hardware effects), as illustrated in Fig.~\ref{fig_calibration_deployment}. Unlike communication systems, which focus on the end-to-end channel of a point-to-point link, positioning and sensing depend on accurately estimating the geometric relationships between the target and multiple anchors. Hence, the deployment strategy and calibration requirements are different for communication, positioning, and sensing tasks. Deployment establishes the foundation for system coverage, ensuring that anchors are optimally placed to minimize signal blockages and maximize spatial diversity~\cite{bi2023physical}. Calibration complements deployment by estimating accurate anchor states, including geometry states (e.g., position and orientation) and hardware impairments (e.g., mutual coupling), which are critical for maintaining high-precision sensing~\cite{ghazalian2024calibration}. Together, deployment and calibration enable ISAC systems to meet the stringent requirements for future applications such as autonomous vehicles, industrial automation, and augmented reality.

Various types of sensing anchors should be considered in the deployment phase, such as active anchors (e.g., base stations), passive anchors (e.g., RISs), and assisting anchors (e.g., other users in the scenario of cooperative sensing). Poorly placed anchors can lead to sensing coverage gaps while providing adequate communication coverage, reduced localization accuracy, and degraded communication reliability. Advanced deployment strategies leverage heuristic or machine learning algorithms to identify optimal anchor locations based on environmental constraints and application requirements~\cite{egwuche2023machine}. Additionally, emerging technologies like fluid and movable antennas provide dynamic deployment capabilities, enabling real-time optimization through feedback~\cite{zhou2024fluid, li2024joint}. In dynamic scenarios, mobile anchors such as drones (which are localized using, e.g., GNSS) can further enhance the adaptability of the system, ensuring the service availability of ISAC systems~\cite{lyu2022joint}. 

Geometry calibration ensures that the spatial states of anchors, such as their position and orientation, are accurately determined (re-calibration may be needed, e.g., after severe weather). These parameters are critical for effective localization in sensing tasks, which reflects the relationship between anchors and targets~\cite{ge2024experimental}. Similar to localization tasks, calibration can be treated as state estimation of the anchor. Specifically, channel parameters measurements like \ac{ToA} and \ac{AOA} can be estimated first, followed by the algorithms such as triangulation or maximum-likelihood estimation~\cite{zheng2023jrcup}. Near-field scenarios, with their complex spherical wavefronts, demand more sophisticated calibration techniques than far-field scenarios~\cite{chen20246g}. Usually, the \ac{MCRB} can be used to assess the impact of geometric mismatches and guide calibration precision based on application requirements~\cite{zheng2023misspecified}.

Hardware calibration is essential for mitigating impairments such as pixel failures, mutual coupling, and phase-dependent amplitude variations, all of which can significantly impact the performance of ISAC systems~\cite{ozturk2024ris, chen2023modeling}. Offline calibration, conducted in controlled environments, establishes a baseline for hardware accuracy, while online calibration addresses real-time impairments to ensure system integrity during operation~\cite{keskin2023monostatic}. Although agent-based calibration methods, which rely on known states, can achieve high accuracy, joint positioning and calibration approaches offer autonomous solutions and are widely used~\cite{pohlmann2022simultaneous}. For complex hardware impairment models, machine learning-based methods have emerged as a preferred choice, providing efficient and adaptive solutions~\cite{mateos2025model}.

\subsection{Hardware Demonstrators for ISAC}

\begin{table*}[ht]
\centering
\caption{Summary of Recent Studies on ISAC hardware implementations, applications, and channel modeling.}
\begin{tabular}{|p{4cm}|p{5cm}|p{1.5cm}|p{5.0cm}|}
\hline
\textbf{Category} & \textbf{Key Contribution} & \textbf{Reference} & \textbf{Highlights} \\
\hline
\multirow{3}{*}{Hardware Implementations} & Multibeam techniques with steerable analog antenna arrays & \cite{zhang2019multibeam} & Complies with modern packet communication systems using multicarrier modulation \\
\cline{2-4}
 & Multi-domain cooperative communication prototype & \cite{yang2024isac} & Supports 5G NR, real-time communication, and accurate sensing \\
\cline{2-4}
 & \ac{mmWave} testbeds for ISAC validation & \cite{tmytek2023mmwave} & Versatile tool for academic and industrial exploration \\
\hline
\multirow{2}{*}{Applications} & Predictive beamforming for vehicular networks & \cite{liu2020radar} & Radar-assisted channel adaptation improves reliability \\
\cline{2-4}
 & Integrating ISAC into IoT systems & \cite{cui2021integrating} & Addresses spectrum sharing and scalability challenges \\
\hline
\multirow{3}{*}{Channel Modeling} & Channel modeling for midbands, \ac{mmWave}, and sub-THz frequencies & \cite{nextgalliance2024jcas} & Emphasizes new models for sensing and communication \\
\cline{2-4}
 & Networking-based ISAC testbed & \cite{ji2023networking} & Explores multi-node cooperative perception and data sharing \\
\cline{2-4}
 & Distributed MIMO measurement system for industrial ISAC & \cite{nelson2024distributed} & Demonstrates multistatic sensing with positioning accuracy below 20 cm \\
\hline
Channel Sounders & Multi-band channel sounder for experimental characterization & \cite{Bomfin2024experimental} & Provides high-resolution measurements of channel conditions \\
\hline
\multirow{2}{*}{Hardware Impairment Mitigation} & Calibration and error correction in ISAC prototypes & \cite{yang2024isac} & Effectively handles non-linearities and imperfections \\
\cline{2-4}
 & Signal processing techniques for \ac{mmWave} testbed & \cite{tmytek2023mmwave} & Ensures robust operation in high-frequency ISAC applications \\
\hline
\end{tabular}
\label{tab:isac_hardware_summary}
\end{table*}

Recent research has provided substantial advancements in the theoretical frameworks and design principles for ISAC as discussed in the previous sections. From another perspective, hardware implementations and prototyping have played a critical role in demonstrating the feasibility of ISAC in practical scenarios, as summarized in Table~\ref{tab:isac_hardware_summary}. In \cite{zhang2019multibeam}, multibeam techniques using steerable analog antenna arrays were introduced in an ISAC setup, complying with modern packet communication systems with multicarrier modulation. A prototype system supporting multi-domain cooperative communication was described in \cite{yang2024isac}, enabling real-time communication and accurate sensing for complex environments. The designed prototype aligns with the 5G New Radio standard, offering scalability for up to 16 UEs in uplink transmission and 10 UEs in downlink transmission. Recent surveys such as \cite{faghih2024hardware} emphasized the importance of integrated platforms for validating ISAC systems. Moreover, the development of \ac{mmWave} testbeds, as highlighted in \cite{tmytek2023mmwave}, has provided a versatile tool for academic and industrial exploration of ISAC technologies.

Applications and use cases of ISAC have shown its versatility in fields such as vehicular networks, the Internet of Things (IoT), and 6G wireless communications. In \cite{liu2020radar}, predictive beamforming techniques were evaluated for vehicular networks, demonstrating improved communication reliability through radar-assisted channel adaptation. Challenges in integrating ISAC into IoT systems, including spectrum sharing and scalability, were addressed in \cite{cui2021integrating}.

Efforts in channel modeling and performance evaluation have provided essential insights into the practical implementation of ISAC. Channel modeling studies in \cite{nextgalliance2024jcas} investigated the midbands, \ac{mmWave}, and sub-terahertz (THz) frequencies for ISAC, emphasizing the need for new models that account for both sensing and communication requirements. In \cite{ji2023networking}, a networking-based ISAC testbed was developed, exploring multi-node cooperative perception and data sharing in networked environments. Here, the demodulation reference signal associated with the physical downlink shared channel is fully utilized to jointly estimate the velocity and distance of targets. Accurate channel modeling plays a pivotal role in the design of ISAC systems, as the shared scatterers arising from multiplexed hardware and environmental overlap can profoundly affect both communication and sensing functionalities. Moreover, \cite{nelson2024distributed} presents a sub-6 GHz \ac{D-MIMO} measurement system tailored for ISAC in industrial environments. The study evaluates channel characteristics, including diversity and link reliability, while also demonstrating multistatic sensing capabilities with positioning accuracy below 20 cm. These findings emphasize the need for new spatially consistent channel models that address nonstationary properties in ISAC hardware setups. A recent study \cite{liu2024shared} has highlighted the importance of capturing the channel sharing degree, which serves as a key metric for realistically evaluating and optimizing the deployment and performance of ISAC systems. These studies underscore the importance of accurate modeling and evaluation in advancing ISAC technologies. 

Along with channel modeling, channel sounders play a pivotal role in ISAC systems, enabling precise characterization of wireless channels for both communication and sensing functionalities. They provide essential data on parameters such as delay spread, Doppler shifts, and angular profiles \cite{maccartney2017flexible}, which are critical for optimizing ISAC performance. In a recent experimental study \cite{Bomfin2024experimental}, a multi-band channel sounder was utilized for experimental channel characterization in the upper mid-band. This work provides detailed insights into the propagation characteristics across different bands, which are crucial for ISAC system design and optimization. By offering high-resolution measurements, this study advances the understanding of the channel conditions under practical deployment scenarios, highlighting the role of channel sounding as a cornerstone for ISAC research.

An important consideration in ISAC systems is the mitigation of hardware impairments, which can significantly impact performance in practical deployments. Such impairments include phase noise, \ac{IQ} imbalance, power amplifier non-linearities, and \ac{ADC}/\ac{DAC} imperfections. In \cite{zhang2019multibeam}, steerable analog antenna arrays were utilized to minimize phase noise and ensure precise beam alignment in hardware-constrained environments. The ISAC prototype described in \cite{yang2024isac} incorporated calibration and error correction algorithms to handle non-linearities and other imperfections effectively. Additionally, the \ac{mmWave} testbed in \cite{tmytek2023mmwave} implemented advanced signal processing techniques to counteract non-linearities and phase noise, ensuring robust operation in high-frequency ISAC applications. These efforts highlight the critical role of hardware impairment mitigation in bridging the gap between theoretical ISAC designs and practical implementations.

As the field progresses, future research directions for ISAC hardware implementations and prototypes focus on developing efficient, low-latency, and high-bandwidth systems capable of seamlessly integrating communication and sensing functionalities. Key areas include designing reconfigurable antenna systems, compact multi-functional transceivers, and energy-efficient circuits to support real-time operations. Prototypes should prioritize adaptability to diverse deployment scenarios, including vehicular networks and smart urban environments. Addressing impairment mitigation will involve sophisticated signal processing algorithms to combat challenges such as phase noise, interference, and hardware non-linearities. Collaborative efforts in integrating machine learning for adaptive impairment correction and enhancing hardware resilience are critical to advancing ISAC systems.

\section{ISAC at Higher Layers -- Protocols and Functions}\label{protocols-and-functions-for-jcas}

For ISAC to work properly in a large set of scenarios, there has to be methods to configure the different parts of the network involved in the various ISAC tasks. Further, there has to be \acp{API} or similar functionality to make data available where needed. Finally, the exposure of data, configuration of involved network nodes and ISAC measurements have to be performed only by authorized parties while keeping all data secure. In addition, privacy of individuals need to be ensured. In the following sections the above mentioned parts are described in detail and put into context, i.e., how are all high layer parts connected.

\subsection{Introduction to Sensing Architecture}\label{sensing-architecture-intro}

A sensing session is initiated by a request from an application; in the following referred to as sensing request. The nodes involved in a sensing session are configured based on the information comprised or determined from a sensing request. For example, the sensing request probably includes an area of interest. The system includes a control function that relies on information in the sensing request, such as the area of interest to configure \acp{SU}, i.e. transmitters and receivers, that perform sensing measurements. It may be so that the request includes details regarding the area of interest, e.g., asking for the position of an object within an area of interest or maybe even characteristics of the object, such as is it a car or a human. Such details further help the system during the configuration of involved units, e.g., letting a sensing process unit that interprets received sensing data know that it should look for an object when processing data.   

\begin{figure}
    \centering
    \includegraphics[width=0.99\linewidth]{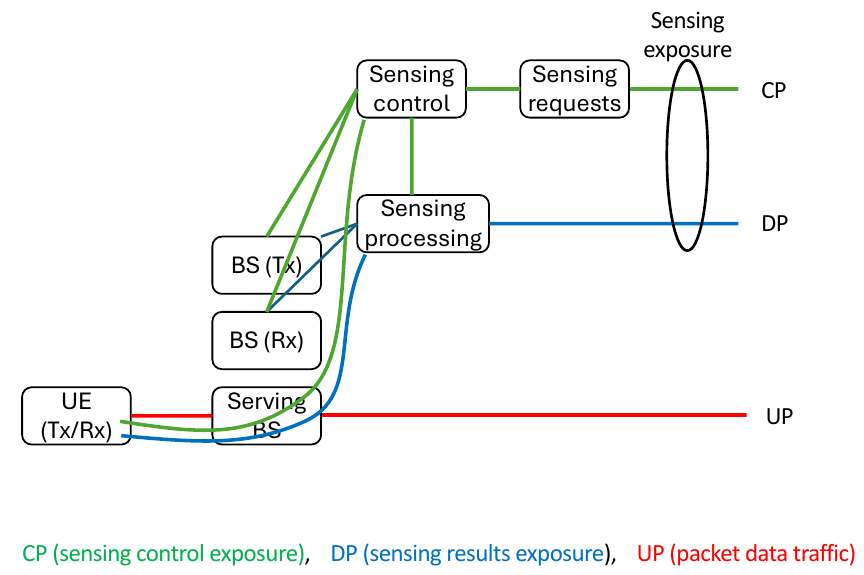}
    \caption{Functional architecture for ISAC with UE involvement. }
    \label{fig:functional_arch}
\end{figure}

A sensing session is characterized by one or more \ac{BS} and/or UE, located in a way that enables sensing of the area of interest, radio resources needed for the sensing measurements, and availability of necessary functions to control and process the sensing data. Thus, the sensing process aims at ensuring the availability (e.g., activating or configuring for sensing), configuring (e.g., sensing monitoring periodicity or the time span, specific radio signals, radio measurements, measurement reporting for sensing) and/or collecting sensing radio measurements from one or more \ac{BS} and/or UE. In the following sections, different functions and procedures are described in a functional architecture. The different nodes involved in the sensing session are described with focus on their parts in the session. The sensing control and sensing processing are sometimes referred to as the \ac{SeMF}. As shown in the figure, the detection is requested via a \ac{CP}, whereas the output / result is provided via a \ac{DP} \cite{hexd33}. \ac{CP} and \ac{DP} separation enables building each with the right technology and capacity, see Fig.~\ref{fig:functional_arch}.

\subsubsection{Sensing Requests}\label{sensing-requests}
To initiate a sensing session, an application sends a sensing request with information on what type of result it expects from this session. Hence, the request includes at least an area of interest and an application \ac{ID}. However, it is likely to include a request to determine if there is an object within the area of interest and if this object has a particular shape, moves, etc. The request could also include priority information to be used when scheduling measurements; e.g., a request from a first responder would be treated with higher priority than a request from a best effort user application. A ``sensing requests" function receives the request and from the application \ac{ID} the function, perhaps with assistance from other core network functions, determines if the application is authorized to request sensing in this location or at all. The application can be external, i.e., some entity requests information from the network, or internal, i.e., the network itself requests the information. The application can be connected to the sensing system via an \ac{API} running on top of an existing exposure framework (e.g., the \ac{NEF}, \ac{CAPIF}, \ac{NWaaP}, etc.), via an existing data distribution network (e.g., \ac{EDCA}) or via a new interface.

\subsubsection{Sensing Control}\label{sensing-control}
The \ac{SCF} is responsible for mapping requests into actual ISAC measurements, translating requested sensing areas into necessary entities (base stations and, possibly UEs), finding the UEs to be involved (mapping area into relevant UEs), and finally configuring the entities to be involved in sensing (base stations and, possibly, UEs). The \ac{SCF} provides information used by the scheduler to coordinate resource sharing between sensing and communication, e.g., sharing radio resources in time, frequency, sharing available antennas, transmission power, processing capabilities, etc. There might be a performance trade-off between the communication capacity and the sensing performance, since more resources allocated for sensing would naturally improve sensing performance, but at the same time would leave less resources available for communication.
In a basic deployment, the nodes depicted in Fig.~\ref{fig:functional_arch} would all be available in a cell and therefore the \ac{SCF} is co-located with the scheduler. Similarly, the cell would include sensing units and the sensing processing node. However, cells can have different size and shape, so alternative deployments will be possible.  For example, the \ac{SCF} can be implemented as a single node, in case the cell is large. Otherwise, the \ac{SCF} can be distributed over multiple nodes, in case there are many small cells where users frequently move between cells. The \ac{SCF} is most likely deployed in the same communications networks as all or some \acp{SU} but even here other deployments could be possible , e.g., when one or more \acp{SU} are deployed in a 5G communications network, while the associated \ac{SCF} is part of a 6G communications network.

\subsubsection{Sensing Processing}\label{sensing-processing}
The \ac{SPF} interprets the sensing measurements and converts them into a sensing result in a format meaningful to an external application. The available local data are handled by this function, for example, the \ac{BS} \ac{ID} and observations. Local data is forwarded from the \ac{SCF} and comprises information from the request, information the \ac{SCF} has on the surrounding environment, and information on the deployed sensing units. In addition, sensor fusion information could be included to improve results. The data can be raw data, measurements, reported events, etc. The amount of signaling overhead between the nodes to exchange the data (e.g., reporting periodicity, raw data vs. event reporting, etc.) and the amount and type of processing per node will depend on the use case and the functional split between the nodes involved in sensing data processing. This function also ensures that the privacy, for example, of bystanders is protected. 
The \ac{SPF} can be a separate entity in the network or can be part of another node, e.g., in a BS or UE. In addition, the deployment of the \ac{SPF} can be distributed in an UE or \ac{RAN} node, in a core node, or in a positioning node. The \ac{SPF} may be implemented together with or as a part of \ac{SCF} depending on the scenario and the actual use case.
The measurement data may be voluminous, especially when data involves \ac{IQ} samples, so there needs to be sufficient transport capacity. Although the characteristics of measurement data may be similar to user data, they are not the same. Therefore, it makes sense to introduce a data plane for the measurement data. Such a data plane could also be used for \ac{AI} and compute services.

\subsubsection{Serving BS}\label{serving-bs}
The serving BS supports the UEs involved in the sensing process with legacy connectivity. In the figure, the BS is depicted as being a separate node; however, in reality, the BS can be involved in the sensing process. The BS to which the UE is connected will, e.g., forwards sensing measurements received by the UE to the \ac{SPF}. As such, the serving BS schedules the UE for the communication service and ensures that the UE is given opportunities to perform its role (\ac{Rx}/\ac{Tx}) in the sensing measurement process.

\subsubsection{Sensing Unit}\label{sensing-units}
Finally, the \ac{SU} consists of sensing capable \ac{RAN} entities and UEs, e.g., base stations and UEs having the capability to \ac{SeRS} used for ISAC as well as reporting measurement results. An \ac{SU} can be capable of at least one of the following:
(i) Radio signal transmission used for sensing, 
(ii) Radio signal reception used for sensing,  or (iii) Radio measurement used for sensing. 
Multiple \acp{SU} with the same or different capabilities can be involved in a sensing session. The involved \acp{SU} may apply different \acp{RAT}, e.g., an \ac{SU}, capable of receiving transmissions from several \acp{RAT}, can receive radio signals for sensing from transmitting \acp{SU} to different \acp{RAT} if this is required to fulfill the sensing request. 

\begin{figure*}
    \centering
    \includegraphics[width=0.99\linewidth]{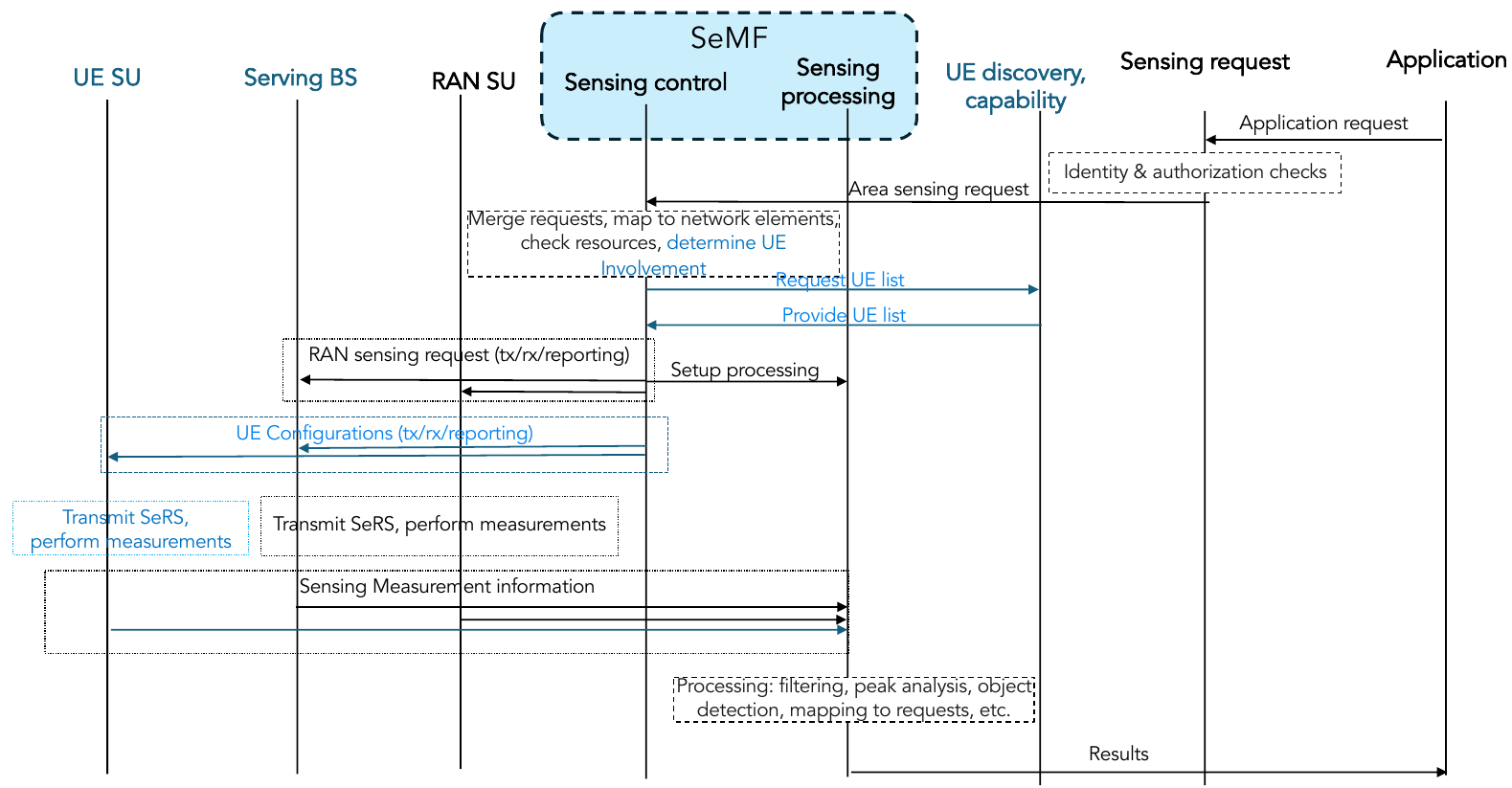}
    \caption{Sequence diagram for a sensing session with UE involvement. (Blue text represents actions due to UE involvement).}
    \label{fig:signaling}
\end{figure*}

\subsection{Sensing Session and the Role of the \ac{SeMF}}\label{role-of-the-semf}

\begin{figure*}
    \centering
    \includegraphics[width=0.8\linewidth]{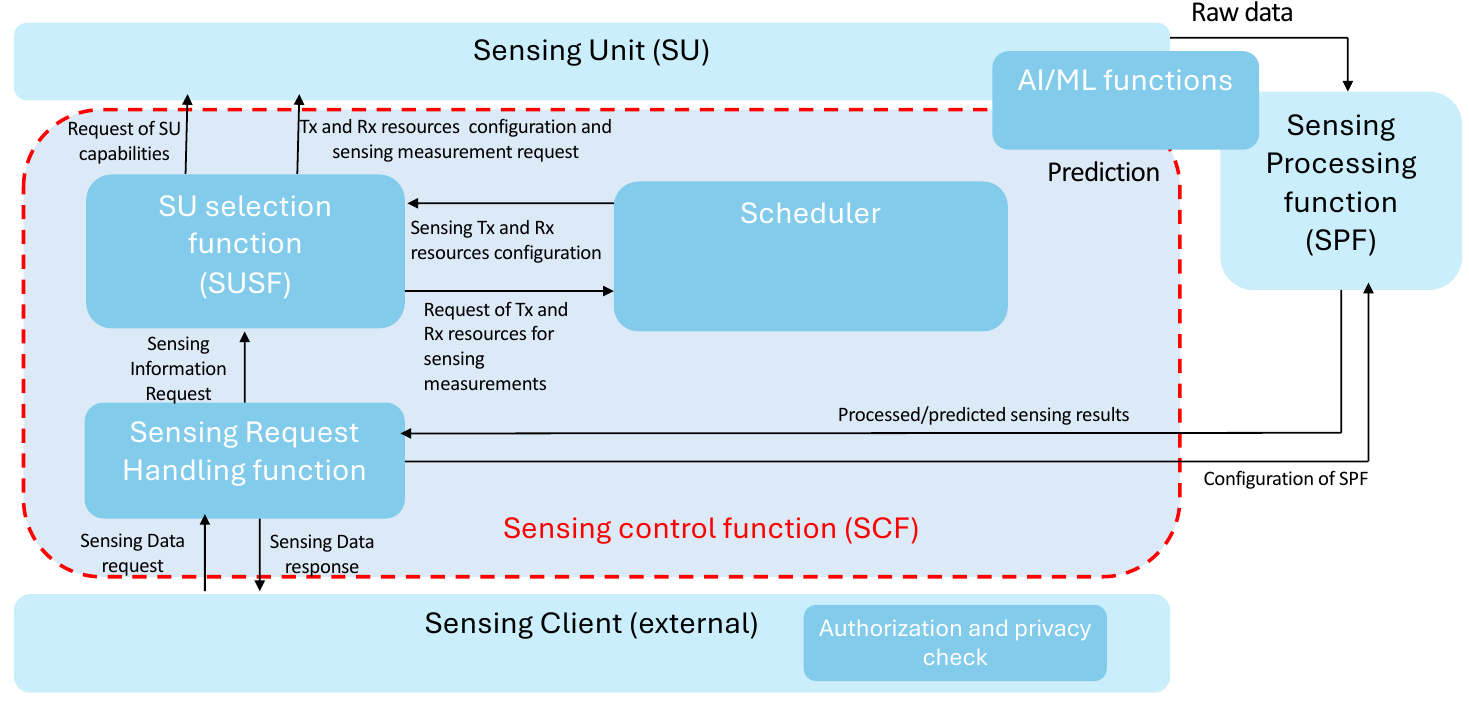}
    \caption{Detailed view of the sensing control function.}
    \label{fig:scf_jcas}
\end{figure*}

The implementation of sensing services in next-generation communication systems requires the introduction of a \ac{SeMF}. This function encompasses two primary components: sensing control and sensing processing. Building upon the functions introduced in the previous section, this section examines the control function and its essential components required for a sensing session. While the \ac{SCF} manages most of the sensing process, certain use cases require input from other system components. For example, when UE devices are involved in the sensing process, additional coordination is necessary. This includes locating optimal UEs based on their positions and coordinating measurement timing through mechanisms such as paging. Fig.~\ref{fig:signaling} illustrates the architecture of a communication system, highlighting the integration of new sensing functions and their role in processing application-based sensing requests. Fig.~\ref{fig:scf_jcas} provides a detailed functional analysis of the \ac{SCF}. Despite we outline the specific capabilities and requirements of the \ac{SCF}, it should be noted that the final implementation may involve adjustments to both the node-level distribution of capabilities and the associated terminology.

The sensing request serves as a crucial component in the process, as it largely determines the configuration of all involved nodes. Each request must contain two essential elements, i.e., an area of interest and an application \ac{ID}. The request may also include additional specifications such as sensing \ac{QoS} requirements or target characteristics (e.g., size, shape, movement patterns, velocity, and other relevant parameters). Each request may require its own configuration, unless the new request covers the same area as another recent request. Before reaching the \ac{SCF}, all requests undergo authentication and validation procedures to ensure security and compliance.

As illustrated in Fig.~\ref{fig:scf_jcas}, the sensing request is initially processed by a request handler. This function maintains comprehensive information about the network conditions, \ac{SU} capabilities, environmental parameters
and available UE that can function as \acp{SU}. The request handler forwards relevant information to the \ac{SPF}. The \ac{SPF} analyzes two key aspects, request requirements \& specifications and system configuration, including deployment geometry and \ac{SU} capabilities. Based on the request geometry, the \ac{SPF} assigns preliminary roles to available \acp{SU}, designating each as either a \ac{Tx} or \ac{Rx}. The request then moves to the \ac{SUSF}, which determines appropriate sensing modes (e.g., monostatic BS-based sensing, bistatic BS-based, UE-assisted sensing) based on multiple factors (e.g., \ac{SU} availability and capabilities, geometric configuration, environmental conditions or specific sensing requirements). The \ac{SUSF} coordinates with the cell scheduler to manage radio resource distribution across all services within the cell.

While this description assumes collocated functions, alternative deployments may be necessary for specific use cases. For example, when tracking an object along a known trajectory, the \ac{SUSF} determines resource allocation requirements and distributes them to all involved BSs. These requirements are then incorporated into the BS schedulers' resource allocation processes. Similar to the collocated case, the \ac{SUSF} provides BSs with two key elements, i.e.,  sensing resource configuration and reporting requirements for sensing-related information and measurements, including target data, signal analysis (periodograms) and environmental mapping (point clouds). For bistatic or multistatic sensing procedures, measurement results are transmitted only to the \acp{SCF} designated for receiving and processing sensing signals.

BSs and UEs selected by the \ac{SUSF} implement the validated configuration and begin sensing procedures based on their assigned roles (i.e., \ac{Tx} of sensing signals and/or \ac{Rx} of sensing signals). BSs and UEs designated as receivers forward sensing measurements to the \ac{SPF}. The \ac{SPF} processes this collected data to generate appropriate output for the requesting client. The output format varies based on the requested service type and may include target identification lists, object tracking information or raw sensing measurements for third-party processing.

\subsection{Computing and Application Layer}\label{computing-and-application-layer-aspects}

The integration of ISAC in 6G necessitates advanced computing frameworks capable of handling the diverse demands of real-time data processing, latency-critical applications, and distributed sensing. This section examines the technical implications for exposing data that is generated from ISAC services to internal/external consumers (e.g., vertical applications), and optimization strategies associated with the placement of the data consumer components, as well as compute offloading for ISAC-specific use cases.

\subsubsection{API Exposure for ISAC Services}

The exposure of network capabilities, including sensing and computational resources, to external applications is critical for realizing ISAC use cases in 6G. 3GPP’s \ac{SEAL} \cite{3gpp_seal} and the \ac{CAPIF}\cite{3gpp_capif} are two frameworks, which may play complementary roles in enabling this functionality. \ac{SEAL} focuses on exposing advanced 5G-specific services, such as location data, sensing functions, and \ac{QoS} management, through tightly integrated interfaces within the 5G service-based architecture. By working closely with the \ac{NEF}, \ac{SEAL} ensures secure, real-time access to network capabilities required for latency-sensitive and resource-intensive applications like navigation and real-time object detection.

\subsubsection{Optimization of ISAC consumer application placement}

ISAC use cases, such as object detection, navigation, and collaborative sensing, involve tasks that must be processed at different layers of the compute continuum—UEs, edge nodes, and centralized cloud resources. ISAC-specific scenarios that call for tailored optimization include object detection and tracking, collaborative sensing, and energy-constrained UEs. In the case of  object detection and tracking, processing high-resolution sensor data for real-time object identification is offloaded to edge nodes to achieve sub-millisecond latency. Coordination across multiple nodes ensures accuracy in multi-sensor deployments. In the case of collaborative sensing and localization, data from distributed UEs is aggregated and processed at edge servers. Dynamic resource allocation ensures real-time updates in scenarios like cooperative autonomous navigation of robots. Finally,  in the case of energy-constrained UE, offloading computationally expensive tasks reduces the energy burden on UEs while ensuring they operate within their thermal and power limits.

Optimizing the placement of these applications involves balancing computation, latency, and energy constraints dynamically across these resources.The placement process may rely on \ac{AI}-based optimization algorithms implemented within orchestration frameworks. These orchestrators continuously monitor metrics, including (i) compute load: the availability of processing and memory resources across the network nodes; (ii) latency metrics: end-to-end latency thresholds required by ISAC applicaitons, especially for real-time scenarios like collision avoidance; (iii) energy efficiency: battery constraints at UEs and energy consumption profiles of edge and cloud nodes; (iv) reliability and resilience: the operational state of computing nodes and their trust scores based on historical performance and dynamic assessments.

The orchestration framework must ensure that computationally intensive functions, such as object classification or sensor data fusion, are executed on edge nodes to minimize latency. Conversely, tasks requiring less stringent latency or greater computational power, like predictive analytics for navigation, may be offloaded to centralized resources.

\subsubsection{Compute offloading for ISAC}

Compute offloading plays a central role in ISAC by transferring processing tasks from resource-constrained UEs to network nodes with higher computational capabilities. This is particularly critical for scenarios involving high-dimensional data, such as multi-sensor fusion, where local processing would result in prohibitive latency or energy costs.

\begin{itemize}
    \item Dynamic workload partitioning: Tasks are divided into modular components using containerized or virtualized frameworks, enabling lightweight execution on heterogeneous platforms.
    \item Synchronization overhead: Synchronizing offloaded modules with remaining local application components involves additional communication. Optimized data encoding and secure transport protocols, such as encrypted session-based communication, mitigate these overheads while maintaining system integrity.
    \item Mobility support: UEs operating in dynamic environments require robust mobility-aware mechanisms. For instance, migrating an ongoing offloaded task between edge nodes while preserving application state is essential for maintaining service continuity in scenarios like autonomous vehicle navigation.
\end{itemize}

\subsubsection{Sensing Power Management}

There could be various use cases for sensing power management, but all require sensing as a network function. Sensing and processing in the BS (if implemented here) might be switched off unless requested by a UE or other BS. UEs contributing to sensing in case of collaborative sensing, might be controlled by the BS or network to enable or disable sensing on UEs. One example: An Autonomous Ground Vehicle at a charging station might not contribute to sensing and therefore the sensing function of the UE will be disabled. In case of multiple UEs are contributing to then sensing function and received information of a set of UEs look similar or the requested accuracy is already sufficient, UEs could be disabled. This in a turn results in less sensing processing power used in the BS or network. In an industrial environment sensing might be scalable on demand to enhance accuracy. Dynamic balancing of data rates or bandwidth for communication and sensing bandwidth might be an option. If only low communication bandwidth is required, the remaining bandwidth could be used for sensing. This provides a trade-off between required data rates and sensing accuracy. Additionally, management of sensing refresh-rate could contribute to power savings. Lower sensing refresh-rates (less processed frames) results in lower power consumption. For the device class RHDRBL this might be limiting due to higher resulting latency. Also, the transmit power could by dynamically adapted to save power. But lower transmit power reduces the maximum achievable distance for sensing and communication.

\subsection{Security and Privacy}\label{privacy-and-security} 

In addition to numerous technological benefits, ISAC also introduces various security and privacy threats. The inclusion of sensing data, which may contain sensitive \ac{PII}, makes ISAC-based applications more vulnerable to privacy threats such as location tracking, identity disclosure, profiling, and the misuse of sensed data \cite{dass2024apf}. Moreover, the addition of new components such as network functions and interfaces makes the ISAC system more prone to security attacks by providing additional avenues to adversaries to attack. 

\subsubsection{Security and Privacy Threats}
The security threats in ISAC pose challenges to the \ac{CIA} of communication as well as sensing data. On the other hand, the privacy threats are mainly due to the involvement of human sensing targets and the associated \ac{PII} collected, which give rise to threats like linkability, identifiability, and observability. Table \ref{tb-secobj} provides the summary of security and privacy threats in ISAC where security threats are identified as per the \ac{STRIDE} framework \cite{stride} while privacy threats are enumerated as per the \ac{LINDDUN} framework \cite{linddun}. Below, we discuss the potential threats in the sensing environment, the network, and the network-to-application interface.
\begin{itemize}
    \item 
\textbf{Threats in the sensing environment:} The wireless interface between the \ac{SU} and the target is susceptible to several passive and active threats due to its open nature \cite{guo2024secure}. Attackers could spoof the \ac{SU} or the target, inject false signals, or tamper with sensing data, leading to inaccurate or misleading results. Additionally, \ac{DoS} attacks can occur when attackers flood the \ac{SU} with excessive requests or jamming signals, causing the \ac{SU} to fail in initiating or maintaining its functions. Compromising the confidentiality and integrity of communications via lack of encryption, weak authentication and access control could lead to unauthorized access. Privacy threats also emerge when sensitive target data, such as location and behavior, is intercepted, allowing attackers to track movements or build profiles \cite{qu2024privacy}. Since the wireless signals can be easily intercepted, attackers could monitor target activities, track movements, or link specific individuals to their behavior \cite{liu2017indoor}. 

\item \textbf{Threats inside the network:} The network serves as the backbone of communication and sensing functions, making it a prime target for various security breaches \cite{park2007surveycoresec}. Attackers may attempt to exploit network vulnerabilities, gain unauthorized access to the network components, or manipulate data. Compromising network authentication or access control could allow attackers to inject malicious data or intercept sensitive information during transmission. Moreover, \ac{DoS} attacks with huge sensing data or requests could disrupt communication or sensing processes in the network. Additionally, due to the logical sharing of the network functions like \ac{SPF} and \ac{SCF} in different network slices, the administering entities could get access to the sensing data and \ac{PII}. Without proper anonymization or pseudonymization techniques, the linkage of these sensitive data could expose users to privacy risks. 

\item \textbf{Threats in the network-to-application interface:} The interface between the \ac{NEF} and applications or the ISAC consumers is vulnerable to a variety of security threats, including data interception, tampering, and impersonation \cite{TR33926}. Attackers may intercept or modify sensing requests and responses, causing incorrect or malicious data to be sent to the application. They could also spoof either the application or \ac{NEF}, gaining unauthorized access to sensitive data or disrupting normal operations. \ac{DoS} attacks in this interface could result in sensing service degradation or complete disruption of functionality. On the other hand, privacy risks in the network-application interface stem from the potential exposure of sensitive information during data exchanges between the network and the application. Attackers could exploit vulnerabilities in this interface to intercept data, track activities, or link personal information to specific users. Attackers may also analyze user behavior or profile the users/targets in the sensing environment by correlating sensing results with the sensing signals. 
\item \textbf{Threats in the UE sensing unit:}
In addition to general security threats such as spoofing, tampering and data disclosure threats \cite{device1800mobile} on UE related to initiation, transmission and processing of sensing data, other potential threats are: unauthorized (without consent) sensing in UE initiated by network or some adversary, unawareness of UE owner about sensing sessions orchestrated on his/her device, non-compliance of sensing activities and processing of sensitive sensing data with the sensing policies and agreements discussed and agreed with UE owner.
\end{itemize}

\begin{table*}
    \centering
    \caption{ISAC Threat Table.}
    \label{tb-secobj}
    \begin{tabular}{|>{\raggedright}p{0.1\linewidth}
|p{0.8\linewidth}|}
        \hline
          \textbf{Threat type}  & \textbf{Description}  \\ \hline
        Spoofing  & Unauthorized entities sending sensing requests, configurations, policies, logs, sensing results etc. pretending to be some legitimate entity \\ \hline  
        
        Tampering & Unauthorized modifications to sensing requests, responses, messages, configurations, policies, logs, results etc. to tamper with the sensing process or to inject malicious data to ISAC system. \\ \hline 
        
        Repudiation  &  An entity denying that it sent (or received) a sensing request or result to (or from) another entity.  \\ \hline
        
        Information disclosure  &  Sensitive information disclosure from sensing requests, messages, responses, configurations, policies, logs, sensing results, sensed raw data e.g., (frequency, size, duration), information about sensing targets, sensing area, sensing applications, ingestion points information or network internal endpoints information, highly valuable data (intellectual property of an operator) about sensing use-case, etc. \\ \hline
        
        \ac{DoS}  &   Modification of messages (unsupported fields and parameters in messages), crafting large sensing requests and sensing data, jamming attacks, physical layer attacks, exhausting UE resources (e.g., battery) by network sensing.\\
        & Some additional attacks (not exactly \ac{DoS} attacks): Ghosting real targets or artificially creating targets in physical environment.                        \\ \hline
        
        Elevation-of-privilege  & Being able to send or execute arbitrary sensing requests and intermediate results with read/write access to files, data records, processes, etc. without authorization. Some other type of threats could be the network being able to change arbitrary configurations at UE sensing unit, etc.   \\ \hline
        
        Linkability  &   Based on the attributes of sensing requests, responses and other messages, the adversary might be able to link the application, network component, and sensing targets or link the relevant data flows in other parts of the system.    \\ \hline
        
        Identifiability  &  Based on the attributes of sensing requests and sensing responses and other different messages, the adversary might be able to identify the application requesting sensing, network component or sensing targets involved in sensing.    \\ \hline
        
        Detectability  &  Observing the data flows traffic (attributes, frequency, message size, etc.) can hint what kind of sensing services and what private information is being processed in a given deployment or sensing area.    \\ \hline
        
        Unawareness  &  Silently starting the sensing without any indication to UE owner, not informing the UE owner about the type of data transmitted in sensing, duration of sensing and the resource usage in UE due to sensing. \\ \hline
        
        Non-compliance  &   Performing any operation deliberately or as a result of tampering or due to incorrect implementation that violates or neglects sensing policies. Assumption: Sensing policies enforces legal and regulatory constraints applicable on sensing.  \\ \hline
        
 \end{tabular}
\end{table*}

\subsubsection{Potential Security Controls}
Here, we discuss the security controls, on a general level, required in ISAC system to mitigate security threats such as spoofing, tampering, data disclosure etc. However, these controls are not exhaustive and do not include controls necessary to mitigate all the threats mentioned in Table \ref{tb-secobj}. Some security controls can be reused from 5G system while others may need further investigation. 
\begin{itemize}
    \item \textbf{Authentication and authorization:} Controls are required to deter spoofing threats by ensuring that communication and information transfer occur with the expected and authorized entity in ISAC system. The traditional 5G authentication methods such as 5G-AKA or EAP-AKA' and IPSec IKEv2 certificates-based authentication\cite{TS33501} can be used to authenticate UE (\ac{SU}) and the network, and BS (\ac{SU}) and the core network respectively. In addition, mobile network operators should take care of verifying and authorizing third parties interacting with the ISAC system. In cases, where UE acts as the sensing unit, explicit authorization of the sensing request is required in ISAC to schedule sensing on UE or use UE resources for the sensing purpose. Also, sensing should be performed in accordance with access control policies and sensing policies that reflect real-life constraints on sensing. The work in \cite{hex24_d24} proposed an architectural element \ac{SPCTM} which provides the framework for managing and applying the sensing policies on the sensing request. Logging and data provenance controls are required to provide accountability about who performed what action, when and for what purpose, etc. We believe that in ISAC, logging at UE side would be beneficial and may assist in collecting evidence to ensure the subscriber that sensing is happening in compliance with the consent and agreement policies set by the subscriber. 

\item \textbf{Integrity and confidentiality:} Protecting ISAC data-in-transit is paramount to ensure integrity and confidentiality of such data. Control data in ISAC can be secured in-transit using the 5G control plane security, which provides confidentiality, integrity protection, and replay protection\cite{TS33501}. In case a new plane, data plane, is designed to transmit sensing data, security controls should be designed to ensure confidentiality and integrity protection to securely transmit such data. Adversaries may eavesdrop on lower layer sensing signals (reflections from \ac{Rx} signals) and can create impression of surroundings or the targets. Further studies are needed to investigate how the sensing signals (\ac{Rx}/\ac{Tx}) can be encrypted and integrity protected to deter tampering and unauthorized disclosure threats. 

\item \textbf{Storage and processing:}
To secure sensing data at-rest (storage) and in-process (processing) at UE and network side, controls should be applied to ensure sensing data protection through entire life cycle (secure collection, storage to secure disposal), data retention policies, secure handling of keys for encryption and integrity protection, strict access control, and monitoring and logging access to the sensitive data. \Ac{TEE} \cite{jauernig2020trusted} technologies can be further investigated to process sensing information in a secure manner.   

\item \textbf{Threat detection:}
Anomaly or misbehavior detection systems, similar to other radar systems \cite{li2017radar, griebel2021anomaly}, could be explored for detecting and protecting against jamming in physical environment. In addition, further studies could investigate methods to protect against physical layer attacks such as adversaries creating false targets or ghosting real targets in the physical environment. The context-aware \cite{context-aware} and channel reciprocity-based key generation mechanisms \cite{FH-PLS} could be used for physical layer security in ISAC. UE should consider implementing transparency solutions and network could utilize such solutions to notify UE owners/subscribers when a sensing session has started on the UE and further inform about the type of data transmitted, duration and UE resources used in sensing.
\end{itemize}

\begin{figure}
    \centering
    \includegraphics[width=0.99\linewidth]{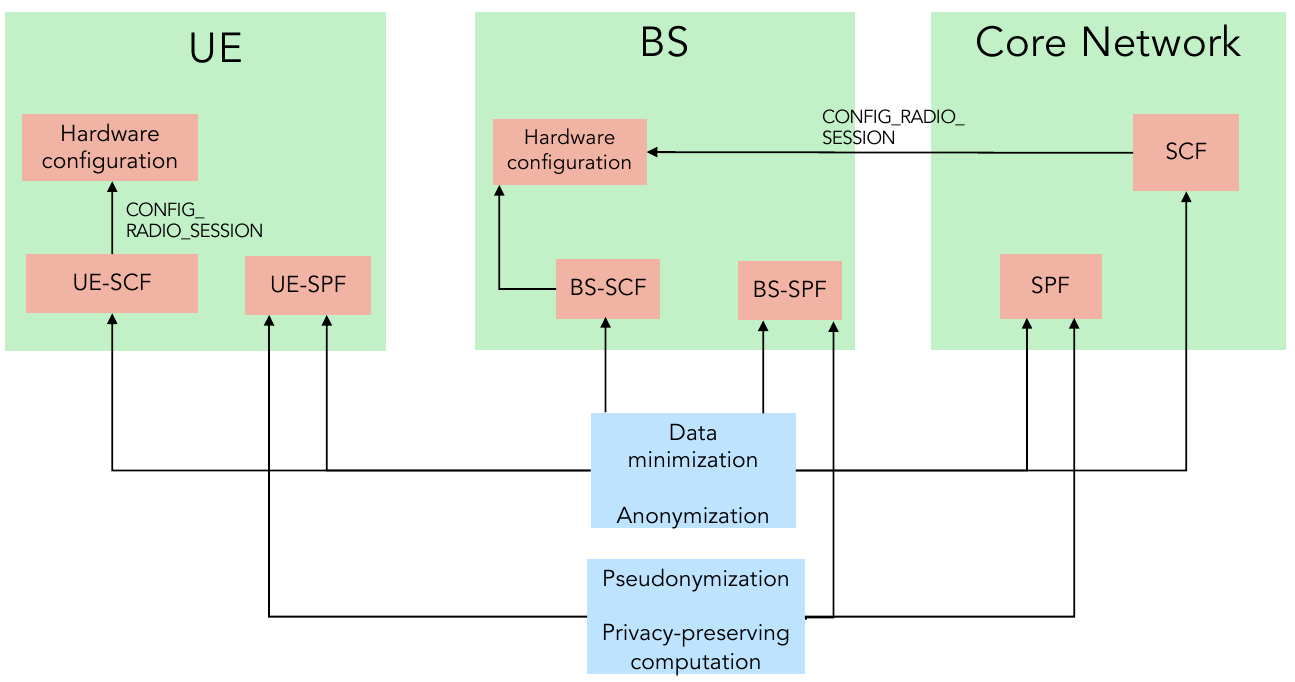}
    \caption{Privacy controls for ISAC.}
    \label{fig:priv_control}
\end{figure}

\subsubsection{Potential Privacy Controls}

Considering the sensitivity of the sensing data and the privacy of sensing targets, several  enhancements have been considered to be introduced to the network. 
\begin{itemize}
    \item \textbf{Dedicated functions:} A new network function called \ac{SPCTM} has been designed \cite{hex24_d24, dass2024apf}. This function serves as an administration tool for enforcing security and privacy controls within the sensing ecosystem. Its primary purpose is to ensure that other network functions involved in sensing activities can adhere to privacy preservation principles effectively along with sensing and consent policies. A dedicated repository called the “Sensing Store” is also proposed to record sensing policies, \ac{UC} data, sensing disclosure logs, consent data, and transparency data. This systematic compilation and storage of data could enhance the overall management and transparency of the sensing ecosystem, while improving compliance with General Data Protection Regulation (GDPR) and International Organization for Standardization (ISO) privacy frameworks. 
When integrating sensing units into \ac{UE}, it is imperative to prioritize security and privacy controls. These controls serve to protect sensitive sensing data, to uphold the UE's interests in consent and policy management and to ensure privacy-preserving and consent abiding sensing in UE. To achieve this, it is suggested to implement functionalities similar to the \ac{SPCTM} on the UE side. This enables the management of sensing policies, obtaining consent for enabling UE sensing services, and ensuring transparent data disclosure to the network.
\item 
\textbf{Data minimization and anonymization techniques:} These include  joint obfuscation, differential privacy, k-anonymization, data redaction, etc. can be applicable to the \ac{SPF} while processing the sensing data. These privacy controls can also be applied at the hardware-level to ensure that the sensing data collected is already privacy-preserved \cite{HW_config}. Beamforming capabilities allow selective coverage and exclusion zones by adjusting beam angles and null steering, which creates low-gain areas while preventing detection in private spaces. Adaptive beam narrowing and dynamic adjustment enable focused sensing on high-priority zones, such as entry points, while reducing coverage of unnecessary areas. Adjustable bandwidth and frequency allocation further help control sensing resolution. Transmission power control adjusts sensing range and detail by modulating power levels, with lower power focusing on closer targets and higher power for precise sensing nearby. Pulse duration and subcarrier modulation also impact resolution that shorter pulses or narrower subcarrier spacing provide finer detail, while longer pulses or wider spacing reduce data volume. 
\item \textbf{Pseudonymization techniques:} These substitute identifiable data with pseudonyms, enabling data linkage across sessions without directly revealing identities \cite{pseudonymization}. In the 5G system, the Subscription Concealed Identifier (SUCI) acts as a pseudonym to protect the Subscription Permanent Identifier (SUPI) of users \cite{SUPI-SUCI}. In ISAC, however, sensitive \ac{PII} must also be safeguarded; therefore, pseudonymization should be applied to the sensing data at the \ac{SPF} to protect the \ac{PII} of sensing targets as well as the entities involved in data collection and processing. Additionally, ISAC systems with distributed sensing processing across UEs and BSs, privacy-preserving computation techniques, such as secure multi-party computation, homomorphic encryption, and federated learning can also be applied at the \ac{SPF} to protect the \ac{PII} during transfer and processing of sensing data.
\end{itemize}

Fig.~\ref{fig:priv_control} represents a privacy-preserving framework for ISAC systems by illustrating key network components and where specific privacy controls can be implemented. The figure divides the system into three main blocks: UE, BS, and the Core Network. The UE and BS include sensing functions similar to those in the core network—specifically, the \ac{SCF} and \ac{SPF}. These are designated as UE-\ac{SCF} and UE-\ac{SPF} for the UE, and BS-\ac{SCF} and BS-\ac{SPF} for the BS.

\section{A Quantitative Cross-Layer Evaluation}\label{a-quantitative-cross-layer-evaluation}
In this section, we evaluate the performance of ISAC systems in various scenarios and across various layers. To be specific, we test bistatic and monostatic ISAC systems in indoor, urban, and rural settings using both simulation tools and \ac{PoC} experimental setups. These evaluations connect lower- and higher-layer degrees of freedom to \acp{KPI} established in Sec.~\ref{KPIs}, namely sensing accuracy, resolution, and latency.
\subsection{Use case Description}\label{use-case-description-and-assumptions}
We introduce the three ISAC use-case scenarios that this study focuses on. For each use-case, we discuss the main goal of sensing, the properties of the targets of interest, the ISAC protocol, the expected \ac{KPI} requirements, and the underlying assumptions.
\begin{itemize}
    \item \textbf{Indoor sensing: Network centric human detection:}
In the indoor ISAC scenario, we focus on the detection and tracking of human activity within confined spaces such as homes, offices, and industrial facilities. Common goals include presence detection, motion tracking, gesture recognition, and posture estimation for use cases such as elder care, intrusion detection, and human-machine interfaces. The targets here are humans (both stationary and mobile) and their fine-grain motion features such as hand gestures or breathing patterns. The ISAC architecture assumes a bistatic setup. The transmitter performs spatial beam sweeping using pilot signals, while the receiver processes reflected signals for sensing tasks. Sensing in indoor settings require high spatial and temporal resolution to perform fine-grain tracking. Low latency is also critical, particularly for real-time applications like security alerts or gesture control. Key assumptions include a relatively stable indoor layout and moderate target mobility. Hardware capabilities such as multi-antenna arrays, real-time signal processing, and beamforming are assumed. 

\item \textbf{Urban sensing: UE-centric intersection hazard detection:}
In the urban setting, we focus on the usecase of improving road safety by detecting and tracking vulnerable road users at intersections. The objective is to enable proactive collision avoidance by providing timely information about pedestrians, bicycles, or vehicles entering an intersection. The use case assumes a bistatic architecture, where the UE periodically or aperiodically emits sensing reference signals based on other UE-triggered requests. The basestation collects reflections and transmits measurements to a centralized \ac{SeMF} for further processing. Key performance requirements follow the 3GPP TS 22.137 specifications: 0.5 m positioning accuracy, 54 km/h object velocity support and 0.1–5 s latency \cite{TS22137}. 
These metrics are crucial to ensure reliable detection in dense, dynamic traffic environments. Assumptions include moderate to high mobility of targets, availability of network infrastructure for coordination, and low signal blockage due to surrounding buildings. Hardware support is assumed on both infrastructure and UE sides, including multi-antenna arrays and fast processing for latency-critical tasks. 

\item \textbf{Rural sensing: Network-centric highway hazard detection:}
In the rural setting, we focus on a highway scenario, were ISAC is employed to detect hazards such as large animals or stationary vehicles on poorly lit or winding roads. These obstacles pose safety threats especially in areas with low infrastructure density and long emergency response times. The sensing architecture is monostatic, where base stations conduct periodic scans of road segments. Due to large cell sizes and sparse infrastructure, sensing accuracy is inherently lower than in urban settings. However, early hazard detection does not require high precision; coarse estimates of obstacle presence are sufficient to alert drivers to reduce speed or brake. Typical \ac{KPI} requirements include tens of meters accuracy and multi-second latency tolerance 
under low-traffic conditions. Assumptions include relatively low background variability and low multipath reflections from surrounding environment (e.g., trees). Infrastructure is expected to support high-power transmission and large coverage. 
\end{itemize}

\subsection{Degrees of Freedom}\label{design-degrees of freedom}

In this study, we consider 15 \acp{DoF} (12 lower-layer \acp{DoF} and 3 higher-layer \acp{DoF}) that can affect the \acp{KPI}. Some of these \acp{DoF} are also affected by the assumptions used from a given use case. In this section, we will first categorize the various \acp{DoF} and then relate them to the \acp{KPI}.
\subsubsection{Categorization of \acp{DoF}}
The \acp{DoF} can be divided into \textit{long-term} \acp{DoF} (BS locations, number of antennas, and number of \ac{RF} chains at the \ac{Tx} and the \ac{Rx}), \textit{short-term radio-based} \acp{DoF} (carrier frequency, sub-carrier spacing, number of sub-carriers, number of \ac{OFDM} symbols, transmit power, and precoder/combiner design), and \textit{short-term networking-based} \acp{DoF} (processing location and processing power). Long-term \acp{DoF} are usually dictated before the deployment of the system and are considered permanent. On the other hand, short-term radio-based and networking-based \acp{DoF} can be configured frequently during operation. Since the combinations between these \acp{DoF} are limitless, some restriction is needed on the \acp{DoF} for a study to be feasible. In this particular study, we divide the \acp{DoF} into fixed and semi-fixed. 
\begin{itemize}
\item \textbf{Fixed \acp{DoF}}: Fixed \acp{DoF} include BSs locations, number of \ac{RF} chains at the BSs and the UEs, transmit power, the beamforming/precoder/combiner vectors, and number of sub-carriers, Those fixed \acp{DoF} will not be changed from one study to the other and are shown in Table~\ref{tab:DoFFixedRadio}.

\item \textbf{Semi-fixed Radio \acp{DoF}}: The semi-fixed radio \acp{DoF} include the carrier frequency, number of antennas at the BS and the UE, number of \ac{OFDM} symbols, and sub-carrier spacing. These five \acp{DoF} are treated together in combinations, with the main \ac{DoF} being the carrier frequency, which will then dictate the number of antennas, number of \ac{OFDM} symbols, and the sub-carrier spacing used. Please note that here, the number of \ac{OFDM} symbols is equal to the number of \ac{Tx} antennas multiplied by the number of \ac{Rx} antennas. In this study, we will test two carrier frequencies, namely 10 GHz (FR3) and 60 GHz (FR2). The corresponding combinations are shown in Table~\ref{tab:DoFSemiFixedRadio}.

\item \textbf{Semi-fixed Networking \acp{DoF}}: The semi-fixed networking \acp{DoF} are related to the sensing processing location (proximity to the sensing nodes and the core) and its computational power. The location will dictate the proximity to the sensing nodes and the communication bandwidth available, which will dictate both the propagation and communication delays. Additionally, the processing location will also influence the computational power available, which will dictate the processing delay. In this case study, we consider three processing locations, namely the extreme edge, the edge, and the core. The propagation delay, effective bandwidth, and computational power associated with each node location are shown in Table~\ref{tab:DoFSemiFixedNetworking}.
\end{itemize}

\begin{table}
\centering
\caption{Fixed simulation degrees of freedom.}
\label{tab:DoFFixedRadio}
\begin{tabular}{|p{0.45\linewidth}|p{0.4\linewidth}|}
\hline
\textbf{\ac{DoF}} & \textbf{Value} \\
\hline
BS locations & Scenario dependent \\
\hline
\#\ac{RF} chains at BSs/UEs& 1 \\
\hline
\ac{Tx} power& 20 dBm \\
\hline
Precoding/combining vectors& \ac{FFT} codebook \\
\hline
\#sub-carriers& 792 \\
\hline
\end{tabular}
\end{table}

\begin{table}
\centering
\caption{Semi-fixed simulation degrees of freedom (radio).}
\label{tab:DoFSemiFixedRadio}
\begin{tabular}{|p{0.4\linewidth}|p{0.2\linewidth}|p{0.2\linewidth}|}
\hline
\textbf{\ac{DoF}} & \textbf{First value} & \textbf{Second value}\\
\hline
Carrier frequency & 10 GHz & 60 GHz \\
\hline
\# antennas at BSs/UEs& 2x2 & 4x4 \\
\hline
\#\ac{OFDM} symbols& 16 & 256 \\
\hline
Sub-carrier spacing& 30 kHz & 120 kHz \\
\hline
\end{tabular}
\end{table}

\begin{table}[t!]
\centering
\caption{Semi-fixed simulation degrees of freedom (network).}
\label{tab:DoFSemiFixedNetworking}
\begin{tabular}{|p{0.15\linewidth}|p{0.2\linewidth}|p{0.2\linewidth}|p{0.2\linewidth}|}
\hline
\textbf{\ac{DoF}} & \textbf{Extreme Edge} & \textbf{Edge} & \textbf{Core}\\
\hline
Propagation latency & 0.015 ms & 0.15 ms & 0.8 ms \\
\hline
Comms. BW& 0.1 Gbps & 1 Gbps & 10 Gbps \\
\hline
Computational power& 10 GFLOP & 100 GFLOP & 300 GLFOP \\
\hline
\end{tabular}
\end{table}

\subsubsection{Relations to \acp{KPI}}
In the following, we relate the \acp{DoF} to the sensing \acp{KPI} highlighted in Sec.~\ref{KPIs}.
\begin{itemize}
    \item 
\textbf{Accuracy and resolution:}
Enhancing sensing accuracy and resolution is usually the main goal for many research works. Theoretical error bounds on sensing accuracy (i.e., target position error bound (TPEB), which is based on the Cramér-Rao lower bound (CRLB) \cite{kay1993fundamentals}) were derived but not included in the paper. However, it is worth noting that the derived CRLB is connected to all 12 lower-layer DoFs discussed earlier. 
As for resolution, we deem a target to be non-resolvable if it cannot be resolved in all range, velocity, and angular domains. More on how non-resolvable targets are treated in this study will follow in the next sections. The range resolution is given by
    $\Delta R = {c}/{(2B)}$, 
where $c$ denotes the speed of light and $B$ is the signal bandwidth. The velocity resolution is defined as 
    $\Delta v = {\lambda}/{(2 T_\text{Tx})}$, 
with $\lambda$ being the wavelength and $T_\text{Tx}=M\Tsym$ is the total waveform transmission duration \cite{koivunen2024multicarrier}. Furthermore, the angular resolution is expressed as 
    $\Delta \phi = {0.89 \lambda}/{D}$ \cite[Eq.~(1.9)]{richards2005fundamentals}, 
where $D$ represents the effective aperture of the antenna array.

\item \textbf{Latency:}  
Achieving low sensing latency is crucial for applications that are either highly dynamic or operate in highly dynamic environments, like autonomous driving, as it affects their sensing accuracy. For instance, if a target is moving at a speed of 36 km/h (i.e., 10 m/s), a sensing latency of 100 ms can cause 1 m of error, regardless of the sensing accuracy \cite{behravan2022positioning}. Sensing latency in ISAC systems encompasses not only the physical layer transmission and propagation delays but also the processing delays associated with data management. The total sensing latency can be expressed as 
    $T_{\text{total}} = T_\text{Tx}+ T_\text{prop} + T_{\text{proc}}$, 
where $T_\text{prop}$ is the propagation delay between the \ac{Tx}, the target, and the \ac{Rx}, and $T_{\text{proc}}$ is the processing latency. Here, the processing latency can be formulated as  $T_{\text{proc}}= T^\text{proc}_\text{Tx} + T^\text{proc}_\text{prop} + T^\text{proc}_\text{comp}$. Here, $T^\text{proc}_\text{Tx}={D_\text{sensing}}/{B_\text{comms}}$ represents the total transmission duration to communicate sensing data between sensing and processing nodes, where $D_\text{sensing}$ is the volume of the sensing data and $B_\text{comms}$ is the effective communication bandwidth available. Next, $T^\text{proc}_\text{prop}$ represents the propagation delay between the sensing and processing nodes (based on proximity). Finally, $T^\text{proc}_\text{prop} = {C}/{F}$ represents delay incurred from the actual processing of the data, where $C$ is the computational load, which in turn depends on the volume of the sensing data and sensing algorithms employed, and $F$ is the processing power of the processing node. In a distributed scenario consisting of $I$ distributed sensing nodes, the processing delay can be generalized to $T_{\text{proc}}= \max_i ( T^\text{proc}_{\text{Tx},i} + T^\text{proc}_{\text{prop},i} + T^\text{proc}_{\text{comp},i})$. All of these aspects will be influenced by higher-layer factors such as the efficiency of application placement strategies, processing architectures, data management, and security, which we do not cover in this study. 
\end{itemize}

\subsection{Experimental Setup}\label{simulation-setup}
The cross-layer evaluation is divided into simulation-based and \ac{PoC}-based experiments. We conduct three simulation-based experiments for the indoor, urban intersection, and rural highway scenarios. The simulation code can be accessed by following the url \url{https://github.com/Hexa-X-II/ISAC-for-6G}. Additionally, we perform two indoor \ac{PoC}-based experiments, encompassing sensing of motion of humans and gesture recognition. In the following, we detail simulation and \ac{PoC} setups.

\subsubsection{Indoor Simulation Setup}

We consider two indoor bistatic sensing scenarios as illustrated in Figs.~\ref{fig:roomQRT60Ghz} and \ref{fig:room60Ghz}. In the first scenario, a \ac{Tx} and an \ac{Rx} are located in the bottom of a rectangular room, facing each other at a 45$\degree$, while the second scenario have the \ac{Tx} and the \ac{Rx} placed at the top and the bottom of the room facing each other. In both scenarios, both the \ac{Tx} and the \ac{Rx} are equipped with a planar antenna array with digital beamforming capabilities. The room is empty except for the \ac{Tx}, \ac{Rx}, and a target to be sensed. The \ac{Tx} illuminates the room and its energy is directed towards the \ac{Rx} through three mechanisms: (i) a \ac{LoS} path, (ii) diffuse scattering off the target, and (iii), specular reflections off flat surfaces present in the room. We make the following assumptions regarding the propagation conditions: \begin{itemize}
    
    \item For (ii), we consider only a single point-target characterized by a position in three-dimensional space and its reflectivity. The reflectivity, in turn, is assumed independent of the angle-of-incidence and modeled as a complex constant with magnitude $\sqrt{\sigma_\mathrm{RCS}}$, where $\sigma_\mathrm{RCS}$ is the target \ac{RCS}, and unknown phase. When illuminated, the target scatters energy in all directions yet we only model the direct path to the \ac{Rx}. The target is stationary or slowly moving such that Doppler effects are negligible.

    \item We restrict the reflecting, flat surfaces in (iii) to include only the four vertical walls of the room. We assume that the \ac{Tx} and \ac{Rx} are situated in the same horizontal plane and their beams are focused center in elevation such that reflections off the floor and ceiling can be ignored. Further, only first-order reflections are considered. 
    
\end{itemize} 


\subsubsection{Outdoor Simulation Setup}
The outdoor scenarios constitute an urban (intersection) setting, shown in Fig.~\ref{fig:intersection10GHz}, and a rural (highway) setting. The urban street intersection scenario comprises a user on the street that is communicating with a \ac{BS} atop the corner of a building, surrounded by three other buildings. In this scenario, the user is at a known location and transmits an up-link reference signal to the \ac{BS} from which a single scatterer is to be located. Both the user and \ac{BS} employs planar arrays and are directive, and follow the received signal model employed in the indoor scenarios with the minor addition of a ground reflection. The second outdoor scenario involves a monostatic radar system overlooking a highway with a single target to be detected. Ground reflections in this case are ignored. Hence, the received signal model is similar to the one discussed earlier without the additional multipath components.

\subsubsection{Higher-layer Simulation Setup}
In this setup, we study the effect of distributed processing of sensing data on the sensing latency, and thus the sensing accuracy. In this experiment, we consider three processing locations, as discussed earlier. For each node, the latency and accuracy performance are examined under various sensing computational loads (varied from 0 to 200 MFLOP) while assuming a target speed of 10 m/s. The computational load, in reality, is dictated by the number of served users, \ac{OFDM} symbols, sub carriers, antennas, and the sensing algorithm used.  For instance, conducting a single sensing request (estimating delay and angle of departure) while utilizing 792 sub-carriers and 64 \ac{OFDM} symbol via 2D \ac{FFT} consumes around 4 MFLOP. This figure can easily increase if more than one sensing request/target is served/sensed simultaneously, more sophisticated techniques are used (e.g., maximum likelihood estimation), or the number of dimensions increases (e.g., using a digital array at the transmitter or the receiver).

\subsubsection{\ac{PoC} Experimental Setup: Human Gesture Recognition}\label{POC}
In the \ac{PoC} experiments for gesture recognition, we set up a bi-static 5G \ac{mmWave} ISAC system using two Sivers Evaluation Kits (EVK6002), functioning as a \ac{Tx} and \ac{Rx}. The EVKs are controlled using a \ac{RF} system-on-chip (RFSoC). The system implements a 5G-NR \ac{OFDM} waveform consisting of a synchronization signal block (SSB) and random data over a single carrier with 792 subcarriers, having a total bandwidth of approximately 760 MHz at 60.48 GHz central frequency. The \ac{Tx} continuously transmits a single frame of 10 milli-seconds according to the 5G standard. Beam sweeping is then performed across 50 \ac{Tx} and 56 \ac{Rx} beams by transmitting 2 \ac{OFDM} symbols per beam. 
To capture the characteristics of each gesture, we compute the \ac{PPBP} across 50 transmit and 56 receive beams. The time series of \acp{PPBP} is input into a \ac{CNN} for gesture classification or pose estimation. For pose estimation, a Microsoft Kinect is used as ground truth to capture the 3D coordinates of 25 joints of human body. The overall dataset comprises 74 minutes of activity.
Human pose estimation \cite{wang2021deep} using wireless signals poses a significant challenge due to the absence of visual information in the \ac{RF} signals. To do this domain translation, from wireless signals to visual skeletons, we use a multi-modal setup with \ac{mmWave} 5G hardware and Microsoft Kinect. In computer vision, going from image to skeletons is a well-solved problem and several solutions exist \cite{qiao2017real,kim2023human}. Using this ground truth, we train a \ac{CNN} to map wireless representations, specifically spatio-temporal \ac{PPBP}, to the corresponding human body skeletons or poses. During the inference phase, the \ac{CNN} predicts human poses solely based on wireless signal input (\ac{PPBP}). We adopt a \ac{CNN} architecture from the work \cite{szegedy2015going} to do this domain translation.

\subsubsection{\ac{PoC} Experimental Setup: Indoor Sensing}
In the \ac{PoC} experiments for indoor sensing, we conducted measurements using 5G-NR-adjusted waveforms in an office environment. The measurements were performed at different bi-static distances, using both a reflector and a human subject as the target. 
The measurement setup, shown in Fig. \ref{fig:all}-(f), 
utilizes two Sivers radios in a bi-static sensing configuration. The carrier frequency was 60 GHz, and a bandwidth of 800 MHz was used to transmit two adjusted 5G NR frames of 10 ms continuously. The waveform consisted of an \ac{OFDM} signal, with resource blocks allocated for SSB and PDSCH symbols, following the 5G NR standard \cite{lin20195g}. Beamforming was achieved by sweeping 50 beams at the transmitter and 56 beams at the receiver in the horizontal plane, resulting in a total of 2800 beam combinations per full sweep. The measurements were conducted in an office space that represents a rich-scattering environment, with numerous unwanted reflections from walls, desks, and cupboards. To mitigate the impact of these reflections, we first performed a background subtraction step by measuring when no one was present in the corridor. By subtracting these background measurements from the full-scale measurements, we isolated and analyzed only the reflections from the human targets in the corridor. To evaluate the performance of our setup, we used two different bi-static distances of 4 and 6 meters within the office space. Human subjects stood in a normal upright position, and 200 frames were recorded for each target. Each frame included 2800 beam combinations, providing a full scan of the office environment. For reference, we also performed measurements with a reflector placed at the same positions as the human targets, in separate rounds.

\subsection{Results and Discussions}\label{results-and-discussions}
\subsubsection{Simulation Results: Indoors}

\begin{figure}[t!]
    \centering
    \includegraphics[trim=70pt 190pt 70pt 190pt,clip,width=1\linewidth]{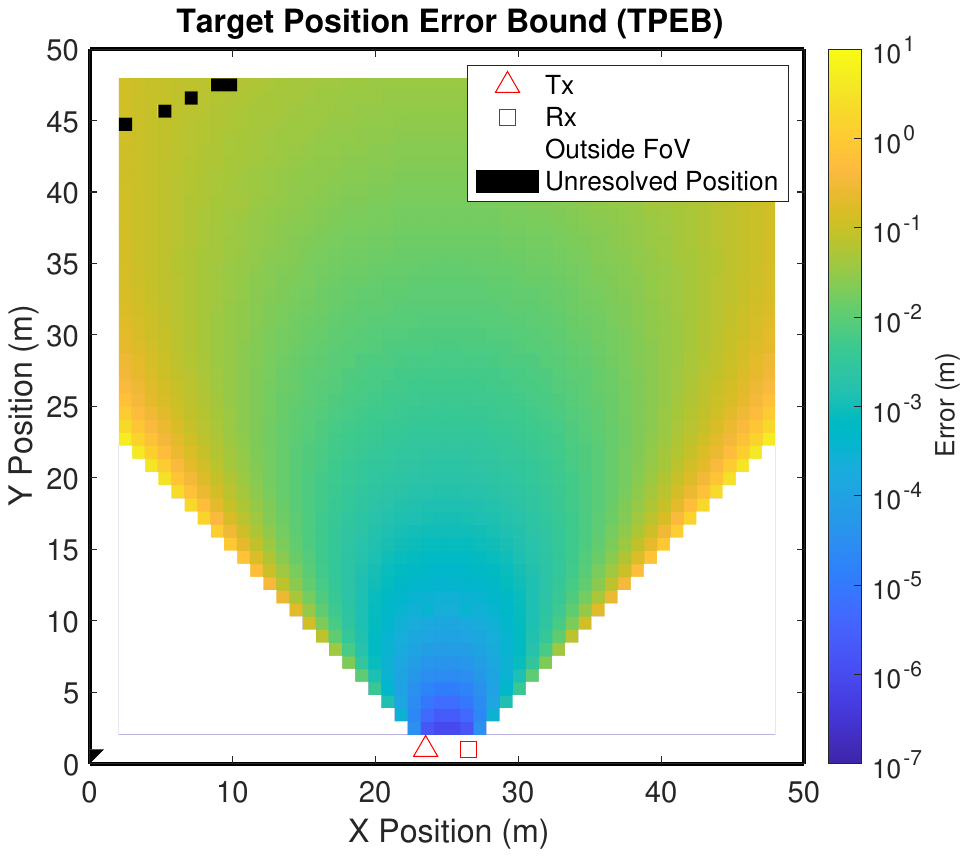}
    \caption{Indoor simulation scenario 1, 60 GHz. \ac{Tx} and \ac{Rx} are placed at the bottom of the room, facing each other at a 45 $\degree$. }
    \label{fig:roomQRT60Ghz}
\end{figure}

\begin{figure}[t!]
    \centering
    \includegraphics[trim=70pt 215pt 70pt 215pt,clip,width=1\linewidth]{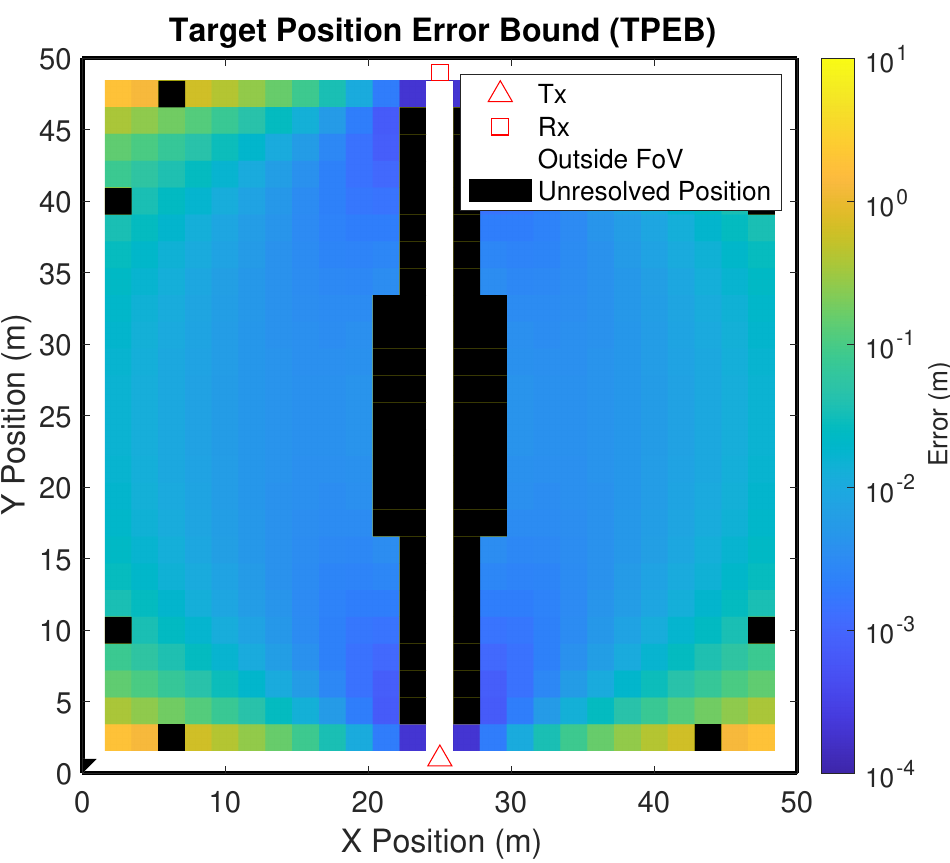}
    \caption{Indoor simulation scenario 2, 60 GHz. \ac{Tx} and \ac{Rx} are placed facing each other at the bottom and top of the room, respectively.}
    \label{fig:room60Ghz}
\end{figure}

Results shown in Figs.~\ref{fig:roomQRT60Ghz}--\ref{fig:room60Ghz} display CRLB-based heatmaps of sensing errors for the two scenarios, respectively, using the 60 GHz setup. The goal of this experiment is to show the effect of \ac{Tx} and \ac{Rx} placement on sensing performance in indoor scenarios. The heatmaps can be divided into three color schemes, namely white, black, and other colors. White regions show areas that cannot be sensed due to 1) not being in \ac{LoS} to either the \ac{Tx} or the \ac{Rx}, or 2) not being in the \ac{FoV} of either the \ac{Tx} or the \ac{Rx}. Since such errors are related to the deployment scenario/geometry and not the radio parameters, they are not taken into consideration when calculating errors. Black regions define areas were targets cannot be resolved in both delay and angular domains from other paths. As discussed earlier, resolvability depend on radio degrees of freedom, and thus, they are included in error calculations, and are given an error equal to the maximum possible in the test region (i.e., the error between the center of the testing region to the furthest corner to it). Finally, other colorful regions denote areas were targets can be feasibly sensed and the the color-map dictates the lowest achievable estimation error.
In Figs.~\ref{fig:room60Ghz}--\ref{fig:roomQRT60Ghz}, it can be seen that 
changing the position of the \ac{Tx} and the \ac{Rx} can dramatically change the regions of resolvability and accuracy of the measurements. For instance, deploying the \ac{Tx} and the \ac{Rx} closer to each other will decrease the area were targets cannot be resolved from the \ac{LoS} path. However, this also leads to  much longer paths for targets that are on the other side of the room, which leads to lower \ac{SNR} at the receiver and lower sensing accuracy. Additionally, placing the \ac{Tx} and the \ac{Rx} close and facing each other with a slight angle causes one-third of the room to be not in the feasible sensing region due to it being out of the field-of-view of the \ac{Tx} or the \ac{Rx}. This underscores the importance of studying the impact of position of the \ac{Tx} and \ac{Rx} on sensing resolvability and accuracy before deployment. 

\subsubsection{Simulation Results: Outdoors}

\begin{figure}[t!]
    \centering
    \includegraphics[trim=75pt 190pt 65pt 190pt,clip,width=1\linewidth]{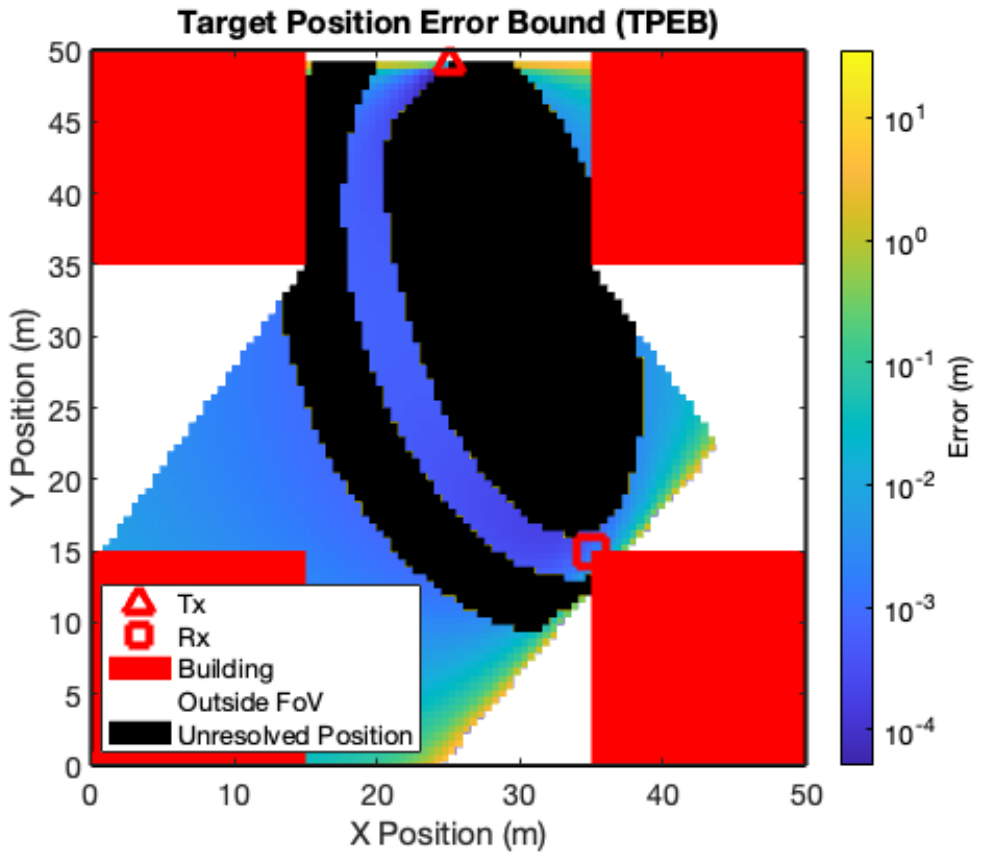}
    \caption{Urban intersection simulation, 10 GHz.}
    \label{fig:intersection10GHz}
\end{figure}

\begin{figure}[t!]
    \centering
    \includegraphics[trim=65pt 215pt 65pt 215pt,clip,width=1\linewidth]{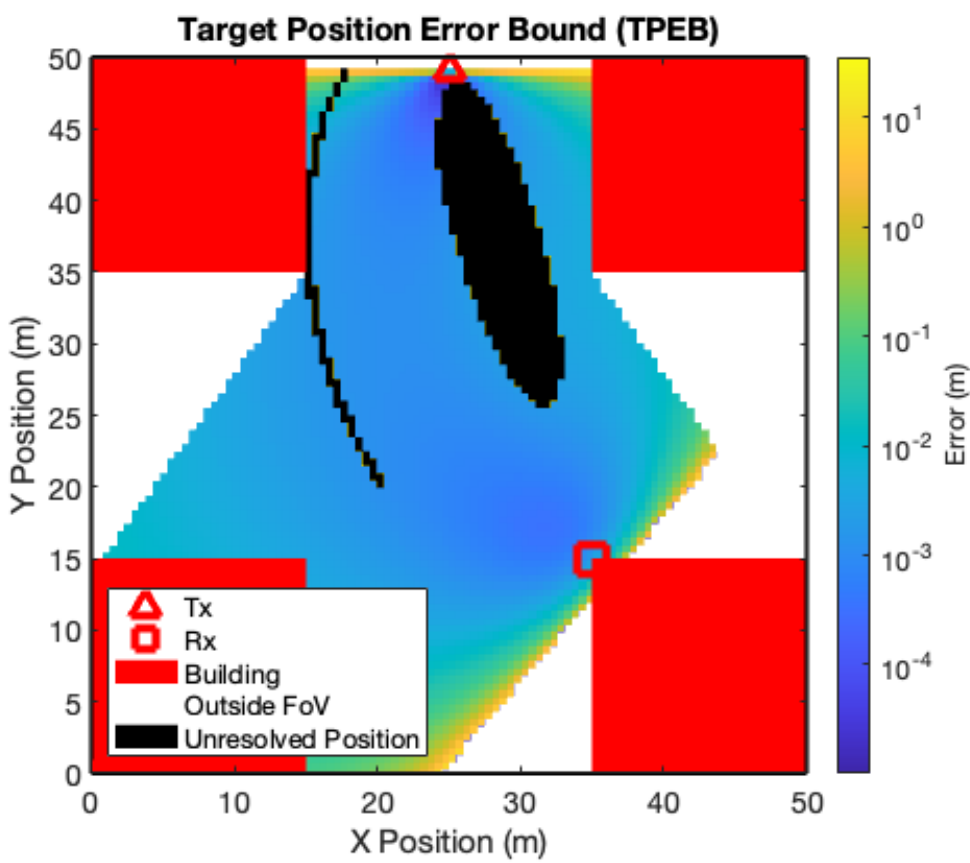}
    \caption{Urban intersection simulation, 60 GHz.}
    \label{fig:intersection60GHz}
\end{figure}

First, we discuss the urban scenario. 
The results shown in Figs.~\ref{fig:intersection10GHz}--\ref{fig:intersection60GHz} showcase the sensing performance of the 10 GHz and the 60 GHz systems, respectively, in an urban intersection. In this scenario, the receiver receives 4 signals comprising the \ac{LoS} signal, a ground reflection, a reflection from the top-left building, and a reflection from the target. The goal here is to assess the effect of the usage of higher frequency, bandwidth, and number of antennas on resolvability in urban environments. It can be seen that in such a typical deployment scenario, where the UE is expected to be far from the BS, resolvability between the target path and the \ac{LoS} path dictates the majority of the feasible sensing area on the ground. Increasing the frequency, bandwidth, and number of antennas helped with minimizing the non-resolvable region significantly. To be specific, the the oval-shaped non-resolvable region in the middle; caused due to non-resolvability with the \ac{LoS} path, shrunk in the 60 GHz setup, as compared to the 10 GHz setup. Similarly, both the width and length and width of the non-resolvable arc on the left-hand side; caused by the reflected path bouncing off the top-left building, were minimized in the higher frequency setup.

\begin{figure}[t!]
    \centering
    \includegraphics[trim=45pt 190pt 65pt 190pt,clip,width=1\linewidth]{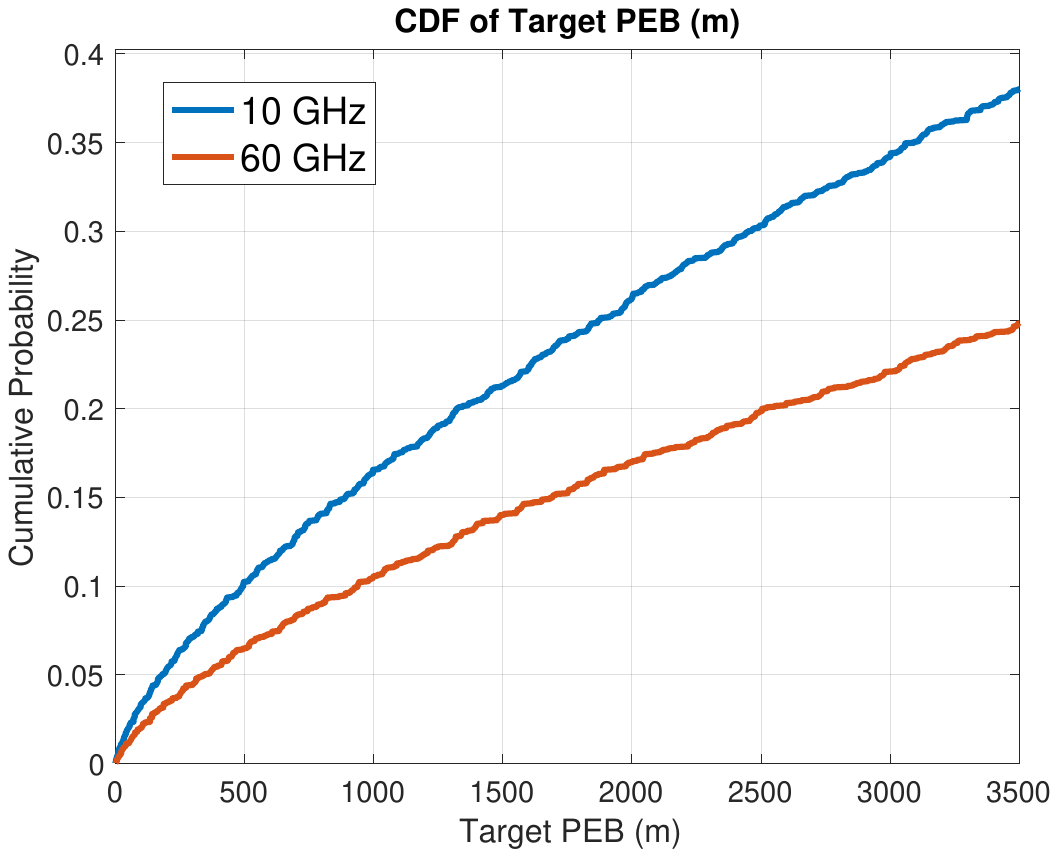}
    \caption{Rural highway simulation, error CDF plot for 10 GHz and 60 GHz systems.}
    \label{fig:ruralCDF}
\end{figure}

Next, we proceed with the rural highway scenario. 
Since the simulated rural environment only consists of a single reflective target without background clutter, all of the simulation region was resolvable. In such case, the higher frequency setup will lose the resolvability advantage gained in both the indoor and the urban intersection scenarios. Thus, any results will reflect the actual effect of frequency, bandwidth, and number of antennas on sensing accuracy in scenarios where resolvability is not a challenge. Fig. \ref{fig:ruralCDF} shows the error CDF of both high and low frequency systems. It can be seen that the usage of lower frequencies results in higher accuracy. Although the higher frequency system had access to higher bandwidth and number of antennas, which should enhance the sensing performance, they were not enough to offset the high pathloss experienced at such high frequencies.

\subsubsection{Simulation results: Higher Layers}
\begin{figure}[t!]
    \centering
    \includegraphics[trim=45pt 220pt 50pt 220pt,clip,width=1\linewidth]{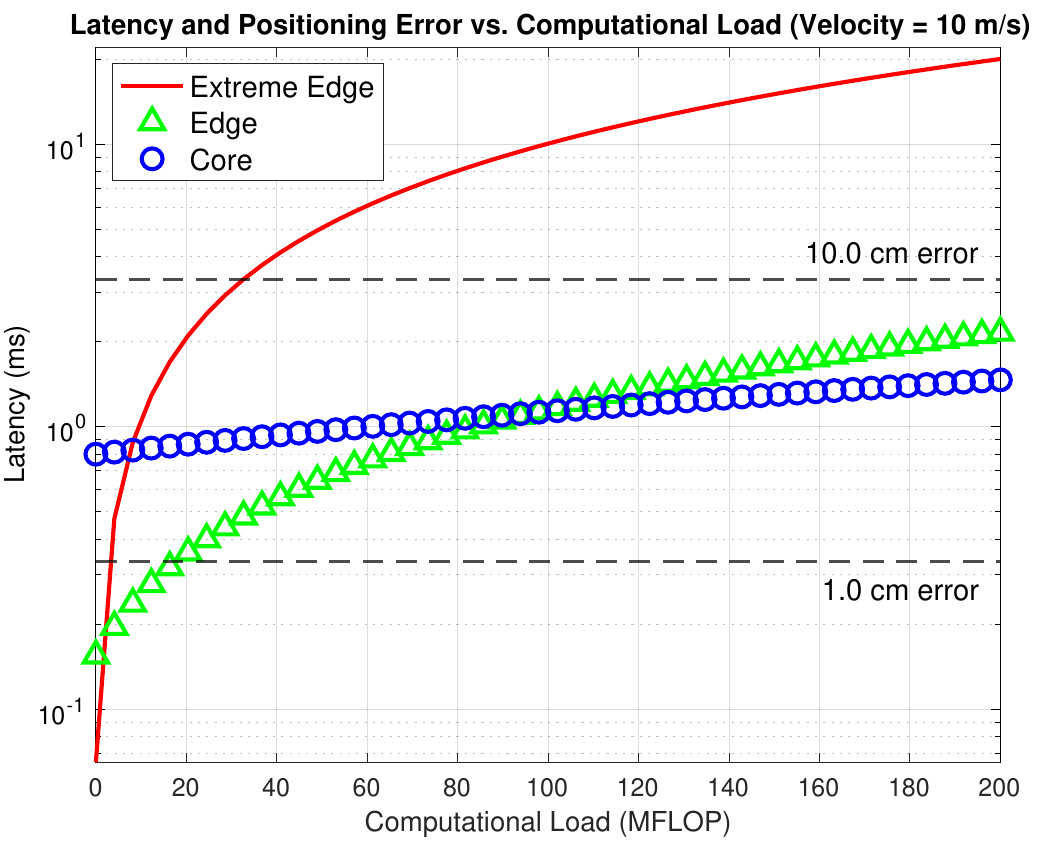}
    \caption{Simulation of the effect of higher-layer computational load distribution on latency and position error at a velocity of 10 m/s.}
    \label{fig:higherLayer}
\end{figure}

Results shown in Fig.~\ref{fig:higherLayer} illustrates the efficacy of utilizing various processing node locations under different computational loads (expressed in \acp{FLOP}) to minimize latency. For instance, serving a single sensing request with computational load below 4 MFLOP should be handled locally or at an extreme edge node, as they have the lowest propagation and communication delays. However, as the load increases, due to increase in number of requests, targets, algorithmic complexity, etc., the sensing process should be offloaded to nodes at the edge, as they cause relatively low propagation delay compared to high computation latency reduction gained from their increased computational power. Lastly, in situations were extreme volume of sensing requests/computational load is encountered (around the 100 MFLOP mark), utilizing the core network's high computational power will offset its relatively large propagation delay and will lead to lower latencies compared to its edge and extreme edge counterparts.

\subsubsection{\ac{PoC} Results: Gesture Recognition}
\begin{figure}[t!]
    \centering
    \includegraphics[width=0.99\linewidth]{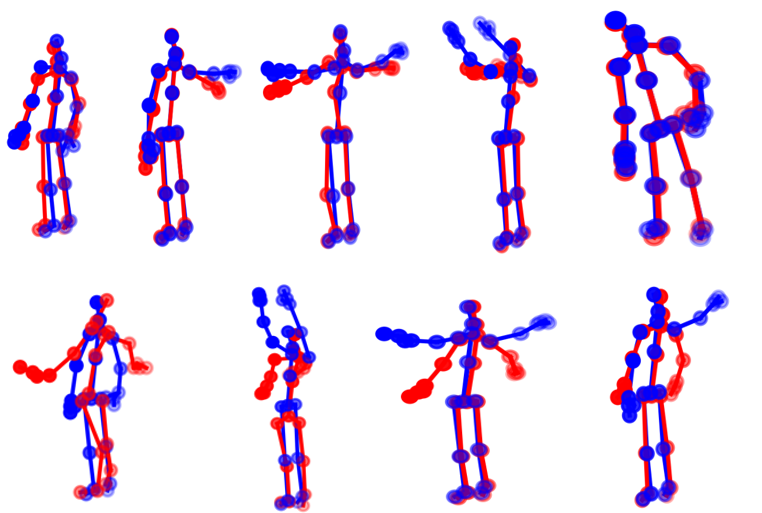}
    \caption{Predicting pose of a human with \ac{mmWave} ISAC: Blue colored skeletons represent ground truth, while the red ones are predicted using \ac{CNN} relying on ISAC signals.}
    \label{fig:skeltons}
\end{figure}

Fig.~\ref{fig:skeltons} shows the skeletons predicted in red, and ground truth in blue. It is evident, that in the first row, the two skeletons closely follow each other, meaning highly accurate pose estimation. We also compute the \ac{MPJPE} \cite{wang2021deep}, which represents the average Euclidean distance between the predicted 3D joint positions and the corresponding ground truth obtained from the Kinect. For the accurately predicted skeletons (see row 1 of Fig.~\ref{fig:skeltons}), the \ac{MPJPE} ranges between 3 and 6 centimeters. However, there are instances of pose errors (see row 2 of Fig.~\ref{fig:skeltons}), where the \ac{MPJPE} exceeds 16 centimeters. The average \ac{MPJPE} computed on the entire test set is 10 centimeters. Besides, we also calculate the percentage of correctly predicted key points (PCK) \cite{wang2021deep}. A key point or a joint is considered as correctly predicted if the distance between the predicted and the true joint is less than 0.2 times the torso diameter, referred to as PCK@0.2. The torso diameter is computed as the Euclidean distance from the left shoulder to the right hip. In this context, we achieve a PCK@0.2 of 77\%. These results highlight the potential of wireless signals to be translated into the visual domain through skeletal body pose estimation, while also indicating room for further improvement in accuracy and robustness.

\subsubsection{\ac{PoC} Results: Indoor Sensing}
\begin{figure*}[!t]
    \centering
    \subfloat[Percentile plots for human and reflector targets]{\includegraphics[width=0.32\textwidth]{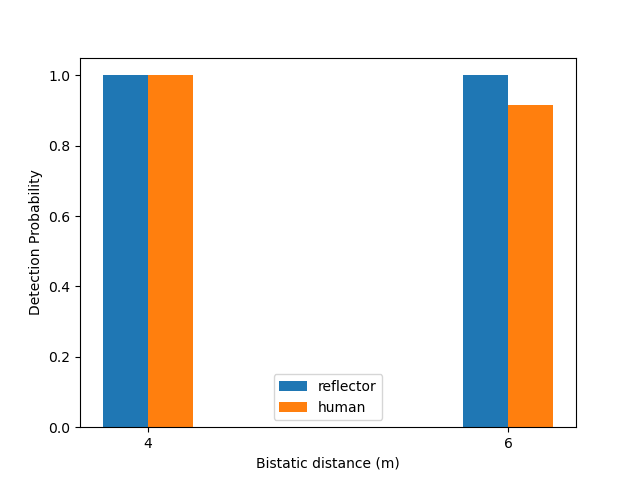}}%
    \label{fig:first}
    \hfil
    \subfloat[Histogram for human target at 4 meters bistatic distance]{\includegraphics[width=0.32\textwidth]{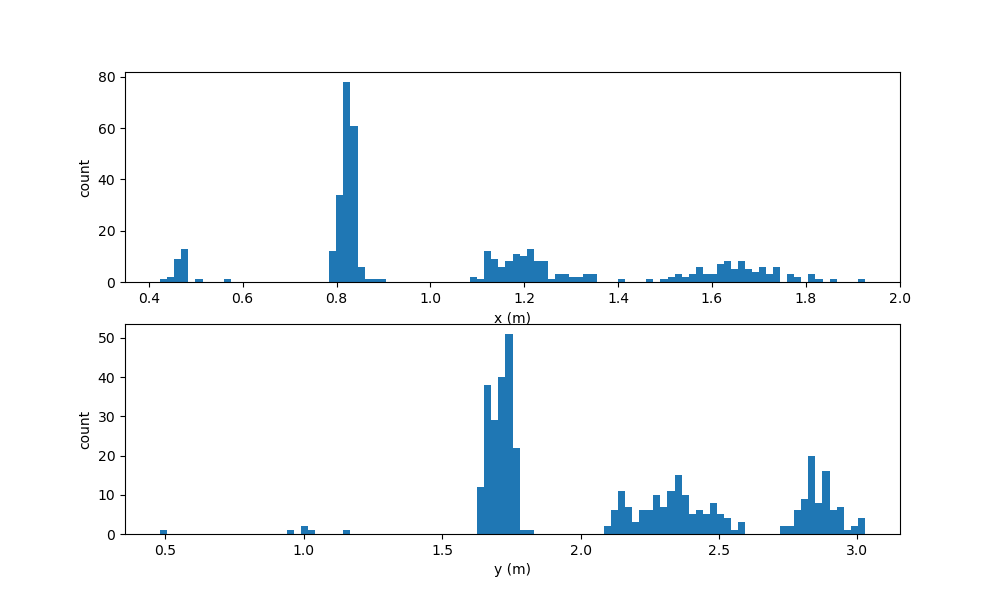}}%
    \label{fig:second}
    \hfil
    \subfloat[Histogram for human target at 6 meters bistatic distance]{\includegraphics[width=0.32\textwidth]{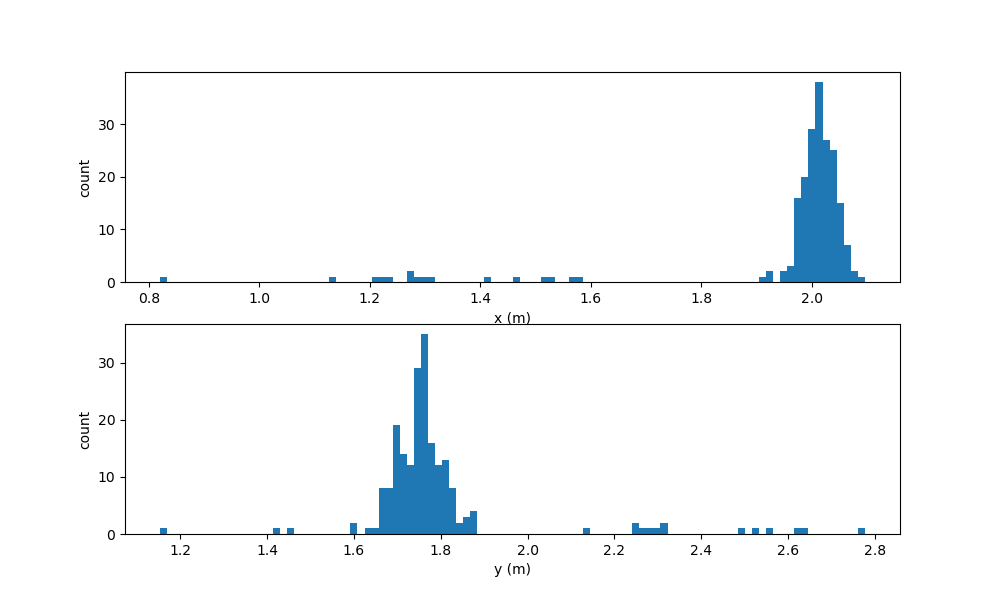}}%
    \label{fig:third}
    
    \vfil
    \subfloat[Histogram for reflector target at 4 meters bistatic distance]{\includegraphics[width=0.32\textwidth]{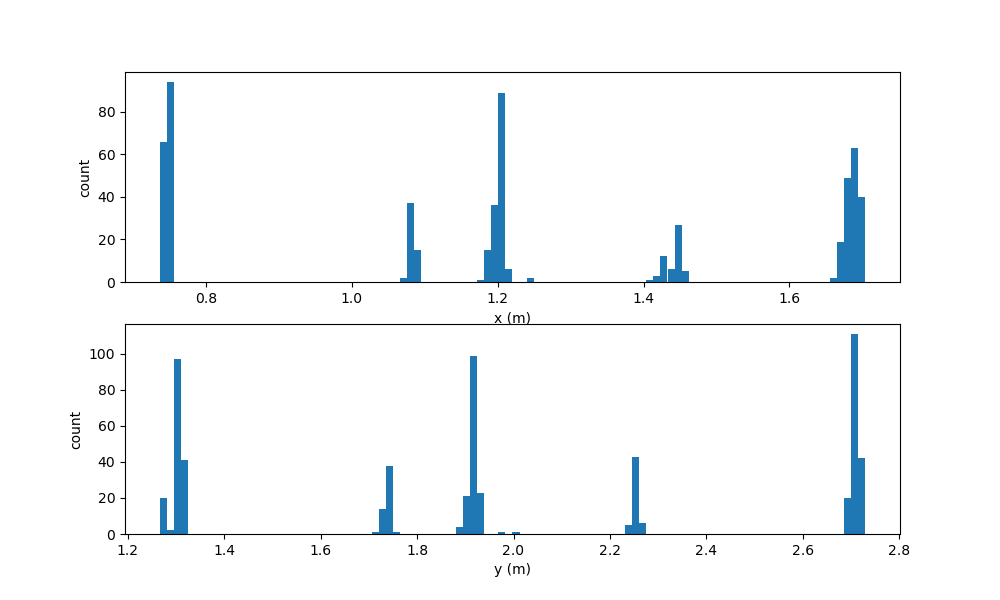}}%
    \label{fig:fourth}
    \hfil
    \subfloat[Histogram for reflector target at 6 meters bistatic distance]{\includegraphics[width=0.32\textwidth]{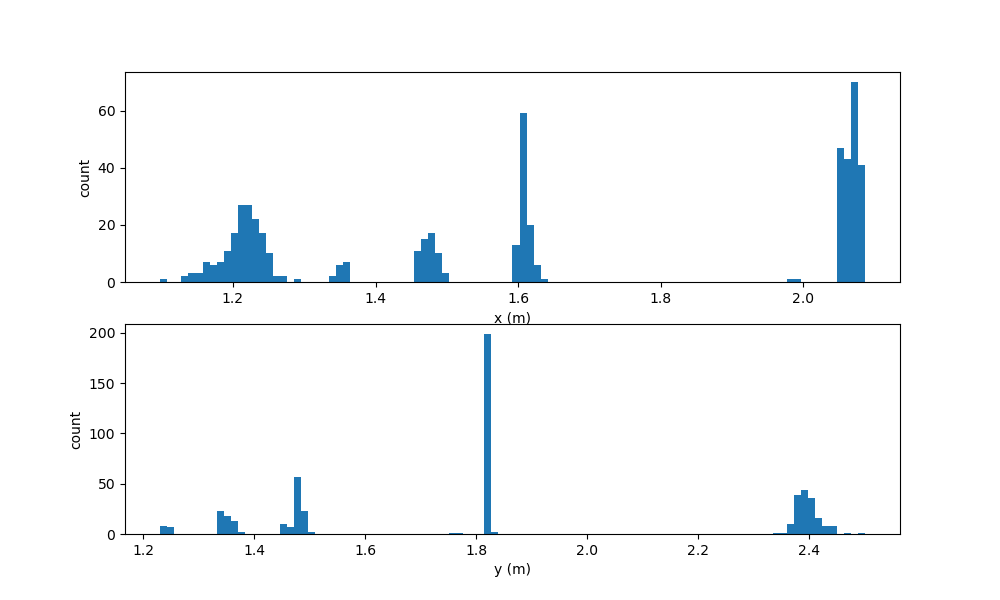}}%
    \label{fig:fifth}
    \hfil
    \subfloat[RF setup]{\includegraphics[width=0.32\textwidth]{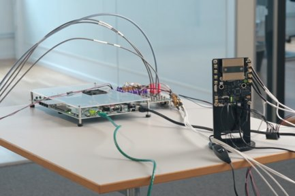}}%
    \label{fig:sixth}

    \caption{\ac{PoC} measurement results using the ISAC testbed. }
    \label{fig:all}
\end{figure*}

 Fig. \ref{fig:all}-(a) shows the percentiles for the reflector and human targets at bi-static distances of 4 and 6 meters. As observed, the detection rate exceeds the 90th percentile for human targets and approaches almost 100\% for the reflector. Additionally, Fig. \ref{fig:all}-(b–e) presents histogram plots of the x and y coordinates of target detections from one measurement frame, visualized in 2D post-processing. These plots still show some unwanted background reflections from the environment. This occurs because the presence of a target slightly alters the environmental reflection response, introducing additional artifacts. These artifacts are removed through post-processing methods applied in the final stage of the analysis. 

\section{Conclusion}\label{conclusion}
This paper presented a comprehensive cross-layer vision for \ac{ISAC} in 6G networks, developed through the collaborative efforts of the Hexa-X-II project. We highlighted ISAC’s potential to elevate wireless systems from data transport platforms to intelligent, context-aware infrastructures. We examined multiple levels of integration, ranging from spectrum sharing to full waveform and resource co-design, and emphasized how deeper integration enhances efficiency and enables novel functionalities, albeit with trade-offs in hardware complexity and system coordination. At the physical layer, we analyzed how key 6G enablers, including FR1–FR3, sub-THz, massive and distributed MIMO, non-terrestrial networks, and reconfigurable intelligent surfaces—support ISAC through improved resolution, coverage, and mobility handling. The discussion extended to hardware challenges such as synchronization, full-duplex operation, and calibration, as well as to higher-layer aspects including protocols, APIs, and security mechanisms necessary for scalable deployment. A cross-layer evaluation framework was proposed to link design parameters with ISAC-specific KPIs and KVIs, enabling system-level optimization.

ISAC is a transformative paradigm for 6G, promising to revolutionize wireless networks through unprecedented integration of sensing and communication capabilities. This unified perspective should serve as a foundational input to ongoing standardization efforts in bodies such as ETSI and 3GPP, guiding future research and development toward practical, high-impact implementations of ISAC technology.

\rev{\subsection*{Future Directions}
As ISAC matures toward commercial 6G deployment, several technical and theoretical challenges remain open. Below, we outline future research directions corresponding to the key topics covered in this paper.
\begin{itemize}
    \item \textbf{ISAC for emerging use cases:}
While this paper has presented key integration levels and use cases (e.g., digital twins, UAV safety, and automotive applications), open questions remain. Future work should develop  KPI/KVI requirements for emerging use cases, including non-human-centric applications (e.g., infrastructure monitoring), and explore economic and regulatory implications of embedding sensing into commercial networks.
\item  \textbf{Physical layer and channel modeling:}
At the the physical layer level, major challenges include accurate modeling of cluttered environments, near-field effects, and mobility-induced Doppler shifts in multistatic, RIS-assisted, and NTN scenarios. More work is needed to extend 3GPP models with realistic target and background clutter channels, capturing, e.g.,  near-field, and extended object effect. 
\item  \textbf{Hardware and synchronization:}
Full-duplex operation, phase coherence across distributed units, and calibration under mobility and hardware impairments remain open hardware bottlenecks. Research should explore low-cost, scalable synchronization schemes for distributed ISAC systems, adaptive calibration strategies using AI/ML to correct hardware impairments in real-time, energy-efficient full-duplex designs with integrated signal leakage suppression, and phase-coherent sensing  for non-isotropic scatterers. 
\item  \textbf{Higher-layer protocols and architectures:}
At the protocol level, research is needed to develop standardized sensing APIs and exposure frameworks with fine-grained privacy control, incorporate sensing functions (e.g., tracking, environment classification) as native 6G services, and design network-level sensing orchestration (e.g., sensing management functions) that coordinate across base stations and spectrum bands.
\item  \textbf{Cross-layer optimization and resource management:}
Future studies should generalize the proposed KPI/KVI framework into optimization schemes that span across layers. Particular areas of interest include multi-objective optimization balancing sensing fidelity and communication throughput,  joint scheduling of pilots and waveform resources under reliability constraints, adaptive cross-layer control loops that react to dynamic scene or traffic changes, and adaptively compress under reduced backhaul capacity. 
\item \textbf{AI and digital twins for ISAC:}
There is a strong opportunity to exploit AI and digital twins to enable robust and adaptive ISAC deployment. Key directions include training data generation using RF digital twins based on site-specific simulations or measurements in realistic environments; online adaptation of AI models in cluttered, dynamic, and non-stationary settings; and the integration of model-based and data-driven approaches to improve robustness, generalization, and explainability. Furthermore, privacy-preserving AI methods, such as federated learning, will be essential for enabling collaborative sensing across multiple operators or domains without compromising sensitive data. A promising avenue for future research lies in multi-modal sensing integration with large AI models, where radio sensing is fused with complementary modalities (e.g., vision) to enhance situational awareness, improve resilience in degraded environments, and reduce ambiguity in target interpretation.
\item \textbf{Open data sets:} 
A  bottleneck in the advancement of ISAC research is the limited availability of standardized, open-access datasets. Currently, most studies rely on proprietary or locally collected datasets that are not publicly released, which hinders reproducibility and impedes fair comparison across methods. In contrast, fields such as computer vision have benefited immensely from shared benchmarks and a centralized leaderboard to track and compare algorithmic performance. Establishing  open datasets for ISAC, ideally spanning various environments, sensing configurations, and frequency bands, combined with open-source code policies would support under-resourced research communities and enable consistent performance evaluation.
\end{itemize}
}

\balance 
\bibliographystyle{IEEEtran}
\bibliography{references}

\end{document}